\documentstyle[12pt]{article}
\def\Dslash{\not\hspace{-0.7ex}D_q\hspace{0.7ex}}

\begin{document}

\rightline{$ \begin{array}{l}
           \mbox{FTUV 94-21 / IFIC 94-19} \\
           \mbox{April  1994, revised April 1995} \\
           \mbox{hep-th 9405161}
            \end{array} $ }
\vspace{2\baselineskip}

\begin{center}
{\LARGE{\bf Quantum groups and deformed special relativity}}
\vspace{0,3cm}
\end{center}
\vspace{1\baselineskip}
\begin{center}
{\large{J.A. de Azc\'{a}rraga$^{*}$, P.P. Kulish\footnote{On
leave of absence from the St.Petersburg's Branch of the Steklov Mathematical
Institute of the Russian Academy of Sciences.}$^{*}$
and F. Rodenas$^{* \dag}$}}
\end{center}
\vspace{0,5cm}
\addtocounter{footnote}{-1}

\begin{center}
{\small{\it $* \quad$ Departamento de F\'{\i}sica Te\'{o}rica and IFIC,\\
Centro Mixto Universidad de Valencia-CSIC\\
E-46100-Burjassot (Valencia), Spain.}}
\end{center}

\begin{center}
{\small{\it $ \dag \quad$ Departamento de Matem\'atica Aplicada,\\
Universidad Polit\'ecnica de Valencia \\
E-46071 Valencia, Spain.}}
\end{center}
\vspace{1cm}

\begin{abstract}

   The structure and properties of possible $q$-Minkowski spaces are
reviewed and the corresponding non-commutative differential calculi
are developed in detail and compared with already existing proposals.
This is done by stressing the covariance properties of these algebras
with respect to the corresponding $q$-deformed Lorentz groups as described by
appropriate reflection equations. This allow us to give an unified
treatment for different $q$-Minkowski algebras.
Some isomorphisms among the space-time
and   derivative algebras are demonstrated, and their representations
are described briefly. Finally, some physical consequences and open
problems are discussed.
\end{abstract}
\newpage

\section{ Introduction}

\indent
The question of the quantization of space has been discussed by physicists from
the very early days of quantum theory. Due to the recent emergence
of a far-reaching generalization  of Lie groups and Lie algebras
\cite{DRINF,JIM,FRT1,WOR1}, known under the
name of quantum groups, it is tempting to
introduce suitable quantum Lorentz $L_q$ and  Poincar\'e $P_q$ groups
to arrive  to the quantum Minkowski space-time
${\cal M}_q$ by extending the classical
construction  ${\cal M} \sim P/L$ to the quantum case.
This program, however, is not completely straightforward. A rigorous
(unique) definition of quantum groups was given \cite{DRINF,JIM,FRT1} only
for the simple
Lie groups and algebras; for inhomogeneous groups
many problems appear. Thinking of the well known classical homomorphism
$SL (2, C) / Z_{2} \approx L$ it was proposed \cite{POWO} to define a quantum
Lorentz group  by using the simplest quantum group,
$SL_{q} (2)$, which is a $q$-deformed analog of the classical $(q = 1)$
commutative algebra of functions on the Lie group $SL (2, C)$. A
quantum Minkowski space ${\cal M}_{q}$ was then introduced by
means of a quadratic
combination of $q$-spinors transforming homogeneously under
the quantum group $SL_{q} (2)$ \cite{WA-ZPC48,WA-A6,SWZ-ZPC52,OSWZ-CMP,SONG}
(the precise definitions of $L_q$ and ${\cal M}_q$ will be given below).

Let us start by writing down some simple algebraic relations to
 introduce some quantum group aspects relevant for our discussion.
The essential feature in the field of {\it  quantum groups} (we shall omit
for a while their dual objects or {\it quantum algebras}, which may
look more familiar
for physicists) is in some sense similar to the relation between classical
and quantum mechanics, where the commutative algebra of functions on phase
space (the algebra of observables) becomes non-commutative after quantization.
This justifies the expression  `quantum' groups, although in
physics the adjective
quantum is reserved for situations in which Planck's constant $\hbar \neq 0$
and, in contrast,  group `quantization' (or {\it deformation}, $q \neq 1$)
does not imply   $\hbar \neq 0$.
In the case of Lie groups, the commutative algebra of functions on the
group manifold is replaced by a non-commutative algebra after
quantization (or $q$-deformation);
in particular, the matrix elements generating the group algebra become
non-commutative.

Let us recall the case of $SL_{q} (2)$
which will be extensively used
below. The quantum group $GL_{q} (2)$ is defined as the associative
{\it algebra} (quantum groups are not really group manifolds)
generated by four elements $a, b, c, d$ satisfying the homogeneous
quadratic relations ($\lambda \equiv q-q^{-1}$, $q \neq 0$)

\begin{equation}\label{ua}
\begin{array}{lll}
ab = qba\;, \quad & bd = qdb \;,\quad & bc = cb \;,\\
ac = qca \;,\quad & cd = qdc \;, \quad & [a,d] = \lambda bc \;.
\end{array}
\end{equation}

\noindent
 The relations  (\ref{ua}), however, include
four generators, while the `classical' $SL (2,C)$ depends on three complex
parameters.
To obtain $SL_q(2)$ one notices that the element

\begin{equation}\label{ub}
\begin{array}{ll}
ad - qbc = da - q^{-1} bc : = det_{q} T, & T =
\left( \begin{array}{cc}
a & b\\
c & d
\end{array} \right)
\end{array}
\end{equation}

\noindent
is a central (commuting) element of the algebra
which defines the $q$-determinant
of the matrix  $T$; then,  the addition of the
constraint $det_{q} T = 1$ to eqs.
(\ref{ua})  consistently reduces the number of generators to three.
If the entries of $T$ satisfy (\ref{ua}), those of $T^n$
satisfy analogous relations with $q$ replaced by $q^n$
(this product should not be confused with the comultiplication,
which preserves (\ref{ua}) \cite{VZW,DAVID}).
In the `classical' limit $q$$=$$1$, (\ref{ua}) just expresses that the algebra
generated
by the elements of $SL(2,C)$ is commutative, and (\ref{ub}) is the usual
determinant.

It is not apparent at first sight why eqs. (\ref{ua})  plus
$det_{q} T$=$1$ define $SL_{q} (2)$, nor how to generalize
them to the $SL_{q} (n)$ case.
This becomes clearer using the $R$-matrix formalism
\cite{FRT1} developed in the
framework  of the quantum inverse scattering method \cite{FADDEEV}.
Indeed, eqs. (\ref{ua}) may be rewritten as `RTT' relations,

\begin{equation}\label{uc}
R_{12} T_{1} T_{2} = T_{2} T_{1} R_{12}
\end{equation}

\noindent
where $T_{1} = T \otimes I$, $T_{2} = I \otimes T$ (see Appendix A for
notation)  and the $ 4
\times 4$ $c$-number matrix $R_{12}$ is given in (\ref{af2}).
In this way, they may
be generalized to $SL_{q} (n)$ by means of the appropriate $n^{2} \times
n^{2}$ $\, R$-matrix \cite{FRT1}.
One could insert any matrix into (\ref{uc}) as $R$. However, the natural
ones (as (\ref{af2})) satisfy
the Yang-Baxter equation  (YBE)
\begin{equation}\label{YBE}
R_{12}R_{13}R_{23}=R_{23}R_{13}R_{12}
\end{equation}
\noindent
which ensures the consistency of (\ref{uc}) (see the end of Appendix A1).
This means that no further relations for the
generators higher than the quadratic relations (\ref{ua})
may be derived from (\ref{uc}) and the requirement of associativity
for the algebra. The latter is postulated from the very beginning and is
independent of (\ref{YBE}).

Since quantum groups are very close to the algebra of functions on a Lie
group, we may expect them to have other characteristics pertaining to the
group multiplication rule, inverse (antipode) and unit elements, etc. In fact,
 they may be characterized as Hopf algebras
\cite{DRINF,JIM,FRT1} (see \cite{TAK,PC,MAJID,DOE}
for a review). Here we shall just underline
some properties which will be relevant for the $q$-Minkowski spaces below.
It is possible to introduce a quantum  or $q$-plane $C_q^{2}$
as an associative {\it algebra} generated by two  elements $x,y$ with
the commutation property
\begin{equation}\label{ud}
 xy = qyx \quad
\end{equation}
\noindent
(notice that a $q$-plane is not a manifold).
If we now define a two-component $q$-vector $X=(x,y)$,
 the commutation properties of the components  of $X' = TX$
satisfy again (\ref{ud})
since it is assumed that the entries of $T$ and  of $X$ commute among
each other.  This compatibility, not evident a priori, permits to look at
quantum groups in general as symmetries of {\it quantum spaces} \cite{MANIN,WZ}
(see also \cite{DAVID}).
In the general case of $SL_{q} (n)$, the analogue
relations for an $n$-component $X$ may be extracted from the $R$-matrix
relation $R_{12}X_1X_2=qX_2X_1$ (for instance, for $C^2_q$,
$X_1X_2=(xx,xy,yx,yy)$ and $X_2X_1=(xx,yx,xy,yy)$). The important property
of this equation is that it is preserved under the coaction $X'=TX$ due to
(\ref{uc}):
$R_{12}X'_1X'_2 = R_{12}T_1T_2X_1X_2=
T_2T_1R_{12}X_1X_2=qT_2T_1X_2X_1=qX'_2X'_1$.

Following the lead provided by the invariance of the above commutation
properties defining
the quantum plane,  we  may now extend this to
a more elaborated situation in which the
generators of a quantum space are put in matrix form $K$ and the action
$\phi$ is given by $\phi : K \mapsto K' = TKT^{-1}$,
where the elements of $T$ commute with
the entries of $K$, $[K_{ij}, T_{mn}] = 0$ ($i,j,m,n $$=$$ 1,2$). A
possibility to organize the commutation properties which define the
algebra generated by the elements $K_{ij}$ is to write
them  \cite{K-SKL,KS,K2}  in the form
of a reflection equation\footnote{The name is due to the fact that
an equation with such a form appeared originally  in the
factorizable scattering theory on the half line  with a  reflection
from a boundary \cite{CHERED} (see \cite{KS}). Originally
the reflection equation was written
in a `spectral parameter'-dependent form \cite{CHERED,SKL}
(see also \cite{LN}); here we shall consider only
constant solutions.} (RE)

\begin{equation}\label{dosj}
R_{12} K_{1} R_{21} K_{2} = K_{2} R_{12} K_{1} R_{21}\quad .
\end{equation}

\noindent
This equation was introduced independently  in
\cite{MAJ-LNM}  in the context  of braided algebras with all $R$-matrices
on one side, and in the form above  in \cite{SMBR2}. It was also discussed
in a general algebraic context in \cite{FM}.

Due to eq. (\ref{uc}) (and its consequence
$R_{21} T_{2} T_{1} = T_{1} T_{2}
R_{21}$) it is not difficult to see that $K' =
TKT^{-1}$ satisfies the same equation (\ref{dosj}). In fact, this
formalism is a convenient
framework to discuss the invariance of  commutation
relations under a certain action  $\phi$.
However, it should be mentioned at this stage
that the non-commutative character of the algebra generated by $K_{ij}$
puts forward some questions as to its physical interpretation. If we consider
the elements of $K$ as non-commutative generators
of the algebra, one has to find their
irreducible representations ${\cal H}_{K}$;
there could be more than one. Also, the non-commutative transformation
coefficients (the generators of the quantum group algebra) have their own
irreducible representation ${\cal H}_{T}$. Hence, after the quantum group
transformation, the new entries of $K' = TKT^{-1}$ act,
generally speaking, as operators
 on a larger space ${\cal H}_{T} \otimes {\cal H}_{K}$: the quantum group
{\it coacts} rather than acts.
This is a rather unusual situation for the symmetries of physical
systems, where the action of a symmetry on the space-time coordinates,
say, does not entail an enlargement of the corresponding algebra
or of the space of physical states (for an alternative approach
see \cite{MSC,RSM}). We will
not develop this important and interesting question here, which we
shall bypass by stressing the isomorphism among the algebras generated
by $K_{ij}$ and $K'_{ij}$; instead, we shall restrict ourselves to the
technical problems of constructing quantum Minkowski space-time algebras
generated by elements transforming covariantly
under the corresponding quantum Lorentz transformations.

Many of the various $q$-Minkowski formulae
presented in this paper were found previously (see, in particular,
\cite{POWO,WA-ZPC48,WA-A6,SWZ-ZPC52,OSWZ-CMP,SONG,PSW,PILLIN,SCHI-MPI93}),
but here they will be obtained in a more  systematic way due to the
flexibility of the $R$-matrix and the reflection equation formalism.
In particular, our approach is particularly suited
to establish possible algebra isomorphisms (Appendix B2) and to
incorporate other  $q$-Minkowski space proposals in an unified manner,
as well as for
discussing the non-commutative differential calculus on other quantum spaces
as, for instance, with $SO_q(3)$ symmetry (see in this respect
\cite{WEICH,WA-ZPC49}).

The plan of the paper is as follows. First, the procedure for defining
a quantum Minkowski space-time algebra ${\cal M}_q$ is given in Sec.2.
Its first part,where the formalism is explained and  some special elements
of the algebra ({\it e.g.}, the $q$-Minkowski length) are introduced,
is devoted to the ${\cal M}_q^{(1)}$
\cite{WA-ZPC48,SWZ-ZPC52,OSWZ-CMP,SONG} case.
Other possibilities ${\cal M}_q^{(i)}$
are  discussed   later in
Sec.2. In Sec.3, arguments of  covariance and consistency
establish the commutation relations among $q$-`coordinates' (generators of
${\cal M}_q$),
$q$-derivatives (id. of ${\cal D}_q$) and $q$-one-forms (id. of $\Lambda_q$).
It is also shown there that in the non-commutative case the existence of
a simultaneous linear hermitean structure (reality conditions)
for coordinates and derivatives is not
guaranteed for all $q$-Minkowski spaces,
recovering previous results \cite{OSWZ-CMP,OZ-MPI25}.
In Sec.4, a convenient set of generators for ${\cal M}_q^{(1)}$
is picked up by taking into account  their $q$-tensor properties with
respect to the $q$-Lorentz transformations.
Sec.5 is devoted to developing the non-commutative
differential calculus for ${\cal M}_q^{(1)}$  \cite{OSWZ-CMP}; the other cases
may be treated similarly.
A number of elaborated questions concerning the mutual interrelations
between quantum Poincar\'e group $P_q$ and algebra ${\cal P}_q$,
the representation theory and the physical interpretation are briefly
considered in Sec.6, where in particular
the mass and momenta spectrum will be calculated.
{}From the discussion it looks that
it may not be easy to reconcile the notion of a quantum Minkowski space
with the standard treatment of the special relativity or relativistic
quantum theory (see also \cite{AKRREL}).
These points, as well as a classical counterpart of
${\cal M}_q$ are discussed shortly at the end.

Dealing with non-commutative and Hopf algebras
to define  $q$-Minkowski spaces requires using a number of
results and explicit expressions from  quantum group theory
\cite{DRINF,JIM,FRT1}. To facilitate the reading of the paper, some
useful facts and formulae are collected in Appendix A; Appendix B
contains the proof of certain important algebra isomorphisms
and algebraic properties for the ${\cal M}_q^{(1)}$ case in   the main text and
other general relations.

There are, certainly, other approaches to define
deformations of kinematical groups and algebras and their realizations
(see, e.g., \cite{POWO,CESAR,DOB,CAR,PUSZ,CAST,SMPoinc,BHOS,PWOR}
and references therein).
 In particular, a widely used
approach  utilizes the contraction procedure \cite{FIR} to obtain the
$\kappa$-Poincar\'e algebra \cite{LNRT,RUEGG}.
The  approaches to $\kappa$-relativity (see \cite{LRZ} and references
therein; cf. \cite{MARU})  based on the
${\kappa}$-Poincar\'e algebra, and the $q$-Minkowski spaces ${\cal M}_q^{(i)}$
to be discussed here  seem, however,  unrelated. In particular, the
deformation parameter $q$ is dimensionless while $1/\kappa$ introduces a
fundamental unit of length. This is an essential difference, and for this
reason, the case of $\kappa$-Minkowski space ${\cal M}_{\kappa}$ \cite{MARU}
and its associated differential calculus \cite{SITARZ}, as well as the
$\kappa$-Newtonian spacetimes  \cite{AZPB}, will not be discussed here.

\setcounter{equation}{0}

\section{Deformed Minkowski space-time algebras}

\indent
The group of transformations preserving the Minkowski metric
$diag (1,$$-1,$ \, $-1,$$-1)$ is the Lorentz group
(we shall consider here the restricted Lorentz group $L$).
Its universal covering
group $SL(2, C)$  possesses two fundamental representations,
$D^{\frac{1}{2}, 0}$ and $D^{0, \frac{1}{2}}$, realized by
2$\times$2 matrices $A$ and $\tilde{A}
= (A^{-1})^{\dagger}$ which act on undotted and dotted spinors respectively,
\begin{equation}\label{dosa}
\xi_{\alpha}' = A_{\alpha}^{\,\;\beta} \xi_{\beta} \quad, \qquad \quad
\bar{\xi}'^{\dot{\alpha}}= \tilde{A}^{\dot{\alpha}} _{\;\, \dot{\beta}}
\bar{\xi}^{\dot{\beta}}\quad.
\end{equation}

\noindent
The vector representation
$(D^{\frac{1}{2}, \frac{1}{2}} =
D^{\frac{1}{2},0} \otimes D^{0, \frac{1}{2}})$
on space-time  coordinates  may be obtained through
\begin{equation}\label{dosb}
K' = A K \tilde{A}^{-1} = A K A^{\dagger} \quad , \quad K=K^{\dagger} \quad.
\end{equation}

\noindent
Writing $K = \sigma^{\mu} x_{\mu}$
$(\sigma^{\mu}=
(\sigma^{0}, \sigma^{i})$; $\sigma^{0} = {\bf I}$
and $\sigma^{i}$ are the Pauli
matrices), $det K = (x^{0})^{2} - \vec{x}^{2}$.
Clearly, $K' =
\sigma^{\mu} x_{\mu}'$ with $x'^{\mu} = \Lambda^{\mu} _{\,\,\nu} x^{\nu}$,
and the
correspondence $\pm A \mapsto \Lambda$$\in$$L$ realizes the covering
homomorphism $SL(2,C)/Z_{2} = SO(1,3)$. The antisymmetric matrix $\epsilon =
i \sigma^{2}$ satisfies the property $\epsilon A \epsilon^{-1} = (A^{-1})^{t}$;
hence $\epsilon A^{*} \epsilon^{-1} = (A^{-1})^{\dagger}$ and
\begin{equation}\label{dosc}
K'^{\epsilon} = \tilde{A} K^{\epsilon} A^{-1} \quad , \quad K^{\epsilon}
\equiv \epsilon K^{*} \epsilon^{-1} \;.
\end{equation}

\noindent
Clearly, since $\epsilon (\sigma^{\mu})^{*} \epsilon^{-1}= (\sigma^{0}, -
\sigma^{i}) \equiv \rho^{\mu}$, another four vector of hermitean
matrices (not related by a similarity transformation to $\sigma^{\mu}$) may be
introduced. Including the appropriate indices, eqs. (\ref{dosb}) and
(\ref{dosc}) read
\begin{equation}\label{dosd}
K'_{\alpha \dot{\delta}} = A_{\alpha} ^{\,\,\beta} K_{\beta \dot{\gamma}}
(\tilde{A}^{-1})^{\dot{\gamma}} _{\,\,\dot{\delta}} \quad ,\quad \quad
K'^{\epsilon \; \dot{\alpha} \delta}
= \tilde{A}^{\dot{\alpha}} _{\,\,\dot{\beta}} K^{\epsilon \; \dot{\beta}
\gamma}
(A^{-1})_{\gamma}^{\,\,\delta}\,,
\end{equation}

\noindent
and $\sigma^{\mu}$ and $\rho^{\mu}$ read $(\sigma^{\mu})_{\alpha \dot{\beta}}$
, $(\rho^{\mu})^{\dot{\alpha} \beta}  $ ; they satisfy

\begin{equation}\label{dosdb}
\frac{1}{2} tr (\rho_{\mu} \sigma_{\nu}) = g_{\mu \nu}\,,
\end{equation}

\noindent
and the Lorentz matrix given by $\pm A$ is

\begin{equation}\label{dose}
\Lambda ^{\mu}_{\,\,\nu} =
\frac{1}{2} tr (\rho^{\mu} A \sigma_{\nu} A^{\dagger})
\,.
\end{equation}

\noindent
If $K$ is now restricted to be traceless, this condition will be preserved
iff $U= (U^{-1})^{\dagger}$ i.e., by the $SU(2)$ subgroup. Then, the
homomorphism $SU(2)/Z_{2} = SO(3)$ is realized by $R^{i} _{\,\,j} = \frac{1}{2}
tr (\sigma^{i} U \sigma_{j} U^{\dagger})$.
As mentioned in the introduction, we wish to explore in this paper the
consequences of $q$-deforming the above relations.

The crucial idea \cite{POWO,WA-ZPC48,WA-A6,SWZ-ZPC52,OSWZ-CMP}
to deform the Lorentz group was to replace the
$SL(2,C)$ matrices $A$ in (\ref{dosa}) by
the generator matrix $M$ of the quantum
group $SL_{q}(2)$. Due to the fact that the hermitean conjugation
$(M^{\dagger} _{ij} = M_{ji} ^{*})$ includes the $*$-operation, an extra copy
$\tilde{M}$ of $SL_{q}(2)$ was introduced, with entries not commuting with
those of $M$. The $R$-matrix form of the commutation relations among the
quantum group generators $(a, b, c, d)$ of $M$ and $(\tilde{a}, \tilde{b},
\tilde{c}, \tilde{d})$ of $\tilde{M}$ are then expressed by
$$
R_{12} M_{1} M_{2} = M_{2} M_{1} R_{12}\;,
$$
\begin{equation}\label{dosf}
 R_{12} \tilde{M}_{1} \tilde{M}_{2} = \tilde{M}_{2} \tilde{M}_{1} R_{12}\;,
\end{equation}
$$
R_{12} M_{1} \tilde{M}_{2} = \tilde{M}_{2} M_{1} R_{12}\;.
$$
\noindent
As in the classical
$(q=1)$ case, the reality condition is expressed \cite{WA-ZPC48,SWZ-ZPC52}
by requiring $\tilde{M}^{-1}
=M^{\dagger}$, which implies\footnote{In terms of Hopf algebras, eqs.
(\ref{dosf}) and (\ref{dosg}) define the complex $*$-Hopf $SL_q(2,C)$
algebra, which
is generated by  two copies $M$ and $\tilde{M}$ of $SL_q(2)$ satisfying
(\ref{dosf}) subjected to the involution $M \mapsto M^{\dagger}=S(\tilde{M})$
where $S$ denotes the antipode (inverse).}
\begin{equation}\label{dosg}
\left(
\begin{array}{ll}
\quad \tilde{d} & -\tilde{b}/q\\
- q \tilde{c} & \quad \tilde{a}
\end{array}
\right)
=
\left(
\begin{array}{ll}
a^{*} & c^{*}\\
b^{*} & d^{*}
\end{array}
\right)\;.
\end{equation}

\noindent
This condition is consistent with all relations
in (\ref{dosf}) provided that the
deformation parameter $q$ is real, $q \in {\bf R}$. The
set of  generators $(a, b, c,
d, \tilde{a}, \tilde{b}, \tilde{c}, \tilde{d})$ satisfying
$det_{q} M$=1=$det_{q} \tilde{M}$, the commutation relations (\ref{dosf}) and
the conditions
(\ref{dosg}) define the quantum Lorentz group\footnote{
Again in terms of Hopf algebras, $ L_{q}^{(1)}$ is defined as the $*$-sub-Hopf
algebra of $SL_q(2,C)$ generated by the elements
$M_{ij} \tilde{M}^{-1}_{ls}$ (see below for
the expression of the $q$-Lorentz matrix $\Lambda$)
with the commutation relations and Hopf structure induced by the $SL_q(2,C)$
$*$-Hopf algebra.} $ L_{q}^{(1)}$.
Other $q$-Lorentz groups $ L_{q}^{(i)}$
exist \cite{WOZA} under rather mild requirements on
the deformation; in fact, there are
other  $M$ and $\tilde{M}$ commuting relations replacing eqs. (\ref{dosf})
if we allow for different
$R$ matrices in it. These, and their corresponding $q$-Minkowski
spaces ${\cal M}_{q}^{(i)}$, will be discussed at
the end of the section.

To introduce the $q$-Minkowski {\it algebra} ${\cal M}_{q}^{(1)}$
associated with $L_q^{(1)}$ it is natural to extend  (\ref{dosb})
to the quantum case (\ref{dosf}) by stating that $K$ generates
a comodule algebra for the coaction $\phi$ defined by

\begin{equation}\label{dosh}
\phi : K \longmapsto
K' = M K \tilde{M}^{-1} \;, \quad K'_{is} = M_{ij} \tilde{M}^{-1}_{ls}
K_{jl} \;, \quad \tilde{M}^{-1}=M^{\dagger} \;,
\end{equation}

\noindent
where it is assumed that the matrix elements of $K$ commute with those of $M$
and $\tilde{M}$ but not among themselves\footnote{A dotted and undotted index
notation may also be introduced in the deformed case. However, since
the dotted and
undotted indices are always located, as in (\ref{dosd}), to make
multiplication of
matrices  always possible, we shall only use latin indices from now on.}.
Much in the same way the commuting properties of $q$-two-vectors
(or, better said here, $q$-spinors) are preserved by the coaction of $M$
and $\tilde{M}$, we now demand that the commuting properties of the entries
of $K$  are preserved by
(\ref{dosh}). More specifically, in order to identify the elements of $K$
\begin{equation}\label{dosi}
K= \left(
\begin{array}{ll}
\alpha & \beta\\
\gamma & \delta
\end{array}
\right)
\end{equation}

\noindent
with the generators of the $q$-Minkowski algebra ${\cal M}_{q}^{(1)}$ we
require,
as in the classical case,

\noindent
a) \, a reality property preserved by (\ref{dosh}),

\noindent
b) \,  a (real) $q$-Minkowski length, defined through the
$q$-determinant $det_qK$ of  $\qquad \qquad \qquad K$,
invariant under the $q$-Lorentz transformation (\ref{dosh}),

\noindent
c) \,  a set of commutation relations for  the elements of $K$
(a `presentation' of the algebra) preserved by
(\ref{dosh}) for (\ref{dosf}).

The reality condition  $K= K^{\dagger}$ is consistent with (\ref{dosh})
since $\tilde{M}^{-1} = M^{\dagger}$ as in the classical case.
The $q$-determinant of $K$ and the $q$-Minkowski metric will be given
below but, using (\ref{dosf}), it is not difficult to check (it will be
discussed in generality later)  that the commutation properties of
the Minkowski algebra generators expressed by
(\ref{dosj}) \cite{MeyerM,AKR,SM2} or, equivalently by
\begin{equation}\label{dosk}
\hat{R} K_{1} \hat{R} K_{1} = K_{1} \hat{R} K_{1} \hat{R} \quad ,
\end{equation}

\noindent
where $\hat{R} \equiv {\cal P}R$ (eqs. (\ref{af}), (\ref{af2})),
are preserved by (\ref{dosh}) and are consistent with the condition
$K $$=$$ K^{\dagger}$ for the $R$-matrix chosen and $q \in R$. In
components, eq. (\ref{dosj}) reads

\begin{equation}\label{dosj2}
R_{ij,kl}K_{kf}R_{lf,gh}K_{gm}= K_{jd} R_{id,cs}K_{ct}R_{st,mh}\;.
\end{equation}

\noindent
For $K$ given by (\ref{dosi}), eq. (\ref{dosj}) is equivalent to
the six basic relations \cite{WA-ZPC48,SWZ-ZPC52,MAJ-LNM}

\begin{equation}\label{dosl}
\begin{array}{ll}
\alpha \beta = q^{-2} \beta \alpha \quad ,\qquad &
[\delta, \beta ] = q^{-1} \lambda \alpha \beta \quad ,\\

\alpha \gamma = q^{2} \gamma \alpha \quad, \qquad &
[\beta , \gamma] = q^{-1} \lambda (\delta - \alpha ) \alpha \quad ,\\

[\alpha , \delta ] = 0 \quad ,\qquad &
 [\gamma , \delta ]
= q^{-1} \lambda \gamma \alpha \quad ,
\end{array}
\end{equation}

\noindent
which characterize the algebra ${\cal M}_{q}^{(1)}$;
we shall  adopt the point of view that this `quantum space' (algebra)
is the primary object on which the non-commutative differential calculus
will be constructed.
Thus, eq. (\ref{dosj}) (or (\ref{dosk})) leads to the following first case:

\noindent
{\it Definition}  (${\cal M}_{q}^{(1)}$)

The quantum Minkowski space-time algebra ${\cal M}_{q}^{(1)}$ is
the associative algebra
generated by the four elements of $K$, subject to the
reality conditions $\alpha = \alpha^{*} \,,\, \delta = \delta^{*}
\,,\, \beta^{*} = \gamma \,,\, \gamma^{*} = \beta$, and satisfying the
commutation relations (\ref{dosj}) (or (\ref{dosl})).

The central (commuting) elements of ${\cal M}_{q}^{(1)}$ may be obtained by
using
the $q$-trace $tr_{q}$ \cite{FRT1,ZU-MPX} which, for  a $2 \times 2$ matrix
$B$ with elements commuting with those of $M$ (as it is the case of
$K$) is defined by

\begin{equation}\label{dosm}
tr_{q} B = tr (DB) = q^{-1} b_{11} + q b_{22} \quad , \qquad D = diag
(q^{-1}, q)\;.
\end{equation}

\noindent
The $q$-trace  is invariant under the coaction
$B \mapsto M B M^{-1}$,
\begin{equation}\label{qtrace}
tr_q(MBM^{-1})=tr_q(B)\;.
\end{equation}
This follows easily by using the preservation
of the $q$-symplectic metric $\epsilon^{q}$ \cite{VZW,DAVID}
(which replaces $i\sigma^{2}$ for
$q = 1$) by the $SL_{q}(2)$ matrices,
\begin{equation}\label{dosn}
M^{t} \epsilon^{q} M = \epsilon^{q} \quad , \quad \epsilon^{q} =
\left[
\begin{array}{cc}
0 & q^{- 1/2}\\
-q^{1/2} & 0
\end{array}
\right]=-( \epsilon^q)^{-1}\;,
\end{equation}

\noindent
since $D = \epsilon^{q} \epsilon^{q\,t} $;
we shall drop the superindex $q$ henceforth.
The matrix $D$ satisfies
\begin{equation}\label{dprop}
M^tD(M^{-1})^t=D \quad, \quad \tilde{M}^t D (\tilde{M}^{-1})^t=D \;.
\end{equation}
\noindent
In the general
$\;R$-matrix case $\;D\;$ may be expressed \, (cf. \, \cite{FRT1}) \,  as
$\;D\;$ $=$ \, $ q^2 tr_{(2)} \{ {\cal P} [(R_{12}^{t_{1}})^{-1}]^{t_{1}} \}$,
where the ordinary trace is taken in the second space and the transpositions
refer to the first space; the factor $q^2$ has been chosen to reproduce
(\ref{dosm}) for the $GL_q(2)$ $R$-matrix (\ref{af2}).

 The centrality of the $q$-trace
follows from (\ref{dosj}), which after left (right)
multiplication by $R_{12}^{-1}$ $ (R_{21}^{-1})$ and
a similarity transformation with  the permutation
operator ${\cal P}$ (eqs. (\ref{af}), (\ref{af2})) may be rewritten as
\begin{equation}\label{dosp}
K_{2} R_{12}  K_{1} R_{12}^{-1} = R_{21}^{-1} K_{1} R_{21} K_{2}\,.
\end{equation}

\noindent
Indeed, since $R_{12}$ and $R_{21}^{-1}$ provide representations of the
$SL_{q}(2)$ quantum group when considered as $2 \times 2$ matrices in the
first space of $C^{2} \times C^{2}$ (although not faithful: $b=0$ for $R_{12}$
and $c=0$ for $R_{21}^{-1}$),  taking the $q$-trace in the first space
it follows that
\begin{equation}\label{dosq}
K c_1 = c_1 K
\end{equation}
\noindent
with
\begin{equation}\label{dosr}
c_{1} \equiv tr_{q} K = tr_{q(1)}(R_{12} K_1 R_{12}^{-1})=
tr_{q(1)}(R_{21}^{-1} K_1
R_{21})= q^{-1} \alpha + q \delta .
\end{equation}

\noindent
By iterating the procedure which lead to (\ref{dosp}) it is found that
\cite{KS,SMBR2}
\begin{equation}\label{dosu}
K_{2} R_{12} K_{1}^{n} R_{12}^{-1} = R_{21}^{-1} K_{1}^{n} R_{21} K_{2}
\end{equation}

\noindent
and hence, after taking $tr_{q (1)}$ in this relation,
\begin{equation}\label{dosv}
K c_{n} = c_{n} K \quad , \quad c_{n} \equiv tr_{q} K^{n} \quad .
\end{equation}

\noindent
The first two central elements  $c_{1}$ and $c_{2}$ are algebraically
independent,
but the $c_{n}$ for $n > 2$ are polynomial functions of them due to the
characteristic equation for $K$\, \cite{K-SKL,K2},
\begin{equation}\label{dosw}
q K^{2} - c_{1} K + \frac{q}{[2]} (q^{-1} c^{2}_{1} - c_{2}) I= 0\,.
\end{equation}

The $q$-determinant $det_{q} K$ of $K$ is obtained by means of the
$q$-antisym\-metri\-zer $P_{-}$ [(\ref{ag2}), (\ref{ah})], which is a
rank one $\,4 \times 4$
projector. It is defined \cite{K-SKL} through
\begin{equation}\label{dosx}
(det_{q} K) P_{-} = -q P_{-} K_{1} \hat{R} K_{1} P_{-} = (\alpha \delta -
q^{2} \gamma \beta ) P_{-}\,,
\end{equation}

\noindent
although one of the projectors $P_{-}$ may be dropped since it is easy to
check that both $P_{-} K_{1} \hat{R} K_{1}$ and  $K_{1} \hat{R} K_{1} P_{-}$
are
proportional to $P_{-}$. It is  central since
\begin{equation}\label{dosy}
det_{q} K = \frac{q^2}{[2]} (q^{-1} c_{1}^{2} - c_{2}) \quad , \quad
\end{equation}

\noindent
and, provided it is not zero,
\begin{equation}\label{dosz}
K^{-1} = (det_{q} K)^{-1}
\left[
\begin{array}{cc}
q^{2} \delta - q \lambda \alpha & -q^{2} \beta\\
-q^{2} \gamma & \alpha
\end{array}
\right]
\quad \cdot
\end{equation}

\noindent
Since it may be expressed in terms of $q$-traces,
$ det_{q} K$ is obviously preserved under a
similarity transformation $K \mapsto M K M^{-1}$ where $M$ and
$M^{-1}$ belong to the same quantum group. But, despite of the fact that the
central elements $c_{n}$ are not invariant with respect the $q$-Lorentz
transformation (\ref{dosh}) because it involves $M$ and $\tilde{M}^{-1}$,
$det_{q} K$ is nevertheless preserved under this coaction.
Using (\ref{dosx}) we find
$$
det_{q} \phi (K) P_- = det_{q} (M K \tilde{M}^{-1}) P_-
= -q P_{-} M_{1} K_{1} M_{2} \hat{R} \tilde{M}_{2}^{-1} K_{1}
\tilde{M}_{1}^{-1} P_{-}
$$
$$
= -q (det_{q} M) P_{-} K_{1} \hat{R} K_{1} P_-
(det_{q} \tilde{M})^{-1}
= (det_{q} M) det_{q} K (det_{q} \tilde{M})^{-1} P_-\;,
$$

\noindent
where we have used $\tilde{M}_{1}^{-1} \hat{R} M_{1} = M_{2} \hat{R}
\tilde{M}_{2}^{-1}$ (from the last equality in (\ref{dosf})) as
well as the  definition (\ref{adet}) of $det_{q} M$. Since
$det_{q} M $=1=$ det_{q} \tilde{M}$,  $det_{q} \phi (K) = det_{q} K $,
and we may identify this real and central invariant element with the square
$l_q$ of the
{\it $q$-Minkowski invariant length} \cite{WA-ZPC48,SWZ-ZPC52,SONG,OSWZ-CMP}
\begin{equation}\label{dosaa}
l_{q} \equiv det_qK = \alpha \delta - q^{2} \gamma \beta \;,
\qquad l_{q} \in {\cal M}_{q}^{(1)}\,.
\end{equation}

\noindent
The $q$-trace $tr_{q} K = c_{1}$ is central but not invariant; it will be
later identified with the time coordinate. It is invariant under the
 $SU_{q}(2)$ `subgroup' as in the classical case since then $M$=$U$,
$\tilde{M}^{-1}$=$M^{\dagger}$=$U^{\dagger}$, the matrices $U$
satisfy the `unitarity' condition $U^{\dagger}$=$U^{-1}$
 and the $SU_q(2)$ coaction is defined by $K \mapsto UKU^{-1}$
(it will be seen in Sec. 5, however, that there is no consistent reduction
to $SU_q(2)$ in the whole ${\cal M}_q^{(1)} \times {\cal D}_q^{(1)}$
algebra). We shall
not need the explicit form of the 4$\times$4 and 3$\times$3 $q$-Lorentz
and $q$-rotation matrices; the interested reader may find them in
\cite{WA-ZPC48,SONG-JPA}.

Let us now find the expression for the $q$-Minkowski metric. First, we notice
that if we define $\hat{R}^{\epsilon}$ by

\begin{equation}\label{dosab}
\hat{R}_{ij,kl}^{\epsilon} = \epsilon^{t}_{js} \hat{R}_{is,kt}
(\epsilon^{t})^{-1}_{tl} \quad (\hat{R}^{\epsilon} = ({\bf 1}
\otimes \epsilon^{t})
\hat{R} ({\bf 1} \otimes (\epsilon^{-1})^{t}))\;,
\end{equation}

\noindent
it follows from the last eq. in (\ref{dosf}), written in the form
$\hat{R} M_1  \tilde{M}_2 $$=$$ \tilde{M}_1 M_2 \hat{R}$ or
$\,\hat{R} M  \otimes \tilde{M} $$=$$ \tilde{M} \otimes M \hat{R}$,  that

\begin{equation}\label{dosac}
\hat{R}^{\epsilon} M_{1} (\tilde{M}_{2}^{-1})^{t} = \tilde{M}_{1}
(M_{2}^{-1})^{t} \hat{R}^{\epsilon} \quad \quad (\hat{R}^{\epsilon} M \otimes
(\tilde{M}^{-1})^{t} = \tilde{M} \otimes (M^{-1})^{t} \hat{R}^{\epsilon})
\end{equation}

\noindent
since $\epsilon^{t} M(\epsilon^{-1})^{t} = (M^{-1})^{t}$, etc.
(cf. (\ref{dosn}); notice that in the $q \neq $1 case, $(M^{-1})^t \neq
(M^t)^{-1}$, although $(M^{-1})^{\dagger} =
(M^{\dagger})^{-1}$).
Now, since the $q$-Lorentz transformation (\ref{dosh}) may be written as
\begin{equation}\label{dosad}
K'_{is} = \Lambda_{is,jl} K_{jl} \,, \qquad \Lambda_{is,jl} = (M \otimes
(\tilde{M}^{-1})^{t})_{is,jl}\,,
\end{equation}

\noindent
(in this form, the reality of $\Lambda$ means that $\Lambda^{*}_{is,jl} =
\Lambda_{si,lj}$ or $\Lambda^{*}={\cal P} \Lambda {\cal P})$.
 As a result, if $K$ transforms
by (\ref{dosh}) say, contravariantly, then
\begin{equation}\label{dosaf}
K_{ij}^{\epsilon} = \hat{R}^{\epsilon}_{ij,kl} K_{kl} \quad \quad
(K^{\epsilon} = \hat{R}^{\epsilon} K)
\end{equation}

\noindent
transforms covariantly,
\begin{equation}\label{dosaf2}
K^{\epsilon '} = \tilde{M} K^{\epsilon} M^{-1} \quad .
\end{equation}

\noindent
{}From (\ref{dosab}) and (\ref{af2}) we find
\begin{equation}\label{dosag}
\hat{R}^{\epsilon}=
\left[
\begin{array}{cccc}
\lambda & 0 & 0 & -q\\
0 & q & 0 & 0\\
0 & 0 & q & 0\\
-q^{-1} & 0 & 0 & 0
\end{array}
\right] \quad , \qquad
(\hat{R}^{\epsilon})^{-1}=
\left[
\begin{array}{cccc}
0 & 0 & 0 & -q\\
0 & q^{-1} & 0 & 0\\
0 & 0 & q^{-1} & 0\\
-q^{-1} & 0 & 0 & - \lambda
\end{array}
\right]\;,
\end{equation}

\noindent
so that (\ref{dosaf}) gives
\begin{equation}\label{dosah2}
K^{\epsilon}=
\left[
\begin{array}{cc}
\lambda \alpha - q \delta & q \beta\\
q \gamma & -q^{-1} \alpha
\end{array}
\right]
\end{equation}
\noindent
and $KK^{\epsilon}$=$-q^{-1}(det_qK)I$ is $q$-Lorentz invariant
(this relation    is trivially checked when
$det_qK \neq 0$  since in this case  $K^{\epsilon} $=$ -q^{-1} (det_{q} K)
K^{-1} $,
eq. (\ref{dosz})).  The scalar product is thus  given by the $q$-trace
of $K^{\epsilon}K$
\begin{equation}\label{dosah}
l_{q}= det_{q} K = \frac{-q}{[2]} tr_{q} (K K^{\epsilon})= \frac{-q}{[2]}
tr_{q} (K^{\epsilon} K) \quad ,\quad [l_q,K]=0\;,
\end{equation}

\noindent
which also follows from (\ref{dosx}) since  $(P_-)_{ij,kl}$=$[2]^{-1}
\epsilon_{ij} \epsilon_{kl}$ (eq. (\ref{ah})) and eq. (\ref{dosab}) give
 $(K_{1} \hat{R} K_{1} P_{-})_{ij,kl}=
(K_{1} K_{1}^{\epsilon} P_{-})_{ij,kl}$ so that  $det_{q} (K){\bf I} = - q
(K K^{\epsilon})$. The square of the invariant length may be expressed in terms
of a $q$-Minkowski tensor $g_{ij,kl}$ as
\begin{equation}\label{ch}
l_q=q^{2} g_{ij,kl} K_{ij} K_{kl}
=\frac{-q}{\left[2\right]} D_{si} \hat{R}_{js,kl}^{\epsilon}
K_{ij} K_{kl}\quad.
\end{equation}
\noindent
Explicitly,
\begin{equation}\label{metric}
g_{ij,kl}
=\frac{-q^{-1}}{\left[2\right]} D_{si} \hat{R}_{js,kl}^{\epsilon}
= \frac{-q^{-1}}{\left[2\right]} \epsilon_{im}
\hat{R}_{jm,kt} \epsilon^{-1}_{lt}\quad, \quad
(g=\frac{-q^{-1}}{\left[2\right]}D_1 {\cal P} \hat{R}_{12}^{ \epsilon}) \quad,
\end{equation}
\noindent
and, on account of (\ref{dosac}) and (\ref{dprop}), $g$  satisfies
\begin{equation}\label{metric2}
\Lambda^t g \Lambda = g \quad, \quad
\Lambda_{rs,ij} g_{rs,mn} \Lambda_{mn,kl}=g_{ij,kl}\;.
\end{equation}

Let us now  analyze in generality the possible
commutation properties of the entries of a matrix $K$ generating a
 $q$-Minkowski algebra ${\cal M}_q$. We may describe them
\cite{AKR} in an unified way by means of a
general RE  (see \cite{KS} and references therein)

\begin{equation}\label{dosak}
R^{(1)} K_{1} R^{(2)} K_{2} = K_{2} R^{(3)} K_{1} R^{(4)}.
\end{equation}

\noindent
Writing the same equation for $K'=MK \tilde{M}^{-1}$ and comparing with
(\ref{dosak}), we conclude that the invariance of the commutation properties of
$K$  under (\ref{dosh}) is achieved if the relations
$$
R^{(1)} M_{1} M_{2} = M_{2} M_{1} R^{(1)} \quad , \quad R^{(2)} M_{2}
\tilde{M}_{1} = \tilde{M}_{1} M_{2} R^{(2)}\,,
$$
\begin{equation}\label{dosal}
R^{(3)} M_{1} \tilde{M}_{2} = \tilde{M}_{2} M_{1} R^{(3)} \quad , \quad
R^{(4)} \tilde{M}_{2} \tilde{M}_{1} = \tilde{M}_{1} \tilde{M}_{2} R^{(4)}\,,
\end{equation}

\noindent
are satisfied (eqs. (\ref{dosak}), (\ref{dosal}) also follow easily from the
bivector -rather, bispinor-  character of $K$ (see \cite{PRAGA})  (in fact,
they might be used to consider other dimensions by extending the arguments
used to deform
$D^{\frac{1}{2}, \frac{1}{2}} =
D^{\frac{1}{2},0} \otimes D^{0, \frac{1}{2}}$ to the
$D^{j, j} =  D^{j,0} \otimes D^{0, j}$ case).
Using the permutation operator ${\cal P}$, the first equation may
be rewritten as
\begin{equation}\label{dosam}
{\cal P} (R^{(1)})^{-1} {\cal P} M_{1} M_{2} = M_{2} M_{1} {\cal P}
(R^{(1)})^{-1} {\cal P}
\end{equation}
\noindent
(and similarly for $R^{(4)}$ and tilded $\tilde{M}'s$). It follows from
(\ref{dosal}) that we may take $R^{(4)}=R^{(1) \, \dagger}$ or
$R^{(4)}={\cal P}(R^{(1)\,-1})^{\dagger}{\cal P}$ and that
$R^{(2)}={\cal P}R^{(3)}{\cal P}$ and
$R^{(2,3)\, \dagger}={\cal P}R^{(2,3)}{\cal P}$.
If $M$ and $\tilde{M}$ are copies of the same quantum group then
$R^{(1)\, \dagger}={\cal P}R^{(1)}{\cal P}$  (or
$R^{(1)\, \dagger}=R^{(1)\, -1}$; this  condition is not satisfied by
(\ref{af2})).

Relations (\ref{dosal}) can be treated simultaneously
by using a four dimensional $q$-Dirac spinorial realization in terms
of the 4$\times$4  matrices
\begin{equation}\label{1e}
{\cal S} \equiv \left( \begin{array}{cc}
                      M & 0 \\
                      0 & \tilde{M}
                  \end{array} \right) \quad , \quad
{\cal K}  \equiv \left( \begin{array}{cc}
                      0 & K \\
           K^{\epsilon} & 0
                  \end{array} \right) \quad , \quad
{\cal K}  \longmapsto {\cal K}'={\cal S} {\cal K} {\cal S} ^{-1} \quad,
\end{equation}
\noindent
which reproduce (\ref{dosh}) and (\ref{dosaf2})
(see the end of Appendix  B), and by introducing
(cf. \cite{DRSWZ}) the ${\cal R}$-matrix
\begin{equation}\label{1f}
{\cal R} \equiv \left( \begin{array}{cccc}
                      R^{(1)} & 0 & 0 & 0 \\
                      0 & R^{(3)}& 0&0 \\
                      0&0& R^{(2)\; -1}&0 \\
                      0&0&0& {\cal P}R^{(4)}{\cal P}
                  \end{array} \right) \quad . \qquad
\end{equation}
\noindent
In this way, the set of relations (\ref{dosal})
defining a deformed Lorentz group can be written in $`RTT'$ form,
\begin{equation}\label{1g}
{\cal R}{\cal S}_1{\cal S}_2={\cal S}_2{\cal S}_1{\cal R} \quad,
\end{equation}
\noindent
where the 16$\times$16 matrices  ${\cal S}_1$ and ${\cal S}_2$ are
defined in block form by
\begin{equation}\label{1h}
{\cal S}_1 \equiv \left( \begin{array}{cccc}
                      M_1 & 0 & 0 & 0 \\
                      0 & M_1 & 0&0 \\
                      0&0& \tilde{M}_1 &0 \\
                      0&0&0& \tilde{M}_1
                  \end{array} \right) \quad , \qquad
{\cal S}_2 \equiv \left( \begin{array}{cccc}
                      M_2 & 0 & 0 & 0 \\
                      0 & \tilde{M}_2 & 0&0 \\
                      0&0& M_2 &0 \\
                      0&0&0& \tilde{M}_2
                  \end{array} \right) \quad .
\end{equation}
\noindent
Relations (\ref{1g}) may be used to define the commutation relations of
the generators
(entries of ${\cal S}$) of a quantum group. Standard arguments (see below
(\ref{YBE})) may be now invoked  to  require
that ${\cal R}$ satisfies the YBE
\begin{equation}\label{1i}
{\cal R}_{12}{\cal R}_{13}{\cal R}_{23}
={\cal R}_{23}{\cal R}_{13}{\cal R}_{12} \quad,
\end{equation}
\noindent
where ${\cal R}_{12}$, ${\cal R}_{13}$ and ${\cal R}_{23}$ are
$4^3 \times 4^3$  matrices acting on $C^4 \otimes C^4 \otimes C^4$.
Multiplying
by blocks the matrices in both sides of eq. (\ref{1i}), it follows that
the 4$\times$4 matrices $R^{(1)}$, $R^{(4)}$ must satisfy the YBE
and $R^{(2)}$, $R^{(3)}$ the  mixed consistency equations (see \cite{FM})
\begin{equation}\label{1j}
R^{(1)}_{12}R^{(3)}_{13}R^{(3)}_{23}=R^{(3)}_{23}R^{(3)}_{13}R^{(1)}_{12}
\quad , \quad
R^{(4)}_{12}R^{(2)}_{13}R^{(2)}_{23}=R^{(2)}_{23}R^{(2)}_{13}R^{(4)}_{12}\;
\end{equation}
\noindent
(these two equations become the same either for
$R^{(1)\, \dagger}={\cal P}R^{(1)}{\cal P}$  or
$R^{(1)\, \dagger}=R^{(1)\, -1}$); notice, for instance, that the
first equation tell us that $R^{(3)}$ is a representation of the  $SL_q(2)$
quantum group if considered as an `RTT' equation, cf. (\ref{uc}).
The possible deformations of the
Lorentz group expressed by (\ref{dosal}) have $R^{(i)}$
matrices satisfying  (\ref{1j}) as a result of (\ref{1i})
(of course, one may arrive to the YB-like eqs. (\ref{1j}) directly from
(\ref{dosal}) by the usual reorderings of the product of three matrices
acting on their respective spaces).

Let us go back to the $q$-Lorentz group $L_q^{(1)}$ defined by eqs.
(\ref{dosf}).  Comparing
equations (\ref{dosal}) with (\ref{dosf}) we find the solutions
\begin{equation}\label{dosan}
R^{(1)} = R_{12} \,\mbox{or}\, R_{21}^{-1} \;, \quad R^{(2)} = R_{21}\;, \quad
R^{(3)} = R_{12}\;, \quad R^{(4)} = R_{21} \,\mbox{or} \, R_{12}^{-1}\,,
\end{equation}

\noindent
with $R_{12}$ given in (\ref{af2}); thus,
${\cal P}R^{(1,4)}{\cal P}=R^{(1,4)\, \dagger}$.
The solution  $R^{(1)} $=$ R_{12}$, $ R^{(4)} $=$ R_{21}$
\cite{AKR}
was the one used in (\ref{dosj}) [(\ref{dosk})].
The possibility $R^{(1)} $=$ R_{12}$,  $ R^{(4)} $=$ R_{12}^{-1}$ implies
replacing (\ref{dosk}) by
\begin{equation}\label{dosap}
\hat{R} K_{1} \hat{R} K_{1} = q^{2} K_{1} \hat{R} K_{1} \hat{R}^{-1}\,.
\end{equation}

\noindent
However, it is shown in  Appendix B1  that (\ref{dosap}) also
leads to (\ref{dosl}) with the restriction $det_{q} K = 0$
and thus it may be discarded.
The other two solutions $R^{(1)} $=$ R_{21}^{-1}$, $ R^{(4)} $=$
R_{12}^{-1}$ and $R^{(1)} $=$ R_{21}^{-1}$, $ R^{(4)} $=$ R_{21}$, are,
respectively,
the same as (\ref{dosk}) and (\ref{dosap}); thus, the assumptions on the
$q$-Lorentz
group reflected by eqs. (\ref{dosf}) lead uniquely to
(\ref{dosj}) or (\ref{dosl}) as
the relations defining the $q$-Minkowski algebra ${\cal M}_{q}^{(1)}$.

The above is not the only possibility. Other solutions may be found
by looking for other consistent sets of matrices $R^{(i)}$ in
(\ref{dosak}), (\ref{dosal}). Let us discuss now two simple examples and
mention a couple of others at the end of the section.
We may introduce another
$q$-Lorentz group $ L_{q}^{(2)}$ generated by the same
non-commuting entries of $M$ and $\tilde{M}$ as reflected by the first two
equations in (\ref{dosf}), with the same $*$-operation, by replacing the third
relation in (\ref{dosf}) by $M_{1} \tilde{M}_{2} = \tilde{M}_{2}
M_{1}$, so that the elements of $M$ and $\tilde{M}$ commute in between.
This corresponds to taking $R^{(2)}= R^{(3)} = I$, and leads to
the possibilities
\begin{equation}\label{dosaq}
R_{12} K^{(2)}_1 K^{(2)}_2 = K^{(2)}_2 K^{(2)}_1 R_{21}\,;
\end{equation}
\begin{equation}\label{dosaq2}
R_{12} K^{(2)}_1 K^{(2)}_2 = q^{2} K^{(2)}_2 K^{(2)}_1 R_{12}^{-1}\;,
\end{equation}

\noindent
(the superindex has been added to distinguish $K^{(2)}$ from
the previous $K \equiv K^{(1)}$).
It is simple to see that eqs. (\ref{dosaq}), (\ref{dosaq2})
are consistent with the reality condition $K^{(2)} = K^{(2)\,\dagger}$.
Eqs. (\ref{dosaq}) (and (\ref{dosaq2}), which corresponds to
$det_qK^{(2)}$=0) lead to the following independent commutation relations
for the entries of $K^{(2)}$ generating the ${\cal M}_q^{(2)}$ algebra
(cf. (\ref{dosl}))
\begin{equation}\label{dosaq3}
\begin{array}{lll}
\alpha^{(2)} \beta^{(2)}=q^{-1} \beta^{(2)} \alpha^{(2)} \;, \quad &
\quad  \alpha^{(2)} \gamma^{(2)}=q \gamma^{(2)} \alpha^{(2)} \;, \quad &
\quad [ \alpha^{(2)}, \delta^{(2)}]= 0 \;, \\
\, & \, & \, \\
{} [ \beta^{(2)}, \gamma^{(2)}]= \lambda \alpha^{(2)} \delta^{(2)}\; , \quad &
\quad \beta^{(2)} \delta^{(2)}= q \delta^{(2)} \beta^{(2)}  \;, \quad &
\quad \gamma^{(2)} \delta^{(2)}=q^{-1} \delta^{(2)} \gamma^{(2)}\;
\end{array}
\end{equation}
\noindent
The identifications $\alpha \leftrightarrow b$, $\beta \leftrightarrow a$,
$\gamma \leftrightarrow d$, $\delta \leftrightarrow c$, make this algebra
isomorphic to the $GL_q(2)$ one in eq. (\ref{ua}) \cite{WA-ZPC48}.
There is an invariant and central element
in the ${\cal M}_q^{(2)}$ algebra which determines the
Minkowski length and metric. The determinant (which is zero
for (\ref{dosaq2})) is given through (see (\ref{gdet}))
\begin{equation}\label{dosaq4}
det_q(K^{(2)})P_- = -qK_1^{(2)} {\cal P} K_1^{(2)}P_- \; , \quad
det_q(K^{(2)})= \alpha^{(2)} \delta^{(2)}-q \gamma^{(2)} \beta^{(2)}\;,
\end{equation}

\noindent
(cf. (\ref{dosaa})).
This definition guarantees that $det_q(K'^{(2)})$=$det_q(K^{(2)})$
for the coaction $\phi^{(2)}$ of $L_q^{(2)}$; notice that in
(\ref{dosx}) $\hat{R}$ is replaced by ${\cal P}$ since now
$M_1 \tilde{M}_2$=$\tilde{M}_2 M_1$. For ${\cal M}_q^{(2)}$, however,
there is no linear central element.
This follows from the fact that eq. (\ref{dosaq})
can be linearly transformed into a `$RTT$' relation [(\ref{uc}), (\ref{ua})]
by the change $T=K^{(2)}\sigma^1$, where $\sigma^1$ is the Pauli matrix,
and that there is no linear
central element for the $GL_q(2)$ algebra.

There is another obvious possibility for the $*$-operation in the case of
(\ref{dosaq}). We can take $K^{\dagger}:$=$ K^{-1} (det_qK)$,
which is consistent
with the coaction (\ref{dosh}) if $M^{\dagger}=M^{-1}$ and
$\tilde{M}^{\dagger}=\tilde{M}^{-1}$ since then
\begin{equation}\label{unit}
K'^{\dagger}=(MK \tilde{M}^{-1})^{\dagger}=
(\tilde{M}^{-1})^{\dagger} K^{\dagger} M^{\dagger}= \tilde{M} K^{-1} M^{-1}
det_qK = K'^{-1} det_qK'
\end{equation}
\noindent
provided that the proportionality coefficient (the $q$-determinant,
eq. (\ref{dosaq4})) is the same. Thus, $M$ and $\tilde{M}$ generate
two unrelated $SU_q(2)$ copies. This was one of the reasons to relate
(\ref{dosaq}) (with this $*$-operation different from the previous one)
with a $q$-deformed euclidean space \cite{WA-ZPC48,MajEucl} (see also
\cite{MAJCLA}) $R_q^4$-covariant with respect the quantum orthogonal group
$SO_q(4) \simeq SU_q(2) \otimes SU_q(2)$ \cite{WA-ZPC48}. Clearly, this
$*$-operation coincides with the classical relation $(\sigma_{\alpha}
x^{\alpha})^{\dagger}=( \sigma_{\beta} x^{\beta})^{-1}
(\sum_{\alpha}x_{\alpha}^2)$ for real coordinates $x^{\alpha}$
($\alpha$=1,2,3,4) with $\sigma_{\alpha}=i \sigma_j$ for $\alpha$=$j$=1,2,3.

A third possibility ${\cal M}_q^{(3)}$ is obtained by setting
equal to unity the $R$
matrices in the two first equations in (\ref{dosf}) but not in  the third.
In this case, the matrix elements of $M$ (and
$\tilde{M}$) are commuting (they define an $SL(2,C)$ group each),
and  the non-commutativity of the entries of $M$
and $\tilde{M}$ is described by a certain matrix $V$ through (cf. (\ref{dosf}))
\begin{equation}\label{dosar}
V M_{1} \tilde{M}_{2} = \tilde{M}_{2} M_{1} V\,,
\end{equation}
\noindent
$\tilde{M}^{-1}$=$M^{\dagger}$. There is no $SU_{q}(2)$
subgroup in this case: the mapping $M \mapsto U$,
$\tilde{M} \mapsto U$ ($U^{\dagger}=U^{-1}$)
identification $M=(M^{-1})^{\dagger} =\tilde{M}$  would imply
$V U_{1} U_{2} =
U_{2} U_{1} V$ contradicting the assumed commutativity of the entries of $M$.
The commuting properties of the
corresponding $K^{(3)} \equiv  Z$, with entries and transformation
properties given by
\begin{equation}\label{dosas}
Z=
\left[
\begin{array}{ll}
z^{1} & z^{4}\\
z^{2} & z^{3}
\end{array}
\right]
= Z^{\dagger} \quad , \qquad Z' =MZ \tilde{M}^{-1} \;,
\end{equation}

\noindent
($z^1$$=$$z^{1\,*}$, $z^{2}$$=$$ z^{4*}$, $ z^{4} $$=$$ z^{2*}$,
$z^{3}$$=$$z^{3*}$) are determined by setting
$R^{(2)}$$=$$ V $$=$$ R^{(3)}$, $R^{(1)}$$=$$ I_4$$=$$ R^{(4)}$ in
eq. (\ref{dosak}),
\begin{equation}\label{dosat}
Z_{1} V Z_{2} = Z_{2} V Z_{1} \quad.
\end{equation}

\noindent
For instance, for $V= diag (p^{2}, 1, 1, p^{2})$ (we have used $p$
rather than $q$ to
stress the trivial deformation  character of this algebra, see below),
we obtain
\begin{equation}\label{dosau}
\begin{array}{lll}
z^1z^{2}=p^{2}z^{2}z^1 \;,\qquad & z^{2}z^{3}=p^{2}z^{3}z^{2} \;,
\qquad & z^{3}z^{4}=p^{2}z^{4}z^{3}\;,\\
z^{4}z^1=p^{2}z^1z^{4} \;,\qquad & z^{4}z^{2}=z^{2}z^{4} \;,
\qquad & z^1z^{3}=z^{3}z^1 \;.
\end{array}
\end{equation}

\noindent
The Minkowski length is again given by the $p$-determinant,
$det_pZ=z^1z^3-p^2z^2z^4$ (see (\ref{gdet})); $trZ$ is here the
ordinary trace (eq. (\ref{gtr}) for $R^{(1)}$=$I_4$ gives $D$=$I_2$),
and it is not central.

We can take this case as an
illustration of the interrelation between the properties of
the $R$-matrices and the reality condition
($*$-structure). For instance, we could take for $V$ an hermitean arbitrary
diagonal matrix, $V=diag(p_1,p_2,p_3,p_4)$. Then $M_2^{\dagger}VM_1
=M_1VM_2^{\dagger}$ implies $M_1^{\dagger}VM_2
=M_2VM_1^{\dagger}$, which using the permutation operator gives
$M_2^{\dagger}{\cal P}V{\cal P}M_1 =M_1{\cal P}V{\cal P}M_2^{\dagger}$.
Thus, ${\cal P}V=V{\cal P}$, which immediately gives the restriction
$p_2=p_3$ fulfilled by the present case.
 This $p$-Minkowski algebra
${\cal M}_{p}^{(3)}$ was obtained in \cite{CHA-DEM} from the analysis of a
deformation of the conformal group $SU(2,2)$ as a real form of a
multiparametric
deformation of $SL(4,C)$, which justifies the previous form for $V$. However,
this algebra and the corresponding deformed Lorentz group have been shown to be
\cite{LRT} a simple transformation (twisting \cite{DTW,RTW}) of the
usual ones. This means, in particular, that it is possible to remove
the non-commuting character of the entries of $Z$, although we shall not
discuss this here (see \cite{AKRREL}).

The properties of the $q$-Minkowski algebras above considered
are summarized below:
\hspace{-0.7cm} \begin{center}
\begin{tabular}{cclc}
 ${\cal M}_q^{(i)}$ & $M$, $\tilde{M}$ relations
&  $ R^{(i)}$ in eq. (\ref{dosal}) & Comments  \\
\hline
\hline
${\cal M}_q^{(1)}$   &
$[M,M]$$ \neq $0$ \neq $$ [ \tilde{M} , \tilde{M}]$;
&  $R^{(1)}$=$R^{(3)}$=$R_{12}$ & $K$ isomorphic to a  \\
 \cite{WA-ZPC48,WA-A6,SWZ-ZPC52}
&   $[M, \tilde{M}] \neq 0$     &  $R^{(2)}$=$R^{(4)}$=$R_{21}$
&  braided algebra \cite{MAJ-LNM,MeyerM} \\
\hline
${\cal M}_q^{(2)}$  &
$[M,M]$$ \neq $0$ \neq $$[ \tilde{M} , \tilde{M}]$;
& $R^{(1)}$=$R_{12}$, $R^{(4)}$=$R_{21}$ & $K$ isomorphic to  \\
  \cite{WA-ZPC48,MajEucl} & $[M, \tilde{M}]=0$
&  $R^{(2)}$=$R^{(3)}$=$I_4$ &  $GL_q(2)$ algebra  \\
\hline
${\cal M}_q^{(3)}$  &
$[M,M]$$ =$0$= $$[ \tilde{M} , \tilde{M}]$; & $R^{(1)}$=$R^{(4)}$=$I_4$ &
Twisted \\
\cite{CHA-DEM} & $[M, \tilde{M}] \neq 0$  &  $R^{(2)}$=$R^{(3)}
$= &
${\cal M}_q$ space \\
\, & \, & $diag(p^2,1,1,p^2)$ & \, \\
\hline
\hline
\end{tabular}
\end{center}

It is clear that other $q$-Minkowski spaces could be found by using the same
general approach.
To conclude, we just mention as other possible examples a `mixture' of
${\cal M}_q^{(2)}$ and ${\cal M}_q^{(3)}$ with
$R^{(1)}={\cal P}R^{(4)}{\cal P}=R_{12}$,
$R^{(2)}={\cal P}R^{(3)}{\cal P}=V$, and a `gauging' of the ${\cal M}_q^{(1)}$
case with $R^{(1)}={\cal P}R^{(4)}{\cal P}=R_{12}$,
$R^{(2)}={\cal P}R^{(3)}{\cal P}=
e^{\alpha \sigma_1^3}R_{21}e^{- \alpha \sigma_1^3}$,
where the subindex in $\sigma_1^3$ refers to the first space.
The quantum determinant, again defined by the general expression
$det_qK \propto P_- K_1 \hat{R}^{(3)} K_1 P_-$ [(\ref{gdet})]
is invariant and central  in both cases, while the time generator,
$x^0 \simeq tr_qK$ (see (\ref{gtr})), is not.

\setcounter{equation}{0}

\section{\bf Deformed derivatives and $q$-De Rham complex}

\indent
The development of a non-commutative differential calculus
(see, e.g., \cite{SLWO,WZ,DIFF,ZU-MPX,MANIN2,SUD})
requires including derivatives and differentials. We shall now do this first
for the $q$-Minkowski space ${\cal M}_q^{(1)}$ [(\ref{dosk})] by extending
the RE to accommodate in them derivatives and differentials.
Consider first an object $Y$ transforming covariantly i.e.,
\begin{equation}\label{ta}
Y \longmapsto Y' = \tilde{M} Y M^{-1} \quad ,
\quad Y =
\left[
\begin{array}{ll}
u & v\\
w & z
\end{array}
\right]\;.
\end{equation}
\noindent
The invariance of the commutation properties of the matrix elements of $Y$
(now described by (\ref{dosak}) with $Y$ replacing $K$) gives,
on account of (\ref{dosf}), the solutions
\begin{equation}\label{tb}
R^{(1)} = R_{12}\, \mbox{or}\, R_{21}^{-1} \,, \quad R^{(2)}=R_{12}^{-1} \,,
\quad R^{(3)}=R_{21}^{-1} \,,\quad  R^{(4)}= R_{21} \,\mbox{or}\, R_{12}^{-1}
\,.
\end{equation}

\noindent
These four possibilities again reduce to two,
\begin{equation}\label{tc}
R_{12} Y_{1} R_{12}^{-1} Y_{2} = Y_{2} R_{21}^{-1} Y_{1} R_{21}
\quad \mbox{or} \quad
\hat{R} Y_{1} \hat{R}^{-1} Y_{1} = Y_{1} \hat{R}^{-1} Y_{1} \hat{R} \;;
\end{equation}
\begin{equation}\label{td}
q^{2} \hat{R}^{-1} Y_{1} \hat{R}^{-1} Y_{1}
                                = Y_{1} \hat{R}^{-1} Y_{1} \hat{R}\;,
\end{equation}

\noindent
of which we shall retain only (\ref{tc}) since (\ref{td}) leads to
the same algebra plus the condition $det_qY$=0 (see (\ref{tf}) below
and Appendix A). Eq. (\ref{tc}) gives
\begin{equation}\label{te}
\begin{array}{lll}
\left[ u, v \right] = q \lambda v z \;, \qquad &
v z = q^{2} z v \;, \qquad & [u, z]= 0 \;,\\
\left[ w, u \right] = q \lambda z w \;, \qquad &  [v, w]= q \lambda (u - z) z
\;, \qquad & w z = q^{-2} z w\;,
\end{array}
\end{equation}

\noindent
for the generators of ${\cal D}_q^{(1)}$.
The (central and $q$-Lorentz invariant) $q$-determinant is defined through
\begin{equation}\label{tf}
(det_{q} Y) P_{-} =(-q^{-1}) P_{-} Y_{1} \hat{R}^{-1} Y_{1} P_{-}
= (uz - q^{-2} v w) P_-
\end{equation}

\noindent
so that, when it is non-zero,
\begin{equation}\label{tg}
Y^{-1} = (det_{q} Y)^{-1}
\left[
\begin{array}{cc}
z & -q^{-2} v\\
-q^{-2} w & q^{-2} u + q^{-1} \lambda z
\end{array}
\right]
\quad .
\end{equation}

\noindent
Since $Y$ is covariant, we may define a contravariant $Y^{\epsilon}$ by
(cf. (\ref{dosaf}))
\begin{equation}\label{th}
Y^{\epsilon}= (\hat{R}^{\epsilon})^{-1} \, Y =
\left[
\begin{array}{cc}
-qz & q^{-1} v\\
q^{-1} w & -q^{-1} u - \lambda z
\end{array}
\right]
\end{equation}

\noindent
(when $det_{q} Y \neq 0 , \; Y^{\epsilon} = -q (det_{q} Y) Y^{-1})$; then,
(cf. (\ref{dosah}))
\begin{equation}\label{ti}
det_{q} Y = - \frac{q^{-1}}{[2]} tr_{q} (Y Y^{\epsilon}) = - \frac{q^{-1}}{[2]}
tr_{q} (Y^{\epsilon} Y) \equiv  \Box_q\;, \quad [ \Box_q , Y]=0\;,
\end{equation}
\noindent
where ${\Box}_q$ becomes the ($L_q$-invariant) $q$-D'Alembertian
once the components of $Y$ are associated with the $q$-derivatives.
Indeed, since we have already associated
$K$ (contravariant) with ${\cal M}_{q}^{(1)}$
and we wish to have the equivalent of the classical Lorentz invariant $x^{\mu}
\partial_{\mu}$ we shall associate $Y$ (covariant)
with the algebra ${\cal D}_{q}^{(1)}$ of the
$q$-Minkowski derivatives; it is clear, however, that one could  proceed
reciprocally.
The commutation properties of the elements of $K$, $Y^{\epsilon}$ and $Y^{-1}$
(when $det_qY \neq 0$) are governed by an equation of the type
(\ref{dosj}-\ref{dosk}); similarly,
those of $Y$, $K^{\epsilon}$ and $K^{-1}$ (when $det_{q} K \neq 0$) obey a
relation like (\ref{tc}) (for instance, inverting (\ref{dosk})  one obtains
$\hat{R} K_{1}^{-1} \hat{R}^{-1} K_{1}^{-1}= K_{1}^{-1} \hat{R}^{-1} K_{1}^{-1}
\hat{R} $,  cf.(\ref{tc})). Thus, the entries of $K$, $Y^{\epsilon}$ and
$Y^{-1}$ (resp. $Y$, $K^{\epsilon}$, $K^{-1}$) satisfy the same commutation
relations, and the algebras they generate are isomorphic. Symbolically,
$K \sim Y^{\epsilon} \sim Y^{-1}$ and $Y \sim K^{\epsilon} \sim K^{-1},$ a
fact that may be explicitly checked using eqs.
(\ref{dosl}), (\ref{te}), (\ref{th}), (\ref{tg})
and (\ref{dosah}), (\ref{dosz}).
Moreover, it is proved in Appendix B2 that the following
isomorphisms  among these algebras hold
\begin{equation}\label{tj}
{\cal M}_{q}^{(1)} \approx {\cal M}_{q^{-1}}^{(1)} \approx {\cal D}_{q}^{(1)}
 \approx {\cal D}_{q^{-1}}^{(1)} \quad,
\end{equation}
\noindent
where the subindex $q^{-1}$ indicates that the corresponding algebras are
defined
by (\ref{dosl}) and (\ref{te}) where $q$ is replaced by $q^{-1}$.

The next step in constructing the non-commutative
$q$-Minkowski differential calculus is to establish the commutation
properties among coordinates
and derivatives. We need  extending  the classical
relation $\partial_{\mu} x^{\nu} = x^{\nu} \partial_{\mu} + \delta_{\mu}^{\nu}
$, $\partial^{\dagger}$=$- \partial$,  to the non-commutative
case in a $q$-Lorentz covariant way. This
requires considering an inhomogeneous RE of the form
(more complicated expressions are possible \cite{AKR})
\begin{equation}\label{tk}
Y_{2} R^{(1)} K_{1} R^{(2)} = R^{(3)} K_{1} R^{(4)} Y_{2} + \eta J,
\end{equation}

\noindent
where $\eta$ is a constant and $\eta J$$\rightarrow$$I_{4}$ in the
$q$$ \rightarrow$$1$ limit, invariant under the transformation
\begin{equation}\label{tl}
J \longmapsto \tilde{M}_{2} M_{1} J \tilde{M}_{1}^{-1} M_{2}^{-1} = J
\,, \qquad  \tilde{M}_{2} M_{1} J = J M_{2} \tilde{M}_{1}\,.
\end{equation}
\noindent
An analysis similar to those of Sec. 2 shows that the invariance of the
non-linear terms in (\ref{tk}) under (\ref{dosh}), (\ref{ta}) is achieved with
\begin{equation}\label{tm}
R^{(1)} = R_{12}\,\mbox{or}\,  R_{21}^{-1} \,, \quad R^{(2)} = R_{21} \,,
\quad  R^{(3)} =
R_{12} \,,\quad  R^{(4)} = R^{-1}_{12} \,\mbox{or}\,  R_{21} \, .
\end{equation}
\noindent
As for $J$,  setting $J\equiv  J' {\cal P}$ in eq. (\ref{tl}) gives
$\tilde{M}_{2}
M_{1} J' = J' M_{1} \tilde{M}_{2}$, hence $J = R_{12} {\cal P}$ (the same
result follows if  we set $J= {\cal P} J'$). This means that there are, in
principle, four basic possibilities consistent with covariance expressing the
commutation properties of coordinates (elements of $K$) and derivatives
(entries of $Y$). Using again $\hat{R} = {\cal P} R$, these read
\begin{equation}\label{tn}
Y_{1} \hat{R} K_{1} \hat{R} = \hat{R} K_{1} \hat{R}^{-1} Y_{1} + \eta_{1}
\hat{R}\;;
\end{equation}
\begin{equation}\label{to}
Y_{1} \hat{R}^{-1} K_{1} \hat{R} = \hat{R} K_{1} \hat{R} Y_{1} + \eta_{2}
\hat{R}\;;
\end{equation}
\begin{equation}\label{tp}
Y_{1} \hat{R}_{1} K_{1} \hat{R} = \hat{R} K_{1} \hat{R} Y_{1} + \eta_{3}
\hat{R}\;;
\end{equation}
\begin{equation}\label{tq}
Y_{1} \hat{R}_{1}^{-1} K_{1} \hat{R} = \hat{R} K_{1} \hat{R}^{-1} Y_{1} +
\eta_{4} \hat{R} \;\;.
\end{equation}

\noindent
Due to the fact that these expressions now involve $K$ and $Y$, they are all
unequivalent. In fact, we do not need assuming that the four $Y's$ appearing
in each of the equations (\ref{tn}-\ref{tq}) are the same; all
that it is demanded is that all they transform as $Y \mapsto \tilde{M} Y
M^{-1}$.

Let us now look at the hermiticity properties of $K$ and $Y$. It is clear that,
since $\hat{R} = \hat{R}^{\dagger}$ ($q$ is real), eqs. (\ref{dosj}) and
(\ref{tc}) are consistent
with the hermiticity of  $K$ and the antihermiticity of $Y$. However, this is
no longer the case
if the  inhomogeneous equations are included. Keeping the physically reasonable
assumption that $K$ is hermitean, eq. (\ref{tn}) gives
\begin{equation}\label{tr}
Y_{1}^{\dagger} \hat{R}^{-1} K_{1} \hat{R} = \hat{R} K_{1} \hat{R}
Y_{1}^{\dagger}
- \eta_{1} \hat{R}
\end{equation}

\noindent
i.e., $Y^{\dagger}$ satisfies the commutation relations  given by
the second inhomogeneous equation (\ref{to}) for $\eta_{2} $$=$$ - \eta_{1}$
(of
course, $Y^{'\dagger} = \tilde{M} Y^{\dagger} M^{-1}$ again since $\tilde{M} =
(M^{-1})^{\dagger})$. Thus, we need accommodating $Y^{\dagger}$ by means of
{\it another} reflection equation, eq. (\ref{to}) for $Y^{\dagger}$. Having
then
selected (\ref{tn}) for $Y$  and (\ref{to}) for $Y^{\dagger}$,
we may now consider the other possibilities
(\ref{tp}), (\ref{tq}). It turns out that these possibilities are inconsistent
with
the previous relations (\ref{dosj}) and (\ref{dosal}),
what may be seen  with a little effort by acting on (\ref{tp}),
(\ref{tq}) with an additional $K$ or $Y$.

In order to have the inhomogeneous term in the simplest form (the analogue of
the $\delta_{\nu}^{\mu}$ of the $q=1$ case) it is convenient to take $\eta_{1}
= q^{2} = \eta$ and to redefine $Y^{\dagger}$ as $\tilde{Y} = -q^{-4}
Y^{\dagger}$. In this way, the full set of equations describing the
commutation relations of the generators of the algebras of coordinates $(K)$,
derivatives $(Y)$ and their hermitean conjugates $(Y^{\dagger}
\propto \tilde{Y})$ are given by
$$
\hat{R} K_{1} \hat{R} K_{1} = K_{1} \hat{R} K_{1} \hat{R}\;,
$$
$$
\hat{R} Y_{1} \hat{R}^{-1} Y_{1} = Y_{1} \hat{R}^{-1} Y_{1} \hat{R}
\quad , \quad
\hat{R}\tilde{Y}_{1} \hat{R}^{-1} \tilde{Y}_{1}
= \tilde{Y}_{1} \hat{R}^{-1} \tilde{Y}_{1} \hat{R}\;;
$$
$$
Y_{1} \hat{R} K_{1} \hat{R} = \hat{R} K_{1} \hat{R}^{-1} Y_{1} + \eta
\hat{R}\,;
\quad \eta = q^{2}\;;
$$
\begin{equation}\label{ts}
\tilde{Y}_{1} \hat{R}^{-1} K_{1} \hat{R} = \hat{R} K_{1} \hat{R} \tilde{Y}_{1}
+ \tilde{\eta}\hat{R}\;, \quad \tilde{\eta}=q^{-2}\;.
\end{equation}

\noindent
Notice that, although we have identified $Y$ with the derivatives and
$\tilde{Y}$
with their hermiteans, the reciprocal assignment is also possible.

To determine now the commutation relations for the $q$-De Rham complex we now
introduce the exterior derivative $d$ following \cite{OSWZ-CMP};
we shall assume that $d^{2}$=$0$
and that it satisfies the Leibniz rule (other
 possibilities  for $q$-differential calculus are occasionally
considered \cite{FP,K-POMI}). To the
four generators
of the ${\cal M}_{q}^{(1)}$ (coordinates) and of ${\cal D}_{q}^{(1)}$
(derivatives) Minkowski
algebras we now add the four entries of $dK$ ($q$-one-forms), which generate
the
algebra of the $q$-forms $\Lambda_{q}^{(1)}$
(the degree of a form is defined as in the classical case). Clearly,  $d$
commutes with the
$q$-Lorentz coaction (\ref{dosh}), so that
\begin{equation}\label{tt}
dK' = M d K \tilde{M}^{-1} \quad .
\end{equation}

\noindent
Applying $d$ to the first equation in (\ref{ts}) we obtain
\begin{equation}\label{tu}
\hat{R} dK_{1} \hat{R} K_{1} + \hat{R} K_{1} \hat{R} d K_{1} = d K_{1} \hat{R}
K_{1} \hat{R} + K_{1} \hat{R} d K_{1} \hat{R} \quad .
\end{equation}

\noindent
We now use Hecke's condition  $\hat{R} = \hat{R}^{-1} + \lambda I$
[(\ref{ae2})]  to replace one $\hat{R}$
in each side in such a way that the terms in $d K_{1} \hat{R} K_{1}$ may be
cancelled. In this way we obtain that a solution to (\ref{tu}) is given by
\begin{equation}\label{tv}
\hat{R} K_{1} \hat{R} d K_{1} = d K_{1} \hat{R} K_{1} \hat{R}^{-1}\;,
\end{equation}

\noindent
from which follows that
\begin{equation}\label{tv2}
\hat{R} d K_{1} \hat{R} d K_{1} = - d K_{1} \hat{R} d K_{1} \hat{R}^{-1}\;.
\end{equation}

\noindent
Again, we may check that these relations are not invariant under
hermitean conjugation, since they lead, respectively,  to
\begin{equation}\label{tw}
\hat{R} d K^{\dagger}_{1} \hat{R} K_{1} = K_{1} \hat{R} dK^{\dagger}_{1}
\hat{R}^{-1}  \;, \quad
\hat{R} d K^{\dagger}_1 \hat{R} d K^{\dagger}_1 = - d K^{\dagger}_{1}
\hat{R} dK^{\dagger}_{1} \hat{R}^{-1}\;.
\end{equation}

\noindent
Notice again that the first equation in (\ref{tw}) is also a legitimate
solution
of (\ref{tu}) for a generic $dK$ ($dK$ and $dK^{\dagger}$ transform in the same
manner); in
fact, it is obtained by replacing two $\hat{R}'s$ in (\ref{tu}) in such a way
 that now the terms  $K_{1} \hat{R} dK^{\dagger}_{1}$ are cancelled.
We  expect  the `$q$-determinant' of $dK$ to vanish; using (\ref{tv2})
we check that
\begin{equation}\label{tw2}
tr_q(dK\;dK^{\epsilon}) =0 \;,
\end{equation}
\noindent
where $dK^{\epsilon}$=$\hat{R}^{\epsilon}dK$ (cf. (\ref{dosaf})) and,
in fact, $P_-dK_1 \hat{R} dK_1 =0$.

Finally, to complete the full set of commutation relations, we need those of
$dK$ and $Y$ (and their hermiteans). They are given in general by
\begin{equation}\label{tx}
Y_{2} R^{(1)} d K_{1} R^{(2)} = R^{(3)} d K_{1} R^{(4)} Y_{2} \;,
\end{equation}

\noindent
which has the same transformation properties as (\ref{tk}) with $J=0$ and hence
the same solutions (\ref{tm}). Thus, we may take
\begin{equation}\label{ty}
Y_{1} \hat{R}^{-1} dK_1 \hat{R} = \hat{R} d K_{1} \hat{R} Y_{1}\;.
\end{equation}

\noindent
Its hermitean conjugated relation
has the form
\begin{equation}\label{tz}
\tilde{Y}_{1} \hat{R} dK^{\dagger}_1 \hat{R} = \hat{R} dK^{\dagger}_{1}
\hat{R}^{-1}
\tilde{Y}_{1} \quad ,
\end{equation}

\noindent
and it corresponds to another possible solution for (\ref{tx}) now written for
$\tilde{Y}$. Notice that the RE (\ref{tx}) is, as always,  characterized by the
transformation properties of its entries, and that $Y$ and $\tilde{Y}$ and
$dK$ and $dK^{\dagger}$, respectively, transform in the same manner due to the
reality condition $\tilde{M} = (M^{-1})^{\dagger}$. The other two solutions
(with $Y $ ($dK$) replaced by $\tilde{Y}$ ($dK^{\dagger}$)) correspond to the
commutation properties of $\tilde{Y}, \, dK$ and $Y, dK^{\dagger}$
\begin{equation}\label{taa}
\tilde{Y}_{1} \hat{R} dK_{1} \hat{R} = \hat{R} dK_{1} \hat{R} \tilde{Y}_{1}
\quad \Longleftrightarrow \quad \hat{R} dK^{\dagger}_1 \hat{R} Y_1
= Y_1 \hat{R} dK^{\dagger}_1 \hat{R}\;;
\end{equation}
\begin{equation}\label{tab}
Y_{1} \hat{R}^{-1} dK^{\dagger}_{1} \hat{R} = \hat{R} dK^{\dagger}_{1}
\hat{R}^{-1}
Y_{1} \quad \Longleftrightarrow \quad \hat{R} dK_1 \hat{R}^{-1} \tilde{Y}_1
= \tilde{Y}_1 \hat{R}^{-1} d K_1 \hat{R}\;.
\end{equation}

\noindent
Eqs. (\ref{ts}), (\ref{tv}-\ref{tw}) and (\ref{ty}-\ref{tab})
\cite{AKR} define
the full differential calculus  on ${\cal M}_q^{(1)}$ \cite{OSWZ-CMP}.
We may now give a compact  expression for $d$. As its invariance suggests,
it has  the form
\begin{equation}\label{tac}
d=  tr_{q} (dKY)\;.
\end{equation}

\noindent
We shall use (\ref{tac}) to obtain further expressions for the non-commutative
differential calculus on ${\cal M}_{q}^{(1)}$ in Sec. 5 but, before doing so,
it will be convenient to discuss in the next section how to select a more
natural basis in
${\cal M}_{q}^{(1)}$ and ${\cal D}_{q}^{(1)}$.

To exhibit the generality of the previous reasonings, we  conclude this section
by applying briefly our framework to
e.g., the ${\cal M}_{p}^{(3)}$ algebra
discussed at the end of Sec.2. For the derivatives (which transform by $\phi:D
\mapsto
D'=(M^{\dagger})^{-1}D M^{-1}$) we find  that
\begin{equation}\label{tae}
D_{1} V^{-1} D_{2} = D_{2} V^{-1} D_{1} \quad ,\quad
D=\left( \begin{array}{cc}
                        \partial_{1} &  \partial_{2} \\
                        \partial_{4} &   \partial_{3}
                         \end{array} \right) \;,
\end{equation}

\noindent
since ${\cal P}V{\cal P}$=$V$;  thus,  ${\cal M}_{p^{-1}}^{(3)}\approx
{\cal D}_{p}^{(3)}$ since $V(p)^{-1}$=$V(p^{-1})$. Here $Z$ and $D$ may be
taken hermitean and antihermitean simultaneously, and there is only one
possibility for the mixed $Z,D$ relation of the form
\begin{equation}\label{taf}
D_{1} Z_{2} V = V Z_{2} D_{1} + {\cal P} V
\end{equation}

\noindent
since $M_{2}^{-1} M_{1}^{\dagger} J M_{1}( M_{2}^{ -1})^{\dagger} = J$ with $J
\propto {\cal P} V$ using (\ref{dosar}).
Clearly, $d=tr(dZD)$. For the basic relations of the $p$-De Rham complex
 we get
\begin{equation}\label{tag}
d Z_{1} V Z_{2} = Z_{2} V d Z_{1} \quad , \quad d Z_{1} V d Z_{2} =
- d Z_{2} V d Z_{1}\quad;
\end{equation}

\noindent
all other relations are found easily. For instance,
\begin{equation}
D_{1} dZ_{2} V = V dZ_{2} D_{1} \quad.
\end{equation}

\setcounter{equation}{0}

\section{\bf $q$-Tensors and covariant `coordinates'}

\indent
 Since neither of the $q$-Minkowski spaces ${\cal M}^{(i)}_q$ are
manifolds (they are non-commutative associative algebras), we cannot define
`coordinates'
for ${\cal M}^{(i)}_q$ in the usual sense. However, there are some
sets of  generators which are more convenient than the
generic ones provided by the entries of $K$, eq. (\ref{dosi}).
To find more suitable set of generators for  ${\cal M}^{(1)}_q$
we now look for
the $q$-equivalent to the undeformed or classical splitting
$K= \sigma_{\mu} x^{\mu}$.
We shall introduce the $q$-sigma matrices by imposing the condition that they
are
$q$-tensors \cite{BITAR,RITS}.
Consider the simplest case of $SU_{q}(2)$. The statement that the
$\sigma^i$ $(i = 1, 2, 3)$ constitute a $q$-tensor (in fact, a $q$-vector
under the $q$-deformed rotation group) means that the adjoint action $\rho (X)$
may be expressed in the form
\begin{equation}\label{ca}
(\rho (X) \vec{\sigma})_{il} = <X, U \vec{\sigma} U^{-1} >_{il} = <X, U_{ij}
U_{kl}^{-1} > \vec{\sigma}_{jk}\;,
\end{equation}

\noindent
where $U \in SU_{q} (2)$ and $X$ is an element of its dual quantum algebra.
Using the product/coproduct duality [(\ref{dua})] and that $\Delta (X)=
\sum_{r} X_{1}^{r}
\otimes X_{2}^{r},$  this may be written as
$$
(\rho (X) \vec{\sigma})_{ij} = \sum_{r} <X_{1}^{r}, U_{ij} > \vec{\sigma}_{jk}
< X_{2}^{r}, U_{kl}^{-1} > =
$$
\begin{equation}\label{cb}
= \sum_{r} < X_{1}^{r} , U_{ij} > \vec{\sigma}_{jk} < S (X_{2}^{r}),
U_{kl} >\;,
\end{equation}

\noindent
since $S (U)_{kl}=U_{kl}^{-1}$ and $< X, S (U) > =
< S (X), U >$, $S$ denoting the antipode. Written in this way
(\ref{cb}) corresponds to the general definition of a tensor operator
$T$ \cite{RITS}
\begin{equation}\label{cc}
\rho (X) T = \sum_{r} (X_{1}^{r})_{w} T S (X_{2}^{r})_{w}\;,
\end{equation}

\noindent
where the subindex $w$ here indicates the representation of $X_{1}^{r}$ and
$S(X_{2}^{r})$. It is simple to see that this expression reduces to the
more familiar one given in terms of commutators with the generators of the
$q$-algebra. Using the  expressions for the coproduct of the
$su_q(2)$ quantum algebra generators
$$
\Delta (J_{3}) = J_{3} \otimes {\bf 1} + {\bf 1} \otimes J_{3} \quad ,
\quad \Delta (q^{J_{3}}) = q^{J_{3}} \otimes q^{J_{3}}\;,
$$
$$
\Delta (J_{\pm}) = J_{\pm} \otimes q^{-J_{3}} + q^{J_{3}} \otimes J_{\pm}\;,
$$
\begin{equation}\label{cd}
S (J_{3}) = - J_{3} \;,\quad S (q^{\pm J_{3}}) = q^{\mp J_3}\;,
\quad
S(J_{\pm}) = -q^{\mp 1} J_{\pm}\;,
\end{equation}

\noindent
eqs. (\ref{cb}) or (\ref{cc}) give for a $SU_{q} (2)$ tensor $T_{m}^{j}$ the
explicit conditions \cite{FENG}
$$
\left[ J_{3}, T_{m}^{j} \right] = m T_{m}^{j}
$$
\begin{equation}\label{ce}
(J_{\pm} T_{m}^{j} - q^{m} T_{m}^{j} J_{\pm}) q^{J_{3}} =
\sqrt{\left[ j \pm m + 1 \right] \left[j - m \right]} T_{m \pm 1}^{j}
\end{equation}

\noindent
which reduce to the usual commutators for $q = 1$. Note, however, that the
generators of the algebra do {\it not} constitute a $q$-vector in the
quantum case.
Using now the representations of $J_{\pm}, J_{3}$ given by
$ \left( \begin{array}{cc}
           0 &1 \\
           0 & 0
          \end{array} \right)$,  $ \left( \begin{array}{cc}
           0 &0 \\
           1 & 0
          \end{array} \right)$,  $ \frac{1}{2} \left( \begin{array}{cc}
           1 &0 \\
           0 & -1
          \end{array} \right)$  (see Appendix A2) it is easy to check that
the `$q$-Pauli' matrices \cite{SONG-JPA} $\sigma_{\pm},
\sigma_{3}$ in the set
$$
\sigma_{0}=
\left[
\begin{array}{cc}
q & 0\\
0 & q
\end{array}
\right]\;; \qquad
\sigma_{+}= \left[2\right]^{1/2}
\left[
\begin{array}{cc}
0 & -q^{-1/2}\\
0 & 0
\end{array}
\right]\;,
$$

\begin{equation}\label{cf}
\sigma_{3}=
\left[
\begin{array}{cc}
-q & 0\\
0 & q^{-1}
\end{array}
\right]\;,\qquad
\sigma_{-}= \left[2\right]^{1/2}
\left[
\begin{array}{cc}
0 & 0\\
-q^{1/2} & 0
\end{array}
\right] \;,
\end{equation}

\noindent
constitute an $SU_{q}(2)$ $q$-vector; the additional $q$-sigma matrix,
$\sigma_{0}$$=$$qI$, is an $SU_{q}(2)$ scalar.
Using (\ref{cf}), the $K$ matrix adopts  the form
\begin{equation}\label{cg}
K = \sigma_{\mu} x^{\mu} =
\left[
\begin{array}{cc}
q(x^{0} - x^{3}) & -\left[2\right]^{1/2} q^{-1/2} x^{+}\\
-\left[2\right]^{1/2}q^{1/2} x^{-} & q x^{0} + q^{-1} x^{3}
\end{array}
\right] \equiv
\left[
\begin{array}{cc}
qD & B\\
A & C/q
\end{array}
\right]\;,
\end{equation}

\noindent
where ($A,B,C,D$) is the basis used in \cite{WA-ZPC48,SWZ-ZPC52,OSWZ-CMP};
thus, the time generator $x^0$=$q^{-1}[2]^{-1}tr_qK$ is central.

The $q$-Minkowski tensor (\ref{metric})
gives in the $(\alpha , \beta , \gamma , \delta )$ basis
\begin{equation}\label{ci}
g = \frac{-q^{-1}}{\left[2\right]}
\left[
\begin{array}{cccc}
\lambda q^{-1} & 0 & 0 & -1\\
0 & 0 & 1 & 0\\
0 & q^{2} & 0 & 0\\
-1 & 0 & 0 & 0
\end{array}
\right]\quad;
\end{equation}

\noindent
in the $x^{\mu}$ ($\mu$=$0,\pm,3$) and $x^I$=$(A, B, C, D)$ basis the metric
reads, respectively,
\begin{equation}\label{cj}
g_{\mu \nu}=
\left[
\begin{array}{cccc}
1 & 0 & 0 & 0\\
0 & 0 & -q^{-1} & 0\\
0 & -q & 0 & 0\\
0 & 0 & 0 & -1
\end{array}
\right] =g^{\mu \nu}
 \quad, \quad
g_{IJ}= \frac{-q^{-1}}{\left[2\right]}
\left[
\begin{array}{cccc}
0 & q^{2} & 0 & 0\\
1 & 0 & 0 & 0\\
0 & 0 & 0 & -1\\
0 & 0 & -1 & q \lambda
\end{array}
\right]
\;.
\end{equation}

\noindent
Similar  expressions are given in \cite{WA-ZPC48,WA-A6,SONG,OSWZ-CMP};
 the overall factor in (\ref{ci}) has been fixed so that in the
`physical' $x^{\mu}$ basis $g_{\mu \nu}$ has determinant one.

Using (\ref{cg}) and (\ref{dosl}), we find that the six basic commutation
relations for the $x's$ are given
by \cite{WA-ZPC48,SONG}
\begin{equation}\label{ck}
\begin{array}{ll}
\left[ x^{0}, x^{\pm}\right] = \left[x^{0}, x^{3} \right] = 0 \quad,& \qquad
x^{-} (x^{0} - x^{3}) = q^{-2} (x^{0} - x^{3}) x^{-}\quad,\\
\left[ x^{+}, x^{-}\right] = \lambda x^{3} (x^{0} - x^{3})\quad, & \qquad
x^{+} (x^{0} - x^{3}) = q^{2} (x^{0} - x^{3}) x^{+} \quad,
\end{array}
\end{equation}

\noindent
and in the ($A, B, C, D$) basis by \cite{WA-ZPC48,SWZ-ZPC52,OSWZ-CMP}
\begin{equation}\label{cl}
\begin{array}{ll}
\left[A, B\right] = -q^{-1} \lambda CD + q \lambda D^{2} \quad, &  \qquad
\left[B, C\right] = -q^{-1} \lambda BD\;,\\
\left[A, C\right] = q \lambda AD \quad ,& \qquad BD = q^{2} DB\;, \\
AD = q^{-2} DA \quad ,& \qquad
\left[D, C\right] = 0 \;.
\end{array}
\end{equation}

Since the metric $g_{\mu \nu}$ is not symmetric, a convention is needed to
rise and lower indices. We  shall adopt the convention
that $g$ acts on coordinates from the left and on the $q$-sigmas from the
right, $K = \sigma_{\mu} x^{\mu} = \sigma^{\nu} g_{\nu \mu} x^{\mu} =
\sigma^{\nu} x_{\nu}$. Thus,
\begin{equation}\label{cm}
\begin{array}{ll}
\sigma^{0} =
\left[
\begin{array}{cc}
q & 0\\
0 & q
\end{array}
\right] = \sigma_{0}\;,\qquad &

\sigma^{+}= \left[2\right]^{1/2}
\left[
\begin{array}{cc}
0 & 0\\
q^{3/2} & 0
\end{array}
\right] =-q \sigma_{-}\;,\; \\

\, &  \, \\

\sigma^{3}=
\left[
\begin{array}{cc}
q & 0\\
0 & -q^{-1}
\end{array}
\right]= - \sigma_{3} \;, \qquad &
\sigma^{-}=\left[2\right]^{1/2}
\left[
\begin{array}{cc}
0 & q^{-3/2}\\
0 & 0
\end{array}
\right] =-q^{-1} \sigma_{+}\;.
\end{array}
\end{equation}

\noindent
Then, $K= \sigma^{ \nu} x_{\nu}$ is given by
\begin{equation}\label{cn}
K=
\left[
\begin{array}{cc}
q x_{0} + q x_{3}  & q^{-3/2} \left[2\right]^{1/2} x_{-}\\
\left[2\right]^{1/2} q^{3/2} x_{+} & q x_{0} -q^{-1} x_{3}
\end{array}
\right]
\quad ,
\end{equation}
\noindent
and $l_q= q^2 g_{\mu \nu} x^{\mu} x^{\nu}= q^2 x_{\mu} x_{\nu}g^{ \nu \mu}$
[(\ref{dosah}) and (\ref{ch})]
\begin{equation}\label{cn2}
\begin{array}{lll}
l_q&=&q^2[(x^0)^2-q^{-1}x^+x^- -qx^-x^+ -(x^3)^2]  \\
\, &=&q^2[(x_0)^2-qx_+x_- -q^{-1}x_-x_+ -(x_3)^2] \;; \\
\,&\,& \, \\
l_q&=& CD-q^2AB \quad.
\end{array}
\end{equation}

A four vector basis $\rho^{\mu}$ for the covariant matrices $Y$ is
immediately obtained from the contravariant $\sigma^{\mu}$ by means of
$\hat{R}^{\epsilon}$. Since all these matrices are defined up to a
proportionality constant, we
introduce a factor $-q^{-1}$ by convenience to define
\begin{equation}\label{co}
\rho_{\mu} = -q^{-1} \hat{R}^{\epsilon} \sigma_{\mu} \quad ,  \quad
\rho^{\mu} = -q^{-1} \hat{R}^{\epsilon} \sigma^{\mu} \quad ,
\end{equation}

\noindent
where $(\hat{R}^{\epsilon} \sigma)_{ij}=
\hat{R}^{\epsilon}_{ij,kl} \sigma_{kl}$. Using (\ref{dosag})
 we obtain $\rho^{0} = \rho_{0} = q^{-1} I$ and
\begin{equation}\label{cp}
\rho_{\mu} = (\rho_{0}, -\sigma_{i}) \quad ,\quad \rho^{\mu} = (\rho^{0},
-\sigma^{i})
\end{equation}

\noindent
and check that the equivalent to (\ref{dosdb}) is satisfied,
\begin{equation}\label{cq}
\frac{1}{\left[2\right]} tr_{q} (\rho_{\mu} \sigma_{\nu}) = g_{\mu \nu} \quad ,
\quad \frac{1}{\left[2\right]} tr_{q} (\rho^{\mu} \sigma^{\nu})= g^{\nu \mu}\;,
\end{equation}
\noindent
where $g_{\mu \nu}$ is given by the first expression in (\ref{cj}). In terms of
the $\rho 's$, we find $Y=\rho^{\mu} \partial_{\mu} = \rho_{\mu}
\partial^{\mu}$
$$
Y=
\left[
\begin{array}{cc}
q^{-1}\partial_{0}-q \partial_{3} & -q^{-3/2} \left[2\right]^{1/2}
\partial_{-}\\
-q^{3/2} \left[2\right]^{1/2} \partial_{+} & q^{-1} (\partial_{0}
+ \partial_{3})
\end{array}
\right]
=
\left[
\begin{array}{cc}
q^{-1} \partial^{0} + q \partial^{3} & q^{-1/2} \left[2\right]^{1/2}
\partial^{+}\\
q^{1/2}\left[2\right]^{1/2} \partial^{-} & q^{-1} (\partial^{0} - \partial^{3})
\end{array}
\right]
$$
\begin{equation} \label{cr}
\equiv
\left[
\begin{array}{cc}
\partial_{D} & q^{-1} \partial_{A}\\
q \partial_{B} & \partial_{C}
\end{array}
\right]
\end{equation}

\noindent
($\partial_{\mu} \equiv \partial / \partial x^{\mu}$,
$\partial / \partial x_0 =\partial^0$, $\partial / \partial x_+
=q^2 \partial^+$, $\partial / \partial x_- =q^{-2} \partial^-$,
$\partial / \partial x_3 =\partial^3$)
and then (\ref{tc}) or (\ref{te})  gives the commutation relations \cite{SONG}
\begin{equation}\label{cs}
\begin{array}{ll}
\left[\partial_{0}, \partial_{\pm}\right] = \left[\partial_{0}, \partial_{3}
\right]= 0 \quad, & \quad \partial_{+} (\partial_{0} + \partial_{3})= q^{-2}
(\partial_{0} + \partial_{3})\partial_{+} \;, \\

\left[\partial_{+}, \partial_{-}\right]= \lambda \partial_{3} (\partial_{3}
+ \partial_{0}) \quad , &  \quad \partial_{-} (\partial_{0} + \partial_{3})=
q^{2}
(\partial_{0} + \partial_{3}) \partial_{-}\;,
\end{array}
\end{equation}

\noindent
which in the $\partial_I$ basis read \cite{OSWZ-CMP}
\begin{equation}\label{ct}
\begin{array}{ll}
\left[\partial_{A},\partial_{B}\right]= q \lambda (\partial_{D} \partial_{C}
- \partial_{C} \partial_{C}) \quad , & \qquad \left[\partial_{C},\partial_{D}
\right]=0\;,\\

\left[\partial_{A}, \partial_{D}\right]= -q^{3} \lambda \partial_{C}
\partial_{A} \quad , & \qquad \partial_{A} \partial_{C} = q^{2} \partial_{C}
\partial_{A}\;,\\

\left[\partial_{B}, \partial_{D}\right] = q \lambda \partial_{C} \partial_{B}
\quad ,&  \qquad \partial_{C} \partial_{B} = q^{2} \partial_{B} \partial_{C}\;.
\end{array}
\end{equation}

\noindent
In these basis, the definition (\ref{tac})  leads to
\begin{equation}\label{ct3}
d=dx^I \partial_I=[2] dx^{\mu} \partial_{\mu} =
[2] dx_{\mu} \partial / \partial x_{\mu}
\quad (x^I=A,B,C,D;\; \mu=0, \pm , 3)\;,
\end{equation}

\noindent
(the factor $[2]$ may be absorbed in the definition of the $\sigma$'s
and $\rho$'s)
and the D'Alembertian operator (\ref{ti}) reads
\begin{equation}\label{ct2}
{\Box}_q= \partial_C \partial_D  - q^{-2}\partial_A \partial_B =
q^{-2} ( \partial_0^2 - q \partial_+ \partial_- -q^{-1} \partial_-
\partial_+ - \partial_3^2)\;.
\end{equation}

\noindent
The relations (\ref{cs}) are, we note in
passing, the same that we would have obtained for the coordinates $x_{\mu}$,
and
those of (\ref{ck}) are the same as those for the derivatives $\partial^{\mu}$.
This fact, not evident a priori, becomes obvious
once we notice that $\hat{R}^{\epsilon}$
transforms $\sigma 's$ into $\rho ' s$, and that accordingly
the entries of $\rho^{\mu}
x_{\mu}$, whose non-commutativity properties are already fixed by those of the
coordinates $x_{\mu}$, have the same commutation properties as
$\rho^{\mu} \partial_{\mu}$ in $Y$.
It may be checked that the $K$-$\tilde{Y}$ and $K$-$Y$ equations
 in (\ref{ts}) reproduce, respectively,
the quadratic coordinates-derivatives relations  in the $x^{\mu}$,
$\tilde{\partial}_{\nu}$ basis \cite{SONG}
(once the $\partial$'s in \cite{SONG} are identified  with the
$\tilde{\partial}$'s in $\tilde{Y}$ and some misprints in formulae (3.9)
of this reference are corrected) and in the $x^{I}$, $\partial_{J}$
basis \cite{OSWZ-CMP}.

The above covariant operators may now be used to construct a
covariant $q$-Dirac operator.
Without making any explicit reference to a basis, it is given by
\begin{equation}\label{cu}
\Dslash =
\left[
\begin{array}{cc}
0 &-q Y^{\epsilon}\\
Y & 0
\end{array}
\right]
\quad
\end{equation}

\noindent
and transforms covariantly under the reducible $q$-Dirac spinorial
representation
\begin{equation}\label{cv}
 {\cal S} (\Lambda ) =
\left[
\begin{array}{ll}
M & 0\\
0 & \tilde{M}
\end{array}
\right]
\end{equation}

\noindent
i.e., $\;{\cal S} (\Lambda ) \Dslash {\cal S}^{-1} (\Lambda ) =\Dslash \!'\;
$. The Dirac gamma matrices may be extracted from (\ref{cu}) easily by
recurring to a
specific basis. In the $\partial_{\mu}$ basis, $\Dslash = \gamma^{\mu}
\partial_{\mu}$ leads to
\begin{equation}\label{cx}
\gamma^{\mu} =
\left[
\begin{array}{cc}
0 & \sigma^{\mu}\\
\rho^{\mu} & 0
\end{array}
\right] \quad,
\end{equation}

\noindent
where the $q$-matrices  $\sigma^{\mu}, \rho^{\mu}$ are given by (\ref{cm}) and
(\ref{cp}); in the
$(A, B, C, D)$ basis they are obtained immediately from (\ref{cr}) and the
analogous expression for $Y^{\epsilon}$, with the result that
\begin{equation}\label{cy}
\Dslash =
\left[
\begin{array}{cccc}
0 & 0 & q^{2} \partial_{C} & -q^{-1} \partial_A \\
0 & 0 & -q \partial_{B} & \partial_{D} + q \lambda \partial_{C}\\
\partial_{D} & q^{-1} \partial_{A} & 0 & 0\\
q \partial_{B} & \partial_{C} & 0 & 0
\end{array}
\right] \quad.
\end{equation}

\noindent
Factoring out the derivatives in (\ref{cy}), the obtained $\gamma^I$ coincide
with those given in \cite{SCHI-MPI93,SCHI-MPI92}. The operator (\ref{cu})
might be used for a $q$-Dirac equation;  by extension (for instance,
via the Bargmann-Wigner procedure) we may obtain a covariant operator
suitable for higher spin $q$-relativistic invariant equations.

  In fact, an immediate and physically necessary application is the analysis of
the deformed relativistic equations. In the preliminary free case, they may
be looked at (specially in momentum space) as the set of constraints defining
 the Hilbert space which is the support of an (almost) irreducible
 representation of the Poincar\'e group (the wavefunctions include
both signs of
the energy in their manifestly covariant formulation).
In the scalar Klein-Gordon case, the constraint is just the mass shell
 condition $p^2=m^2$. In the $q$-deformed case
the description of the one particle states \cite{PSW} and of the invariant
equations is more
complicated. It is possible to define the $q$-operators corresponding
to the covariant kernels of the $q$=1 relativistic equations,
like the Klein-Gordon, Dirac or even
Weinberg-Joos \cite{SONG,SCHI-MPI92,PILLIN,MEQ} or Bargmann-Wigner ones in the
arbitrary spin case  by using
their covariance as their defining property
as it may be done in $q$=1 case (see, e.g., \cite{ABR}).
We may then look to the solutions of the $q$-relativistic equations as the
kernels  of the corresponding covariant $q$-operators. The study of
$q$-deformed
field theory, however, is not without difficulties. For instance, as the
previous formalism clearly shows, we cannot have hermitean/antihermitean
coordinates and derivatives simultaneously; if the coordinates are taken to be
hermitean, the linear conjugation structure of the derivatives is lost
\cite{OSWZ-CMP,OZ-MPI25}. Moreover,
the space of solutions (`$q$-wavefunctions') is mathematically
not well defined, i.e., it is not clear what kind
of subspace or subalgebra of ${\cal M}_q$ they define, nor their
relation to the irreducible representations of the quantum Poincar\'e group.
Further, an analysis of simple non-relativistic second-quantized models shows
\cite{JW} that there may be a problem with locality in $q$-deformed theories.
All these problems require further study.

\setcounter{equation}{0}

\section{Non-commutative differential calculus and invariant operators:
the case of ${\cal M}_q^{(1)}$}

\indent
In the previous sections, the coordinates $K$, the $q$-derivatives $Y$ and the
$q$-differentials  $dK$ for the $q$-Minkowski space have been introduced.
 Using these basic elements, higher rank tensors and invariant
differential operators can be constructed respectively by tensoring or by
contraction.
The contraction (scalar product) of a covariant vector with a contravariant
one is given by the $q$-trace of the product of the
corresponding matrices. Examples of this procedure are the invariant
operators $l_{q}, \Box_{q}$ and $d$, already introduced.
Another one is the $q$-analogue of the dilatation operator,
\begin{equation}\label{tad}
s=tr_q(KY)\;.
\end{equation}

\noindent
Using the flexibility  of RE formalism, we shall now derive  a complete list
of the relations involving all these invariant operators and the generators
of the  ${\cal M}_{q}^{(1)}$, ${\cal D}_{q}^{(1)}$ and $\Lambda_{q}^{(1)}$
algebras; the other cases could be treated similarly. Since in this section
and in Sec.6 we  refer mostly to the ${\cal M}_q^{(1)}$ case,
we shall drop the  superindex henceforth.
First of all, since the time variable $x^{0}$ and the corresponding derivative
$\partial_{0}$ are singularized by being the linear central elements
of ${\cal M}_{q}$ and ${\cal D}_{q}$, let us deduce their commutation
relations
with the generators of ${\cal D}_q$ and ${\cal M}_q$.
For those between $x^{0}$=$q^{-1}[2]^{-1}tr_qK$ and $Y$
it is enough to take the $q$-trace of
the  eq. before the last in (\ref{ts}),
\begin{equation}\label{xd}
Y_2 R_{12}K_1R_{21}= R_{12}K_1R_{12}^{-1}Y_2 + q^2 R_{12}{\cal P}\quad,
\end{equation}
\noindent
with respect to the first space. For
$\partial_{0}$=$q[2]^{-1}tr_qY$ and $K$, eq. (\ref{xd}) is
multiplied  by $R_{12}^{-1}$ from the  left and by $R_{21}^{-1}$
from the right before taking the $q$-trace with respect to the second
space. In this way we get
\begin{equation}\label{cia}
Y c_{1} = c_{1} Y + q^{4} I - q^{2} \lambda Y K  \;,
\end{equation}
\begin{equation}\label{cib}
\partial_{0} K = K \partial_{0} + I - q^{2} \lambda K Y \;,
\end{equation}

\noindent
where the invariance of the $q$-trace and the first two of the
relations (see (\ref{ae2}), (\ref{af3}))
\begin{equation}\label{cic}
R_{21}^{-1} = R_{12} - \lambda {\cal P}_{12} = {\cal P}_{12} (q^{-1} I -
[2] P_{-\,12}) \;,
\end{equation}
\begin{equation}\label{cid}
tr_{q(1)} (R_{12} {\cal P}_{12})^{\pm1} = q^{\pm 2} I_{2} \quad,
\qquad tr_{q(2)} ({\cal P}_{12}R_{12})^{\pm1} = q^{\pm 2} I_{1} \quad,
\end{equation}

\noindent
were used. Inspection of formulae (\ref{cia}) and (\ref{cib})
demonstrates that setting $c_{1}$=0, $tr_qY \propto \partial_{0}$=0,
does not produces a consistent  reduction to a three dimensional space
algebra
since $tr_qY c_1=c_1 tr_qY + [2]- q^{-2} \lambda s$.

 Next, we compute the commutation relations involving the quadratic central
elements with generators. The $q$-Minkowski length $l_{q}$
satisfies the following relations with the generators of
${\cal D}_{q}$ and $\Lambda_{q}$:
\begin{equation}\label{cie}
Y \, l_{q} = q^{-2} l_{q} \, Y - q^{2} K^{\epsilon} \;,
\end{equation}
\begin{equation}\label{cif}
l_{q} \, dK = q^{-2} dK \, l_{q} \quad.
\end{equation}

\noindent
To obtain (\ref{cie}), $Y$ must go through $l_{q} P_{-\,12}$$=$
$-qP_{-\,12} K_{1} \hat{R}_{12} K_{1}$, what may be done multiplying
eq. (\ref{xd}) by
$R_{31} R_{32} K_{3} R_{21}^{-1} {\cal P}_{13}$ from the right and by
$P_{-\,13}$ from the left. Using then (\ref{xd}) once more in the r.h.s., the
Yang-Baxter equation (\ref{YBE}) to reorder the $R$-matrices, the $R$-matrix
$q$-determinant  $P_{-\,13} R_{12}
R_{32} = q P_{-\,13}$ and finally the expression (\ref{bh})
for the covariant vector
$K^{\epsilon}$,
eq.  (\ref{cie}) is obtained. In a similar way, eq. (\ref{cif}) follows
from  (\ref{tv}) multiplying it by $R_{32} K_{3} R_{23} R_{13}$ from the left,
using (\ref{YBE}) and (\ref{tv}) for $dK_{2}, K_{3}$ and
finally multiplying by $P_{-\,31}$ from the left.

Iterating (\ref{cie}) any power of $l_q$ can be differentiated
\begin{equation}\label{cig2}
Y \, l_{q}^n = q^{-2n} l_{q}^n \, Y - q^{2}[n;q^{-2}]K^{\epsilon}
l_q^{n-1} \;,
\end{equation}
\noindent
where
\begin{equation}\label{nq}
[n;q] \equiv \frac{q^n-1}{q-1}\quad,\quad [1;q]=1 \quad.
\end{equation}

For the relations involving  the D'Alembertian $\Box_{q}$ and  the generators
of ${\cal M}_{q}$ and $\Lambda_{q}$, we obtain
\begin{equation}\label{cih}
\Box_{q}\, K = q^{-2} K\, \Box_{q} - Y^{\epsilon} \;,
\end{equation}
\begin{equation}\label{cii}
\Box_{q} \,dK = q^{2} dK \, \Box_{q} \;.
\end{equation}
\noindent
Expression (\ref{cih}) (which can be seen as the `dual' analog of (\ref{cie}))
follows
from  eq. (\ref{xd}) multiplying it by $Y_{3} R_{32}^{-1} R_{13}$
from the left, using (\ref{YBE}), and again  (\ref{xd}) in the
r.h.s., and finally multiplying
the resulting equation by $R_{21}^{-1} {\cal P}_{23}$ from the left and by
$P_{-(32)}$ from the right; for the contravariant vector $Y^{\epsilon}$,
eq. (\ref{bi})
is used. Eq. (\ref{cii}) follows from (\ref{ty}) in a similar way.

Let us now look at $\Box_{q}$ and
$l_{q}$, eqs. (\ref{ti}) and (\ref{dosah}). Applying
$\Box_{q}$ to the product $K^{\epsilon}\, K$ leads to
\begin{equation}\label{cik}
\Box_{q} K^{\epsilon}z K = q^{-4} K^{\epsilon}z K \Box_{q} - q^{-2}
K^{\epsilon}z Y^{\epsilon} - YzK \;
\end{equation}

\noindent
by using the transformed (\ref{cih}),
\begin{equation}\label{cil}
\Box_{q} K^{\epsilon} = q^{-2} K^{\epsilon} \Box_{q} - Y \;.
\end{equation}

\noindent
To relate $\Box_q$$l_q$ to $l_q$$\Box_q$
we need the intermediate expression
\begin{equation}\label{cin}
Y_{1}\,K_{1} = q^{-2} tr_{q(2)} (R_{21} K_{2} \hat{R}_{21}^{-1} Y_{2}
R_{21}^{-1}) + [2] I_{1}\quad,
\end{equation}

\noindent
which follows from (\ref{xd}) by using the second equality in (\ref{cid}).
Its $q$-trace is
\begin{equation}\label{cio}
tr_{q} (Y\,K) = q^{-4} s + [2]^{2}\;.
\end{equation}

\noindent
Then, using  (\ref{cio})  and\footnote{The equality $tr_{q} (AB) =
tr_{q} (A^{\epsilon} B^{\epsilon})$ obviously holds for any pair
$A,B,$ if one is covariant and the other contravariant.}
$tr_{q} (K^{\epsilon} \, Y^{\epsilon})$=$ tr_{q} (K\,Y) \equiv s$, the
$q$-trace of  (\ref{cik}) gives final expression
\begin{equation}\label{cip}
\Box_{q} \, l_{q} = q^{-4} l_{q} \, \Box_{q} + q^{-2} s + (q^{2} + 1)\;.
\end{equation}

\noindent
Using the relation between $sl_q$ and $l_qs$ to be proved below [(\ref{ciz})]
and iterating (\ref{cip}) we find
\begin{equation}\label{cip2}
\Box_{q} \, l_{q}^n = q^{-4n} l_{q}^n \, \Box_{q} + a_n l_q^{n-1}\,s
+(q^2+1) b_n l_q^{n-1} \;,
\end{equation}

\noindent
where the coefficients $a_n$ and $b_n$ are determined by the recurrence
equations
\begin{equation}\label{cip3}
a_{n+1}=q^{-2}(q^{-4n}+a_n) \quad, \quad b_{n+1}=q^{-4n}+a_n+b_n \quad,
\quad n \geq 1 \quad,
\end{equation}

\noindent
($a_1=q^{-2}$, $b_1=1$) which give (see (\ref{nq}))
\begin{equation}\label{cip4}
a_{n}=q^{-2n}[n;q^{-2}] \quad, \quad
b_{n}=(q^{-2}+1)^{-1}[n+1;q^{-2}][n;q^{-2}] \;.
\end{equation}

Let us now consider the  exterior derivative $d$=$tr_{q} (dK \, Y)$.
Its action on coordinates, $q$-derivatives and $q$-differentials is given
by
\begin{equation}\label{ciq}
d \cdot K = (dK) +K\;d \;,
\end{equation}
\begin{equation}\label{cir}
Y\,d = q^{2} d\,Y + q \lambda dK^{\epsilon} \Box_{q} \;,
\end{equation}
\begin{equation}\label{cis}
d\,(dK) = -(dK)\,d \quad.
\end{equation}

\noindent
The first one is the expression of the Leibniz rule and (\ref{cis}) reflects
the nilpotency of $d, d^{2}=0$. These relations are easily obtained using the
invariance of the $q$-trace, the appropriate RE and the
properties of the $R$-matrix. For example, for (\ref{ciq}) we write
\begin{equation}
d \cdot K_{2} = tr_{q(1)} (dK_{1} Y_{1}) K_{2}
= tr_{q(1)} (R_{21}^{-1} dK_{1} Y_{1} R_{21}) K_{2}
\end{equation}

\noindent
and, using (\ref{xd}) and (\ref{tv}) in the form
$dK_1R_{21}K_2R_{21}^{-1}=R_{21}K_2R_{12}dK_1$, we obtain
\begin{equation}
\begin{array}{ccl}
d \cdot K_2 &=& tr_{q(1)} ( K_{2} R_{12} dK_{1} Y_{1} R_{12}^{-1} )
+ q^{2} tr_{q(1)} (R_{21}^{-1} {\cal P}_{12} dK_{2} ) \\
  &=& K_{2} tr_{q(1)} (R_{12} dK_{1} Y_{1} R_{12}^{-1}) + q^{2} tr_{q(1)}
(R_{21}^{-1} {\cal P}_{12}) dK_{2},
\end{array}
\end{equation}

\noindent
from which  and from (\ref{cid}) eq. (\ref{ciq}) immediately follows.

The relation (\ref{cir}) is obtained in a similar way. The RE
(\ref{tc}), (\ref{ty}), and the relations (cf. (\ref{cic}))
\begin{equation}\label{cit}
R_{21} R_{12} = q^{2} I - \lambda [2] P_{-\,12} \quad,
\end{equation}
\begin{equation}\label{ciu}
d K_{1}^{\epsilon} = [2] tr_{q(2)} ( \hat{R}_{12} dK_{1} P_{-\,12} ) \quad,
\end{equation}

\noindent
(cf. (\ref{bh})) are enough to get (\ref{cir}), starting from
\begin{equation}
Y_{2} d = tr_{q(1)} ( Y_{2} R_{21}^{-1} dK_{1} Y_{1} R_{21} ) \quad .
\end{equation}
\noindent
Eq. (\ref{cis}) is  obtained from $d\;dK_{2}=tr_{q(1)}
(R_{12} dK_{1} Y_{1} R_{12}^{-1} dK_{2})$, using only (\ref{ty}) and
(\ref{tv2}).

Relations (\ref{cir}) and (\ref{cis}) can be used to show explicitly that
$d^{2} = 0$, since
\begin{equation}
d^{2} = d\;d =d\; tr_{q} (dK\;Y) = tr_{q} (d\; dK\;Y)\quad.
\end{equation}

\noindent
Using now (\ref{cir}) and (\ref{cis}) to move $d$ through $dK\;Y$
\begin{equation}
d^{2}=-q^{-2} tr_{q} (dK\;Y)\;d +
q^{-1} \lambda tr_{q} (dK \; dK^{\epsilon}) \,\Box_{q}\;,
\end{equation}
\noindent
and $tr_{q} (dK\;dK^{\epsilon})=0$ [(\ref{tw2})], one finds that
$d^{2}= -q^{-2} \,d^2$, or that $d^{2}=0$ ($q$ is real).

The previous set of relations allow us to compute in a direct way
the commuting properties of
$d$ with any invariant operator. The action of $d$ on
the quadratic central elements $ \Box_q$ and $l_q$ gives
\begin{equation}\label{civ}
d \,\Box_{q} = q^{-2} \Box_{q} \,d \;,
\end{equation}
\begin{equation}\label{ciw}
d l_{q} = l_{q} d - q^2 W \;, \qquad
W \equiv tr_{q} (dK\,
K^{\epsilon})\quad.
\end{equation}

\noindent
Eq. (\ref{civ}) is obtained from the expression (\ref{ti}) for
$\Box_{q}$ and by using twice (\ref{cir})
(a simpler possibility is to apply  $d=tr_{q} (dKzY)$ to
(\ref{cii})). In the same way, to find (\ref{ciw}) we may use (\ref{ciq}) twice
and that

\begin{equation}\label{cix}
tr_{q} (K\;dK^{\epsilon}) = q^{2} tr_{q} (dK\;K^{\epsilon})\equiv q^{2} W \;,
\end{equation}
\noindent
(a property which follows from (\ref{tv}))
or,
equivalently, eqs. (\ref{cie}) and (\ref{cif}) to move $l_{q}$ through $d$.

The relation of $d$ with the $q$-analogue [(\ref{tad})]
of the   invariant dilatation operator
is easily obtained using (\ref{ciq}) and (\ref{cir}):

\begin{equation}
\begin{array}{ccl}
d\;s &=& d \,tr_{q} (K\,Y)
=  tr_{q} (dK\;Y)   +tr_{q} (K\,d\,Y) \\

   & =& tr_{q} (dK\;Y) +  q^{-2} tr_{q} (K\,Y) \,d - q^{-1} \lambda
tr_{q} (Kz\,dK^{\epsilon}) \Box_{q} \quad
\end{array}
\end{equation}

\noindent
so that, recalling (\ref{cix}) we get
\begin{equation}\label{ciy}
d\,s = q^{-2} s\,d - q \lambda W \, \Box_{q} + d \;.
\end{equation}

The operator $s= tr_{q}(KY)$ is related (see below) to a central element for
the
complete  algebra ${\cal M}_{q} \times {\cal D}_{q} \times \Lambda_{q} $,
and it is useful in the irreducible representation description
(cf. the $q$-oscillator case \cite{K-TMF,RID}).
Therefore, it is important to have the complete set of relations of $s$ with
the generators of ${\cal M}_{q} ,\, {\cal D}_{q}$ and $\Lambda_{q}$, and
with the central elements $l_{q}$ and $\Box_{q}$. The latter ones may be
obtained in a  simple way. Using eq. (\ref{cie}) for $l_{q}$ and
(\ref{cih}) for $\Box_{q}$, the resulting relations are given by
\begin{equation}\label{ciz}
s \, l_{q} = q^{-2} l_q \, s + (q^{2} + 1) l_{q} \;,
\end{equation}
\begin{equation}\label{ciaa}
\Box_{q} z s = q^{-2} s \Box_{q} + (1 + q^{2}) \Box_{q} \;.
\end{equation}

\noindent
The operator $s$ commutes with the elements of $ \Lambda_q$, for
\begin{equation}\label{ciab}
s z dK = dK z s \quad.
\end{equation}
\noindent
Indeed, the invariance of the $q$-trace  permits us to
write
\begin{equation}\label{ciab2}
s z dK_{2} = tr_{q(1)} (R_{12} K_{1} Y_{1} R_{12}^{-1}) dK_{2} \quad ,
\end{equation}

\noindent
and eqs. (\ref{ty})  and (\ref{tv}) transform the r.h.s. of (\ref{ciab2})
into
$tr_{q(1)}$$(dK_{2}$ $ R_{12}$$K_{1} Y_{1} R_{12}^{-1})$=$dKzs$
again by the invariance of the $q$-trace.

The relations among $s$ and the coordinates $K$ and the $q$-derivatives $Y$
are more complicated. Multiplying eq. (\ref{xd}) by $R_{21} K_{2}$
from the left and by $R_{21}^{-1}$ from the right, taking the $q$-trace in the
second space and using (\ref{dosj}) and the identities
(\ref{cic}), (\ref{cid}) one gets
\begin{equation}\label{ciac}
s K = K s - q^{2} \lambda K Y K + q^{4} K \;.
\end{equation}
\noindent
For the covariant combination $KYK$, it follows
using eqs.  (\ref{cin}), (\ref{dosj}), (\ref{xd}),
(\ref{bi}) and the defining relation (\ref{dosx}) for $l_{q}$ are used,  that
\begin{equation}\label{ciad}
K\, Y\, K = q^{-1} s\, K + q K + q^{-1} l_{q}\, Y^{\epsilon} \;.
\end{equation}

\noindent
Thus, it follows from (\ref{ciac}) and (\ref{ciad}) that

\begin{equation}\label{ciae}
s K = q^{-2} K s - q^{-1}\lambda  l_{q} Y^{\epsilon} +K \;.
\end{equation}

The relation between $s$ and $Y$ is found in the same way. Multiplying
eq. (\ref{xd}) by $R_{21}^{-1} Y_{1} R_{21}$ from the right, using
(\ref{tc}) and taking the $q$-trace in the first space with the help
of (\ref{cic}), (\ref{cid}), one gets
\begin{equation}\label{ciaf}
s \,Y = Y \,s + \lambda q^{2} Y \,K \,Y - q^{4} Y \;.
\end{equation}
\noindent
To compute  $YKY$ we use (\ref{cin}), $\Box_{q}$ [(\ref{tf})],
(\ref{cic}), (\ref{cid})
and the expression (\ref{bh}) for $K^{\epsilon}$  with the result
\begin{equation}\label{ciag}
Y \,K \,Y = q^{-3} s\, Y + q^{-1} K^{\epsilon} \,\Box_{q} + [2] Y \;;
\end{equation}
\noindent
then, from (\ref{ciaf}) and (\ref{ciag}) we obtain
\begin{equation}\label{ciah}
Y\, s = q^{-2} s \,Y - q \lambda K^{\epsilon} \,\Box_{q} + Y \;.
\end{equation}

Finally, we mention that it  is possible to find a central element in
${\cal M}_{q} \times {\cal D}_{q} \times \Lambda_{q}$ using the operators
$s,\, l_{q}, \, \Box_{q}$ and a grading operator $N$ defined by the
relations
\begin{equation}\label{ciai}
[N, K] = K \quad , \quad [N, dK]=0 \quad, \quad [N, Y]= -Y \quad ,
\end{equation}
\noindent
which may be introduced since
${\cal M}_{q}$ and ${\cal D}_{q}$ are graded algebras.
Indeed, the element $z$ (cf \cite{OSWZ-CMP})
\begin{equation}\label{ciaj}
z=q^{2N}\tilde{s}\quad, \quad
\tilde{s} \equiv (q^{-2} -1)s + (q^{-2} -1)^{2} l_{q} \Box_{q}+ 1 \quad ,
\end{equation}

\noindent
is central in ${\cal M}_{q} \times {\cal D}_{q} \times \Lambda_q $.
Using the set of
relations among $s,\, l_{q}$ and $\Box_{q}$ and the generators $K, \,Y$ and
$dK$,  i.e., (\ref{cie}), (\ref{cif}),
(\ref{cih}), (\ref{cii}), (\ref{ciab}), (\ref{ciae}) and (\ref{ciah}),
it is found that $\tilde{s}$ is a scaling operator
\begin{equation}\label{scaling}
\tilde{s} K= q^{-2}K \tilde{s} \quad, \quad
\tilde{s} Y= q^{2}Y \tilde{s}
\end{equation}

\noindent
and $\tilde{s} dK$=$dK \tilde{s}$. Since, in contrast, eq. (\ref{ciai})
gives $q^{2N}K$=$q^2Kq^{2N}$ and $q^{2N}Y$=$q^{-2}Yq^{2N}$
the centrality of $z$ follows.

\vspace{1\baselineskip}

As an example of how this discussion may be extended to other
$q$-spacetimes, let us consider the $q$-dilatation operator $s$ for the
${\cal M}_q^{(3)}$ case. With $Z$ and $D$ given in (\ref{dosas}) and
(\ref{tae}), it is given by $s$=$tr(ZD)$ (cf. (\ref{tad})) and satisfies
\begin{equation}
sZ=Z(s+1) \quad,\quad sD=D(s-1) \quad;
\end{equation}
\noindent
which exhibit once again (cf. (\ref{ciae}), (\ref{ciah})) the `triviality'
(or twisted character) of ${\cal M}_q^{(3)}$. These relations are easily
obtained using (\ref{dosat}) and (\ref{taf}). For instance,
\begin{equation}
\begin{array}{ll}
sZ & =tr_{(1)}(Z_1D_1)Z_2=tr_{(1)}(Z_1(VZ_2D_1V^{-1} + {\cal P}))\\
\, & = tr_{(1)}(Z_2VZ_1D_1V^{-1} +Z_1 {\cal P})=
Z_2tr_{(1)}(VZ_1D_1V^{-1})  + Z_2 = Z(s+1)\\
\end{array}
\end{equation}
\noindent
since $tr_{(1)}{\cal P}=I_2$.

\setcounter{equation}{0}

\section{Comments on representations of quantum space-time
algebras and other  problems}

\indent
   In classical and quantum relativistic theory the Poincar\'e group
transformations may be realized in terms
of a complete set of observables.
The generators of the infinitesimal Lorentz transformations are functions of
$x^\mu$ and $p_\nu$,
\begin{equation}\label{sa}
 M_{\mu\nu} =x_\mu p_\nu -x_\nu p_\mu \;,
\end{equation}
\noindent
so that $M_{\mu\nu}$ and $p_\nu$ define a basis of the ten dimensional
Lie algebra ${\cal P}$ of the Poincar\'e group $P$. In the $q$-deformed case
we have encountered the analogue of the momenta $p_\mu$, provided by the four
quantum
derivatives $Y_{ij}$ giving the $q$-translation {\it algebra}
${\cal D}_q$, and the  quantum Lorentz {\it group} transformations,
described by the six independent entries of $M$, ${\tilde M}
=(M^{-1})^\dagger$ and the corresponding coaction
(\ref{dosh}), (\ref{ta}). Due to the quantum group structure, the relation
between the quantum group and the corresponding quantum algebra is
expressed by the duality among these Hopf algebras: the elements of one
are linear functionals for the other. The duality relation
is an abstract one, but its explicit realization requires selecting
a basis of the Hopf algebras in question, which (in our case) are
constructed from a finite number of generators. Once we change the
generators (often nonlinearly) of a  Hopf algebra ${\cal A}$ there is
no easy and canonical way to find the corresponding change in the dual
Hopf algebra ${\cal A}^*$ generators. If the quantum Poincar\'e group
$P_q$ and the corresponding quantum algebra $(P_q)^*
={U}_q({\cal P})$ must have ten generators each, then from the previous
sections we can extract six generators of $ L_q\subset P_q$
and four generators of ${\cal D}_q \subset { U}_q({\cal P})$
(the subindex in $U_q$ denotes quantization of the universal enveloping
algebra).
A  set of generators of the quantum Poincar\'e algebra
 was found in \cite{OSWZ-MPI51,OSWZ-CMP} and
it was transformed into another
set in \cite{SCHI-MPI93}. However, the duality relation with the $P_q$
and ${U}_q({\cal P})$ basis remains
to be clarified. For instance, the $L_q$ covariance of the  braided
coadditions of four-vectors requires the addition of a dilatation element
\cite{MeyerM,SMPoinc} (see also \cite{MAJCLA}). This leads to a
$q$-Poincar\'e group which
is not included  in \cite{PWOR} because this classification does not include
quantum Poincar\'e groups with dilatations (the $P_q$ in \cite{OSWZ-CMP} is
also  not included  in the scheme of \cite{PWOR}).

   Once the complete set of the $q$-Poincar\'e algebra generators is
obtained, one may adopt the Newton-Wigner's point of view and identify
the $q$-deformed
elementary systems (particles) with the unitary irreducible representations
of ${U}_q({\cal P})$. These were constructed in \cite{PSW} for the scalar
particle. Nevertheless, there are still some difficulties that require
further study. To illustrate them, it is sufficient to consider
the representations of the $q$-translation algebra ${\cal D}_q$
as a subalgebra of ${U}_q({\cal P})$.
Since ${\cal D}_q$ is isomorphic to ${\cal M}_q$, eq. (\ref{tj}),
we may consider the
irreducible unitary representations of this algebra \cite{PPK}. In the
non-deformed case ($q$=1) this problem is trivial: since the algebra of
translations or coordinates is commutative, its irreducible representations are
one-dimensional. For $q \neq 1$, however, this is no longer the case.

Due to the fact that one central element of ${\cal M}_q$
is linear in its generators one can
change the basis  (\ref{dosi}) to $\alpha , \beta , \gamma$ and
$\tau$,
$q \tau \equiv  c_{1} = q^{-1} \alpha + q \delta$. Then, there are
three non-trivial
commutation relations ($\hat{\lambda} \equiv \lambda /q \equiv (1-q^{-2})$;
$\hat{ \lambda}>0$ for $q>1$)
$$
\alpha \gamma = q^{2} \gamma \alpha \; ,
\quad \alpha \beta = q^{-2} \beta \alpha \;,
$$
\begin{equation}\label{sd}
\beta \gamma = q^{2} \gamma \beta + \hat{\lambda} (l_q-\alpha^{2})\;,
\end{equation}

\noindent
which follow from (\ref{dosl}) plus the centrality of $\tau$ and of
(cf. (\ref{dosaa}))
\begin{equation}\label{se}
l_q=\alpha \tau - \alpha^{2}/q^{2} - q^{2} \gamma \beta = q^{2} (\alpha \tau -
\alpha^{2} - \beta \gamma ) \;.
\end{equation}

\noindent
To analyze the irreducible representations in a Hilbert space with positive
metric the following consequences of (\ref{sd}) are useful:
\begin{equation}\label{sf}
\beta \gamma^{n} = (q^{2} \gamma )^{n} \beta + \hat{\lambda} [n; q^{2}]
\gamma^{n-1} (l_q-q^{2(n-1)} \alpha^{2})\;,
\end{equation}

\begin{equation}\label{sg}
\gamma \beta^{n} = (q^{-2} \beta)^{n} \gamma - \hat{\lambda}/q^{2} [n;q^{-2}]
\beta^{n-1} (l_q-q^{-2(n-1)} \alpha^{2})\;.
\end{equation}

\noindent
These relations may be proved by induction or by assuming,
looking at (\ref{sd}) that, e.g.,
\begin{equation}\label{sg2}
\beta \gamma^{n} = (q^{2} \gamma )^{n} \beta + \hat{\lambda}
\gamma^{n-1} (a_n l_q- b_n \alpha^{2})\;,
\end{equation}

\noindent
which gives the recurrence relations $a_{n+1}=q^{2n}+a_n$ and $b_{n+1}=
q^{2n}+q^4b_n$, which have the solutions $a_n=[n;q^2]$ and
$b_n=q^{2(n-1)}[n;q^2]$.

Since  ${\cal M}_{q} \approx  {\cal M}_{q^{-1}}$ \,[(\ref{tj})]
we may assume  $q>1$. The irreducible representations (irreps)
are parametrized by the
different values of the central elements $l_q$ (denoted by $l$) and $\tau$.
The representations fall into two broad categories, $l \leq 0$ or
$l>0$, but there are subclasses in each of them, which are listed below.

\vspace{1\baselineskip}
\noindent
{\bf 1.}  $\alpha = 0$, then the other three generators $ \beta , \gamma ,
\delta$
commute among themselves as it follows from (\ref{dosl}). Hence, $\delta$ is
real and arbitrary while $\beta =
\gamma^*$ is an arbitrary complex number, $\tau = \delta , l= -q^{2} |
\gamma |^{2}$. This irrep is not faithful.
It gives a one-dimensional irrep of  (\ref{dosl}).

\vspace{1\baselineskip}
\noindent
{\bf 2.}  $l- \alpha^{2} =0 \; ,\;
\delta=\alpha \: , \beta=\gamma = 0, \tau = \alpha (1 + q^{-2}).$
This is also a one-dimensional representation, which is not faithful and
corresponds to the stationary point $\alpha I$ of the coaction $K \mapsto
UKU^{\dagger}$
of the quantum
`subgroup' $SU_{q}(2)$ of $ L_{q}$.

\vspace{1\baselineskip}
\noindent
{\bf 3.}  $l > \alpha^{2}_{0} > 0$, where $\alpha_{0}$ is the vacuum eigenvalue
of
$\alpha$ and $\beta |0\rangle=0$. Then from (\ref{sf}) for unnormalized
eigenvectors
$|n \rangle=\gamma^{n} |0 \rangle$ of $\alpha$ one gets
\begin{equation}\label{sh}
\langle n|n\rangle=\langle 0|(\gamma^*)^n \gamma^n | 0 \rangle=
\langle 0| \beta^n \gamma^n | 0 \rangle=
 (\hat{\lambda})^{n} [n;q^{2}]! \Pi_{k=1}^{n} (l-q^{2(k-1)}
\alpha_{0}^{2})\;,
\end{equation}

\noindent
where $[n;q]!=[n;q][n-1;q]...1$. Clearly, $\langle 1|1 \rangle =
\hat{\lambda} (l- \alpha_0^2)>0$, but
for $\alpha_{0} \neq 0$ and $q>1$ the norm will be negative if the integer $n$
is sufficiently large. Because we are looking for irreps in a Hilbert space
with states of positive norm there must be some $n$, $N=d-1$, such that $\|
|N+1\rangle \| \sim
(l-q^{2N} \alpha_{0}^{2})=0$, or

\begin{equation}\label{si}
\gamma |N\rangle=\gamma |d-1\rangle=0 \; ,
\end{equation}

\noindent
where $d$ is the dimension of the irrep and hence

\begin{equation}\label{sj}
l=q^{2d} \alpha_{0}^{2}/q^{2} \quad ,\quad \tau=(q^{2d}+1) \alpha_{0}/q^{2}\;,
\end{equation}
\noindent
where the last expression follows from computing $l_q |N\rangle =
q^2(q^{2N}\tau \alpha_0-q^{4N}\alpha_0^2)|N\rangle$ $= l|N\rangle$
using (\ref{se}).

\vspace{1\baselineskip}
\noindent
{\bf 4.} $\alpha_{0}^{2}>l>0$,
hence $(l - \alpha_{0}^{2})<0$. From (\ref{sf}) one now
concludes that $\beta$ cannot be an annihilation operator. So we have to use
(\ref{sg}) supposing that $\gamma$ is now the annihilation operator,
 $\gamma |0\rangle=0$. Then for $|n\rangle= \beta^{n} |0\rangle$ one gets

\begin{equation}\label{sk}
\langle n | n\rangle = (\hat{\lambda}/q^{2})^{n} [n; q^{-2}]!
\Pi^{n}_{k=1}(q^{-2(k-1)}
\alpha_{0}^{2} -l)\quad.
\end{equation}

\noindent
Using the same positivity arguments of the previous case we now obtain,
with $d=N+1$ as before,
\begin{equation}\label{sl}
l=q^{-2(d-1)}\alpha_{0}^{2}\quad, \quad \tau=(q^{-2d}+1) \alpha_{0}\quad.
\end{equation}

\vspace{1\baselineskip}
\noindent
{\bf 5.}  $ l \leq 0 \; , \; \alpha \neq 0$, hence $(l-\alpha^{2})<0$ and one
has to
use (\ref{sg}) with $\gamma$ as the annihilation
operator $\gamma | 0\rangle=0$. In this way  one
obtains for $|n\rangle=\beta^{n} |0\rangle$ again  (\ref{sk}),
which is now positive for any integer $n$. This irrep is infinite dimensional.

\vspace{1\baselineskip}
\noindent
There are some  relations of the $q$-Minkowski algebra ${\cal M}_q$
with other  $q$-algebras: $su_{q}(2)$, the $q$-oscillator algebra ${\cal A}(q)$
and the $q$-sphere $S_q^2$.
For instance,  once $\tau$ and $l$ are fixed the
relations (\ref{sd}), (\ref{se}) coincide with the defining relations of the
quantum sphere
algebra \cite{PODL,NOUMI}.

The algebra ${\cal M}_q$ is isomorphic to the $q$-derivative or $q$-momentum
algebra ${\cal D}_q$, hence the irreducible representations coincide with those
found in \cite{PSW} for the $q$-deformed Poincar\'e algebra, which has the
algebra
${\cal D}_q$ as a subalgebra.
Once we identify ${\cal M}_q \approx  {\cal D}_q$, we can set $m_q^2$=$l_q$
and $\tau$=$[2]p_0$; the energy is a central element. However, although
unitarity permits any real value
for $m_q^2$ and $p_0$,
the physical meaning of the central elements of ${\cal D}_q$ requires that the
eigenvalues of the energy $p_0$ and the square of  mass  $m_q^2$  satisfy the
obvious physical restrictions $p_0^2 \geq m_q^2 \geq 0$.
If $m_q^2$=$l>$0, one may eliminate $\alpha_0$ from (\ref{sj}) or (\ref{sl}).
In both cases we obtain
\begin{equation}\label{pcerod}
p_0^2=m_q^2 \frac{(q^d+q^{-d})^2}{q^2[2]^2}\;.
\end{equation}
\noindent
In the classical $q \rightarrow 1$ limit, the values of all generators are
fixed while the dimension $d \rightarrow \infty$ in such a way that the
factor $(q^d+q^{-d})^2$ is also fixed and the usual formula
$p_0^2=m^2+ \vec{p}^2$ is reproduced.

\vspace{1\baselineskip}

The next step in the representation theory is related to the construction
 of a representation in the tensor product of two
irreducible representations. It depends on existence of a bialgebra
(or a Hopf algebra) structure for the algebra ${\cal M}_q$ or a homomorphism
from ${\cal M}_q$ to ${\cal M}_q \otimes {\cal M}_q$. The  existence of such a
map could be
interpreted physically as the $q$-Lorentz group covariance for two- (in
general  multi-)
particle system. There are a few proposals for a possible `coproduct'
$\Delta  : {\cal M}_q \rightarrow  {\cal M}_q \otimes {\cal M}_q$.
These proposals use:

\vspace{1\baselineskip}
\noindent
{\bf 1.}  the relation of ${\cal M}_q$ to the quantum algebra $su_q(2)$,
extending it
to isomorphism (modulo some additional requirements) and introducing the
bialgebra structure through the factorization \cite{ALEK} (here and below
the indices (1), (2) refer to the factors in ${\cal M}_q \otimes {\cal M}_q$)
\begin{equation}\label{sm}
K=L^{(+)}(L^{(-)})^{-1}=L^{(+)}_{(1)}K_{(2)}(L^{(-)}_{(1)})^{-1}\;.
\end{equation}

\vspace{1\baselineskip}
\noindent
{\bf 2.} the  appropriate non-commutativity (`braid statistics') of the factors
in a
`braided tensor product'
${\cal M}_q \otimes {\cal M}_q$ \cite{MAJ-LNM,SMBR2,KS}.
It is not difficult to check
that the matrix product of two matrices
\begin{equation}\label{sn}
\Delta(K)=K \dot{ \otimes} K=K^{(1)}K^{(2)} \quad , \quad \Delta(K)_{ij}
=K_{il} \otimes  K_{lj} \quad,
\end{equation}

\noindent
satisfying the commutation relations given by \cite{SM2,ISVL}
\begin{equation}\label{sn2}
\hat{R}^{-1}K_1^{(2)}\hat{R}K_1^{(1)}=K_1^{(1)}\hat{R}^{-1}K_1^{(2)}\hat{R}\quad,
\end{equation}
\noindent
satisfy the RE (\ref{dosj})  and that its entries
generate an algebra isomorphic to  ${\cal M}_q$. Eq. (\ref{sn2}),
however, is not  preserved under the coaction (\ref{dosh}), (\ref{dosf}).

\vspace{1\baselineskip}
\noindent
{\bf 3.} the  coaddition  of the two 2$\times$2 matrices expressed by
\begin{equation}\label{so}
\Delta (K)=K^{(1)}+K^{(2)} \equiv K \otimes 1 + 1 \otimes K\quad.
\end{equation}

\noindent
If the non-commutativity   between the
generators of   $K^{(1)}$ and  $K^{(2)}$ is given e.g., by
(\cite{MeyerM,SM2})
\begin{equation}\label{sbrai}
\hat{R}K_1^{(2)}\hat{R}K_1^{(1)}=K_1^{(1)}\hat{R}K_1^{(2)}\hat{R}^{-1}\quad,
\end{equation}
\noindent
the corresponding set of commutation relations permits
 that $ \Delta (K)$  satisfies (\ref{dosj}).
Eq. (\ref{sbrai}) is easily obtained by imposing that the matrix $K_1^{(1)}
+K_1^{(2)}$ satisfies (\ref{dosk}) and using that $\hat{R}$ is of Hecke type.

\vspace{1\baselineskip}
\noindent
{\bf 4.}  an additional matrix ${\cal O}$ including the $q$-Lorentz algebra
generators acting on the first factor of  ${\cal M}_q \otimes {\cal M}_q$
such that the matrix
\begin{equation}\label{sp}
\Delta (K)=K^{(1)}+{\cal O}^{(1)} \otimes K^{(2)}
\end{equation}

\noindent
will satisfy the RE (\ref{dosj}), while the entries of $K^{(1)}$ and $K^{(2)}$
commute \cite{OSWZ-CMP} (a kind of `undressing' of the preceding case).

Though last two cases look physically reasonable, they, together with {\bf 2},
are
not symmetric with respect to the permutation of factors
(notice that (\ref{sbrai}) is not symmetric) and not all irreps
of $K^{(1)}$ and $K^{(2)}$ are compatible.

\vspace{1\baselineskip}

 The form of the coproduct depends on the chosen
basis. The coproduct of the
$q$-Lorentz algebra part of ${U}_q({\cal P})$ looks simple in the $sl_q(2)$
basis \cite{FRT1} $L^{\pm},{\tilde L}^{\pm}$  related to the matrices
$M,\,{\tilde M}$. However, the basis of \cite{OSWZ-CMP} is related to the
$SU_q(2)$ `subgroup' of $ L_q$. The coproduct for the $q$-derivatives
subalgebra ${\cal D}_q$ in this basis looks like \cite{OSWZ-CMP}
\begin{equation}\label{sb}
\Delta(Y_{ij})=Y_{ij}\otimes I +\sum_{kl} l_{ij,kl}\otimes Y_{kl}\quad,
\end{equation}
\noindent
where the $l_{ij,kl}$ are made up of the $q$-Lorentz algebra generators
and the scaling operator (which is outside the $q$-Lorentz algebra).
One of the obvious property of this set of elements of the dual algebra
$( L_q)^*$ of $L_q$
follows from the covariance requirement, which must be preserved by the
coproduct:
\begin{equation}\label{sc}
Y\mapsto {\tilde M}YM^{-1}  \Rightarrow
\Delta(Y)\mapsto {\tilde M}\Delta(Y) M^{-1} \quad .
\end{equation}
\noindent
This form of the coproduct demonstrates that the representation theory
of the $q$-momentum subalgebra ${\cal D}_q \subset {U}_q ({\cal P})$ is
not closed: one cannot consider the tensor product of arbitrary irreducible
representations $V_1, V_2$ of ${\cal D}_q$, related to $Y\otimes I$
and $I\otimes Y$. It is necessary to take the $V_1$, reducible
generally speaking, that permits an irreducible representation of the
whole set, $Y_{ij}$ and $l_{ij.kl}$. This raises the question of the
physical interpretation and its consequences: the observables of two
particles enter into the coproduct  in an asymmetric way.

If we introduce two sets of the triangular $L^{\pm}$, $\tilde{L}^{\pm}$
matrices corresponding to $M$ and $\tilde{M}$, then their
entries (six of them) will define a basis of generators of the quantum Lorentz
algebra  $(L_q)^*$ dual to $L_q$.  The commutation relations of
the entries of $L^{\pm}$
are the standard ones ($sl_q(2)$, see Appendix A) and the same
is true for $\tilde{L}^{\pm}$. They could commute in between $[L^{\pm}_{ij},
\tilde{L}^{\pm}_{kl}]$, for the coproducts of $M$ and $\tilde{M}$ are the
usual ones, $\Delta (M)=M \otimes M$. However, in order to have the usual
coproduct for
$L^{\pm}$, $\tilde{L}^{\pm}$ due to the non-commutativity of $M$
and $\tilde{M}$ (\ref{dosf}) one can introduce a `mild'  non-commutativity
between the entries of  $L^{\pm}$ and $\tilde{L}^{\pm}$ too. Then, the
corresponding duality relations are:
\begin{equation}\label{sc2}
\begin{array}{ll}
<L^{\pm}_1,M_2>=R^{\pm}_{12}\;,\quad & \qquad
< \tilde{L}^{\pm}_1, \tilde{M}_2>=R^{\pm}_{12}\;,\\

<L^{\pm}_1, \tilde{M}_2>=A_{12}\;, \quad & \qquad
<\tilde{L}^{\pm}_1, M_2>= \tilde{A}_{12}\;.
\end{array}
\end{equation}

\noindent
Then, the action of such operators on the algebra ${\cal M}_q$ (the entries
of $L^{\pm}$ and $\tilde{L}^{\pm}$ are now operators on ${\cal M}_q$)
could also be written in matrix form. Adding a hat to stress the operator
character of $\hat{L}^{\pm}$ we find
\begin{equation}\label{lopa}
\begin{array}{lll}
(\hat{L}^{\pm}_1 K_2) &=& <L_1^{\pm}, \cdot >\phi(K_2) =
<L_1^{\pm}, M_2K_2 \tilde{M}^{-1}_2> \\

\, &=& < \Delta (L_1^{\pm}), M_2K_2 \tilde{M}^{-1}_2>
=<L_1^{\pm}, M_2>K_2 <L_1^{\pm}, \tilde{M}^{-1}_2> \\

\, &=& R_{12}^{\pm} K_2 A_{12}^{-1} \quad ,
\end{array}
\end{equation}
\noindent
where we have used the duality between product and  coproduct
in the third equal sign, after which $M_2 K_2 \tilde{M}^{-1}_2 $
really means $(M_2)_{ij} \otimes ( \tilde{M}_2^{-1})_{kl}K_{jk}$.
Thus, we obtain
\begin{equation}\label{lopb}
\hat{L}^{\pm}_1 K_2 = R_{12}^{\pm} K_2 A_{12}^{-1} \hat{L}^{\pm}_1 \quad.
\end{equation}

It would be interesting to study which of the above constructed
representations may
be extended to representations of a larger algebra
${\cal M}_q \times {\cal D}_q$ including ${\cal D}_{q}$ as
well as ${\cal M}_{q}$. This algebra is defined by 8 generators (the entries of
$K, Y$) and the relations (\ref{dosj}), (\ref{tc}) and (\ref{xd}). Introducing
explicitly the matrix elements $\partial_I$
[(\ref{cr})]  of $Y$,
one may check that $\partial_{B}$ and $\partial_{C}$ together with
${\cal M}_{q}$ generate a closed subalgebra. Most of the constructed irreps can
be easily extended to this subalgebra. However, these extensions usually have
a singular $q$ dependence for $q$$ \rightarrow$$1$. For instance,
for the one-dimensional representation $\alpha=0$
(in which the representation is, in fact, of the whole
${\cal M}_q\times {\cal D}_q$ algebra) one finds
\begin{equation}\label{sq}
 \partial_{C}=0 \quad,
\partial_{A}= q^{4}/ \gamma (q^{2}-1),\quad
\partial_{B}=q^{2}/ \beta (q^{2}-1),\quad
\partial_{D}=-q \delta /(q^{2}-1).
\end{equation}

It is well
established that covariant algebras such as $SU_{q} (2)$ or the
$q$-sphere $S_{q}^2$ can
be represented as a direct sum of subspaces invariant with respect to the
corresponding quantum groups  coactions: $U' \rightarrow U U'$
for $SU_q(2)$ or $K \rightarrow
UKU^{t}$ for the $q$-sphere (cf. \cite{PODL,NOUMI}). This expansion is related
to the fusion
procedure of the quantum inverse scattering method.
These invariant subspaces in the case of the $SL_{q}
(2)$ covariant RE algebra are generated by the entries of the $(2j + 1) \times
(2j + 1)$ matrices $K^{(j)}$, which correspond to the higher spin
representations of the quantum group from the coaction, e.g.

\begin{equation}\label{sr}
K^{(1)} = P_{+} K_{1} R_{21} K_{2}{\cal P}_{12}=K_{1} \hat{R}_{12} K_{1} P_{+}.
\end{equation}
\noindent
This is associated with the  construction of  $q$-deformed relativistic wave
equations \cite{PILLIN}. There are even some universal $K$ matrices
\cite{GER}  related to a specific representation of ${\cal M}_q^{(1)}$.

\setcounter{equation}{0}

\section{Concluding remarks}

The main aim of this paper was to analyze the
$q$-deformed
Minkowski space-time and
the associated non-commutative differential calculus by
using the
$R$-matrix and the reflection equation formalism. This
permitted us to establish  in a systematic and economic
way many features
of the quantum Minkowski space algebras including
the  complete definition
of ${\cal M}_q$,  the corresponding De Rham complex $ \Lambda_q$
and the algebra of
$q$-derivatives ${\cal D}_q$; the covariance properties of these algebras
under the quantum
Lorentz group transformation (coaction), and the action of
the quantum Lorentz algebra (by duality). A special
basis of generators of ${\cal M}_q^{(1)}$ was defined by using the
$q$-adjoint (co)action
of the quantum (group) algebra; this allowed us to
introduce $q$-sigma
and $q$-gamma matrices as appropriate $q$-tensors.
The possible ambiguities in the definition of
$q$-Minkowski space and consequently in its differential
calculus as well as  some important $q$-algebra isomorphism were   discussed
and   the irreducible representations  of ${\cal M}_q^{(1)}$ algebra
were found.
Also, in the course of the discussion of this case,
a few invariant
(scalar) operators were defined by means of the
$q$-trace, and their
commutation relations among themselves  and with the generators
of   ${\cal M}_q^{(1)}$ and ${\cal D}_q^{(1)}$  were established.

All the previous analyses of the $q$-deformed
space-time were directly formulated in the $q \neq 1$ framework, without
considering a
classical counterpart. The exact $q$-relations may be used, however,
for some constructions in the classical theory.
For instance, it is known that  the quasiclassical limit of
the main ingredients of the
quantum inverse scattering method gives rise
to the classical $r$-matrix and the classical Yang-Baxter equation. If
in the present case we
introduced Planck's constant just by multiplying the defining relations of
the $q$-algebras by $\hbar$, and then we took the independent
limits $q \rightarrow 1,
\hbar \rightarrow 0$, the resulting relations would be nothing but the
standard Poisson brackets for the commuting
coordinates and momenta of the scalar relativistic particle,
$\{ x_{\mu},p_{\nu} \}= g_{\mu \nu}$.
If, on the other hand, the Planck's constant and the
deformation parameter are directly related (e.g.
by setting $q= exp (\gamma \hbar)$, which requires introducing
 an additional dimensional constant in the theory), the Poisson brackets
in the quasiclassical limit are highly nontrivial \cite{FM,K-SKL}.
For instance, writing $R \simeq I+ \gamma \hbar r$ in the quasiclassical limit,
the RE (\ref{dosj}) gives, neglecting the terms in $\hbar^2$,
$$
- \frac{1}{\hbar} [K_1,K_2]= \gamma ( K_1r_{21}K_2 - K_2 r_{12} K_1)+
\gamma (r_{12}K_1K_2- K_2K_1r_{21})
$$
\noindent
so that the Poisson brackets for the classical entries of $K$ (coordinates)
read
\begin{equation}\label{poisson}
\{ K_{1},K_{2} \}= \gamma ([K_{1} \hat{r}_{12} K_{1},{\cal P}_{12}]  +
[\hat{r}_{21}, {\cal P}_{12} K_{1} K_{2} ] )\quad.
\end{equation}

\noindent
In this case  the Poincar\'e group would not be purely geometrical: it
would be dynamical (a Lie-Poisson group
\cite{DRH,DRINF,SEM,TAK}) because its
parameters would have nontrivial Poisson brackets. An
application of the Dirac theory for  constrained
systems results in non-autonomous equations, though with conserved momentum.
This gives rise to additional questions of interpretation if one wished to
preserve the usual mathematical structure of a physical theory (see
\cite{AKRREL} in this respect). An opposite possibility would be to
preserve the trajectories  of the system  generated by the geometric
group action, but change the Hamiltonian description.

To conclude, we wish to come back to  other topics that  were not discussed
in the paper in  detail. The covariance of
a two (multi-) particle system requires
a map from ${\cal M}_q$ to
${\cal M}_q \otimes {\cal M}_q$ which is an algebra
homomorphism (`coproduct'). Two
variants among those given in the
text have reasonable physical behaviour,
reproducing in the classical limit
the sum of the particle coordinates,
although these coordinates do not
commute in between in the case $q \neq 1$.
Both recipes have the drawback
of being asymmetric under interchange of
two particles. These properties    result in a more complicated
representation theory of these quantum algebras.
Another subject just mentioned in the text is the construction of
(free) $q$-relativistic wave
equations \cite{SONG,SCHI-MPI92,WEICH,MEQ}.
For instance, to discuss a physical meaning for the  formal solutions of the
$q$-Klein-Gordon and/or $q$-Dirac equations
in ${\cal M}_q^{(1)}$ we have to construct
irreducible representations of the whole algebra
${\cal M}_q^{(1)} \otimes {\cal D}_q^{(1)}$ (coordinates
and momenta). Then, the $q$-Dirac equation could be defined as
an operator in the corresponding Hilbert space with
an orthodox
interpretation of its spectrum and eigenvectors (wavefunctions).
Another point which requires clarification is the relation between
the relativistic wave equations  in configuration and
momentum space   which in the  classical theory
are connected by the Fourier
transform (for ${\cal M}_q^{(3)}$ see \cite{AKRREL}). A possibility
to discuss the $q$-Fourier transform  is by using the
$*$-quantization which operates in algebra of functions on a phase space
(cf. \cite{FLATO} and refs. therein); other possibilities are
\cite{ZUMF,KEMPF}. Nevertheless,
the difficulties already mentioned prevent us, at present,
to speculate on  a possible
quantum field theory having a  $q$-Minkowski algebra
 as a base space-time.

\vspace{1\baselineskip}
\noindent
{\bf Acknowledgements:} This research has been partially sponsored by
a CICYT (Spain) research grant.
P.P.K. and F.R. wish to thank the DGICYT, Spain,
for financial support; P.P.K. also wishes to thank the hospitality
of the Dept. of Theor. Physics of Valencia University. Helpful discussions with
D. Ellinas and comments of R. Sasaki are also acknowledged.

\appendix

\vspace{1\baselineskip}
\noindent
{\Large {\bf Appendices}}

\setcounter{equation}{0}

\section{Some facts and formulae on quantum groups}

{\bf  A1 \hspace{0.5cm} Notation and useful expressions}

We list here some expressions and conventions that are useful in the main
text. `$RTT$' relations as those in (\ref{uc}), (\ref{dosf}) follow the
usual conventions
i.e., the 4$ \times $4 matrices $T_1$, $T_2$ are the tensor products
\begin{equation}\label{aa}
T_{1} = T \otimes I \quad , \quad T_{2} = I \otimes T \quad .
\end{equation}

\noindent
The tensor product of two matrices, $C= A \otimes B$, reads in components
\begin{equation}\label{ab}
C_{ij,kl} = A_{ik} B_{jl} \quad ,
\end{equation}

\noindent
so that the comma separates the row and column indices of the two matrices.
Thus, $(A_{1})_{ij,kl} = A_{ik} \delta_{jl}\,$; $(A_{2})_{ij,kl}
= A_{jl} \delta_{ik}\,$. The
transposition in the first and second spaces is given by
\begin{equation}\label{ac}
C^{t_{1}}_{ij,kl} = C_{kj,il} \quad , \quad C^{t_{2}}_{ij,kl} = C_{il,kj}
\quad, \end{equation}

\noindent
i.e., $C^{t_{1}} = A^{t} \otimes B$ (resp. $C^{t_{2}}= A \otimes
B^{t}$) is given by a matrix in which the blocks 12 and 21 are
interchanged (each of the four blocks is replaced by its transpose). Of course,
$C_{ij,kl}^{t_{1}t_{2}} = C^{t}_{ij,kl} = C_{kl,ij}$ is the ordinary
transposition. Similarly, the traces in the first and second spaces are given
by
\begin{equation}\label{ad}
(tr_{(1)} C)_{jl} = C_{ij,il} \quad , \quad (tr_{(2)} C)_{ik} = C_{ij,kj}
\quad .
\end{equation}

\noindent
They correspond, respectively, to replacing the $4 \times 4$ matrix
$C$ by the $2 \times 2$ matrix resulting from
adding its two diagonal boxes or by the $2 \times 2$
matrix obtained by taking the trace of
each of its four boxes. If $\,C= A \otimes B$, $tr_{(1)}C= (tr A)B$ and
$tr_{(2)} C= A(tr B)$.

The action of the permutation matrix ${\cal P}_{12} \equiv {\cal P} $
is defined by $({\cal P} C {\cal P})_{ij,kl}=C_{ji,lk}$
(${\cal P} (A \otimes B) {\cal P} = B \otimes A$ if the entries
of $A$ and $B$ commute); thus
\begin{equation}\label{ae}
({\cal P} A_{1} {\cal P})_{ij,kl} = (A_1)_{ji,lk}= A_{jl}
\delta_{ik} = (A_{2})_{ij,kl} \quad ;
\end{equation}

\noindent
$({\cal P} C)_{ij,kl} = C_{ji,kl} \;,\; (C {\cal P})_{ij,kl} =
C_{ij,lk}$. Explicitly, ${\cal P}$=${\cal P}^{-1}$ is given by
\begin{equation}\label{af}
{\cal P}=
\left[
\begin{array}{llll}
1 & \, & \, & \, \\
\, & 0 & 1 & \, \\
\, & 1 & 0 & \, \\
\, & \, & \, & 1
\end{array}
\right]
\quad , \quad {\cal P}_{ij,kl}= \delta_{il} \delta_{jk} \quad;
\end{equation}

\noindent
acting from the left (right) it interchanges the second and third rows
(columns).

   For $SL_q(2)$, the $R_{12} (q) \equiv R_{12}
\equiv R$ and ${\cal P} R_{12} \equiv
\hat{R}_{12}\equiv\hat{R}$ matrices are given by
\begin{equation}\label{af2}
R=
\left[
\begin{array}{llll}
q & \, & \, & \, \\
\, & 1 & 0 & \, \\
\, & \lambda & 1 & \, \\
\, & \, & \, & q
\end{array}
\right]\;,
\quad
 \hat{R}=\left[
\begin{array}{llll}
q & \, & \, & \, \\
\, & \lambda & 1 & \, \\
\, & 1 & 0 & \, \\
\, & \, & \, & q
\end{array}
\right]\;, \quad {\cal P}R{\cal P}=R^t \;,
\end{equation}
\begin{equation}\label{ag}
R_{12} (q^{-1})= R_{12}^{-1} (q) \;,\qquad
\hat{R}_{12}^{-1} (q)= \hat{R}_{21} (q^{-1}) \; ;
\end{equation}

\noindent
where $\lambda \equiv q - q^{-1}$;
$\hat{R}_{21}= {\cal P} \hat{R}_{12} {\cal P}$. Due to the special form of $R$,
${\cal P} R_{12} {\cal P}=R_{21}=R_{12}^t$, but the last equality does not
hold for a general 4$ \times $4 matrix.
$\hat{R}$ satisfies Hecke's condition
\begin{equation}\label{ae2}
\hat{R}^{2} - \lambda \hat{R} - I = 0 \quad , \quad (\hat{R} - q)
(\hat{R} + q^{-1}) =0
\end{equation}

\noindent
and
\begin{equation}\label{af3}
\hat{R} = q P_{+} -q^{-1} P_{-} \;, \quad \hat{R}^{-1} = q^{-1} P_{+}
- q P_{-}\;,\quad
[\hat{R} , P_{\pm} ] =0\; , \quad P_{\pm} \hat{R} P_{\mp} = 0 \;,
\end{equation}

\noindent
where  the projectors $P_{\pm \,12} \equiv P_{\pm}$ are given by
\begin{equation}\label{ag2}
\begin{array}{l}
P_{+} = \displaystyle{\frac{1}{[2]} }
\left[
\begin{array}{cccc}
[2] & 0 & 0 & 0\\
0 & q & 1 & 0\\
0 & 1 & q^{-1} & 0\\
0 & 0 & 0 & [2]
\end{array}
\right] = \displaystyle{\frac{I+q \hat{R}}{1+q^2}} \quad , \\
P_{-}= \displaystyle{\frac{1}{[2]}}
\left[
\begin{array}{cccc}
0 & 0 & 0 & 0\\
0 & q^{-1} & -1 & 0\\
0 & -1 & q & 0\\
0 & 0 & 0 & 0
\end{array}
\right] =  \displaystyle{\frac{I-q^{-1} \hat{R}}{1+q^2}} \quad ,
\end{array}
\end{equation}

\noindent
with $[2] \equiv ( q + q^{-1})$. It is often convenient
to express $P_{-}$ in the form
\begin{equation}\label{ah}
(P_{-})_{ij,kl} = \frac{1}{[2]} \epsilon_{ij}^{q} \epsilon^{q}_{kl} \;,\qquad
([x] \equiv  \frac{q^{x} - q^{-x}}{q - q^{-1}})  \quad,
\end{equation}

\noindent
where $\epsilon^{q}$$=$$-(\epsilon^{q})^{-1} $$\neq $$
(\epsilon^{q})^{t}$  is given in (\ref{dosn}).
 The determinant of an ordinary $2 \times 2$ matrix  may be defined
as the proportionality
coefficient in  $(det M)P_{-} = P_{-} M_{1} M_{2}$ where $P_{-}$ is obtained
from (\ref{ag2}) setting $q$=1. The analogous  definition in  the
$q \neq 1$ case
\begin{equation}\label{adet}
(det_q M)P_{-} := P_{-} M_{1} M_{2}\quad, \qquad
(det_q M^{-1})P_{-} = M_{2}^{-1} M_{1}^{-1}P_-\;,
\end{equation}
\noindent
($det_q M^{-1}= (det_q M)^{-1}$)
leads to the  expression for $det_q M$ given in (\ref{ub}).
For the $K$ matrix, the definition of $det_qK$ is given by (\ref{dosx}),
(\ref{dosy}).

The YBE (\ref{YBE}) is sometimes introduced by reordering $T_1T_2T_3$ to
$T_3T_2T_1$ by two different paths using the `RTT' relation and the
associativity property of the algebra. In this way
one is lead to $(R_{12}R_{13}R_{23}-R_{23}R_{13}R_{12})T_1T_2T_3=
T_3T_2T_1(R_{12}R_{13}R_{23}-R_{23}R_{13}R_{12})$. Thus, eq. (\ref{YBE})
is {\it consistent} with eq. ({\ref{uc}), but it is not implied by it. To
see this explicitly, consider eq. (\ref{uc}) rewritten in the form
$\hat{R}T_1T_2=T_1T_2 \hat{R}$ using (\ref{af}). Then, due to (\ref{af3})
we get ($q^2+1 \neq 0$)
\begin{equation}\label{proj}
P_+T_1T_2P_-=0 \quad , \quad P_-T_1T_2P_+=0 \quad.
\end{equation}
\noindent
The first equation implies $ab-qba$=0, $cd-qdc$=0 and $[a,d]$=$qbc-q^{-1}cb$,
while the second gives $ac-qca$=0, $bd-qdb$=0 and $[a,d]$=$qcb-q^{-1}bc$.
In all, these equations reproduce (\ref{ua}). These equations also follow
from $P_+T_1T_2$=$ T_1T_2P_+$, {\it i.e.} from an `RTT' relation with $P_+$
as an $\hat{R}$-matrix,  although ${\cal P}P_+$ is not invertible
and does not satisfy the YBE (\ref{YBE}) (see also \cite{WZ}).

\vspace{1\baselineskip}

\noindent
{\bf  A2 \hspace{0.5cm} Hopf algebra duality}

We recall here some expressions on the quantum groups and quantum algebras
duality. Let $H$ and $H^*$ be a pair of dual Hopf algebras. Then, there
is a {\it pairing} map  $< \cdot , \cdot >$:
$H^* \times  H \mapsto C$, consistent with the commutation relations
in each algebra, which satisfies:

\begin{equation}\label{dua}
\begin{array}{l}
1) \quad <XY,a>= < X \otimes Y, \Delta (a)> \quad, \qquad
<1,a>= \varepsilon (a) \;,  \\

2) \quad <X,ab>= < \Delta (X), a \otimes b> \quad, \qquad
<X,1>= \varepsilon (X) \;, \\

3) \quad <S(X), a>=<X, S(a)> \quad,
\end{array}
\end{equation}
\noindent
where $X,Y \in H^*$, $a,b \in H$ and $ \Delta, \varepsilon, S$ are the usual
notations for the coproduct, counit and antipode in Hopf algebras
(see, e.g., \cite{TAK,PC,MAJID}).
In particular, if $H$ and $H^*$ are respectively a quantum group {\it \`a la}
FRT \cite{FRT1}  and its quantum dual algebra, the generators of $H$ are the
entries of
a matrix $T$ satisfying a `RTT' relation and the generators of $H^*$ are
arranged
in two triangular matrices $L^{\pm}$ satisfying
\begin{equation}\label{lmatr}
R^+L_1^{\pm}L_2^{\pm}=L_2^{\pm}L_1^{\pm}R^+ \quad,\quad R^+L_1^+L_2^-=
L_2^-L_1^+R^+
\end{equation}
\noindent
where $R^+= {\cal P}R{\cal P}=R_{21}$.
In this case, the pairing is defined
in terms of the corresponding quantum group $R$-matrix by the expressions
\begin{equation}\label{dub}
<L^{\pm},T>= R^{\pm} \quad, \qquad <L^{\pm}, 1> =I=<1, T> \quad,
\end{equation}
\noindent
where   $R^-=R^{-1}$. This definition is extended
to higher order monomials by
\begin{equation}\label{duc}
\begin{array}{l}
<1\;,\; T_1\,T_2 \,...\,T_k>= I^{ \otimes k}
= <L^{\pm}_1\,L^{\pm}_2\,...\,L^{\pm}_k\;,\; 1> \quad, \\

<L^{\pm}_1\,L^{\pm}_2\,...\,L^{\pm}_k\;,\; T_{k+1}>= R^{\pm}_{1\, k+1}\,
R^{\pm}_{2\, k+1} \,...\,R^{\pm}_{k\, k+1} \quad , \\

<L^{\pm}_1 \;,\; T_2\,T_3 \,...\,T_{k+1}> =
R^{\pm}_{12}\,R^{\pm}_{13}\,...\,R^{\pm}_{1\, k+1}\quad.
\end{array}
\end{equation}
\noindent
This is consistent with the commutation relations defining the
quantum group and its dual algebra and satisfies the properties (\ref{dua}).

The pairing may be used to define the fundamental representation
of the generators of the quantum algebra; for each
$l_{ij}^{\pm}$ (entries of $L^{\pm}$) the following map is defined
\begin{equation}\label{dud}
<l_{ij}^{\pm}, \cdot > : H \longmapsto C \quad, \qquad
<l_{ij}^{\pm}, t_{kl}>= R^{\pm}_{ik,jl} \quad,
\end{equation}
\noindent
then, $<l_{ij}^{\pm}, T>$ is a matrix which constitutes the fundamental
representation of the generators $l_{ij}^{\pm}$; in general, $<X, T>$
is the representation of $X$. This representation for the generators
$J_{\pm}$, $J_3$ of $su_q(2)$ was used in Sec. 4 to check that the
given $q$-sigmas constitute
a $q$-tensor operator according to (\ref{ce}).

\setcounter{equation}{0}

\section{Proof of some properties of $q$-Minkowski algebras}

{\bf  B1 \hspace{0.5cm} The algebras defined by eq. (\ref{dosap}),
(\ref{td})}

We now analyze the algebra (\ref{dosap}), and why it may be
discarded. First,
it is easy to explain the presence of the factor $q^{2}$, which is due to the
imbalance of the $\hat{R}$ and $\hat{R}^{-1}$ factors; notice that eqs.
(\ref{dosal}) allow for a proportionality constant in the definition of
$R^{(i)}$.
Consider (\ref{dosap}) written in the form
\begin{equation}\label{aq}
\hat{R} K_{1} \hat{R} K_{1} = \rho K_{1} \hat{R} K_{1} \hat{R}^{-1}\quad,
\end{equation}

\noindent
where $\rho$ is a constant to be determined. We may now
multiply this equation by $P_{\pm}(\quad ) P_{\pm}$ and by $P_{\pm} (\quad )
P_{\mp}$. Using eq. (\ref{af3}) we obtain that $P_{+} (\quad )P_{+}$ gives
\begin{equation}\label{ar}
q P_{+} K_{1} \hat{R} K_{1} P_{+} = \rho P_{+} K_{1} \hat{R} K_{1} q^{-1}
P_{+}
\end{equation}

\noindent
which fixes $\rho = q^{2}$. Then, $P_{-}(\quad )P_{-}$ gives
\begin{equation}\label{as}
-q^{-1} P_{-} K_{1} \hat{R} K_{1} P_{-} = q^{2} P_{-} K_{1} \hat{R} K_{1}
P_{-} (-q) \quad ,
\end{equation}

\noindent
which implies
\begin{equation}\label{at}
(q^{3} - q^{-1}) P_{-} det_{q} K = 0
\end{equation}

\noindent
  so that for this algebra ($q^4 \neq 1$) $det_{q} K=0$. The commutation
properties of the
entries of $K$, however, are the same as for (\ref{dosj}).
Using $q^2 \hat{R}^{-1}$=$ \hat{R}-
q \lambda [2]P_-$, eq. (\ref{dosap}) reproduces (\ref{dosk}),
\begin{equation}\label{at2}
\hat{R} K_{1} \hat{R} K_{1} = K_1 \hat{R} K_{1} \hat{R}
-q \lambda [2]K_1 \hat{R} K_{1}P_- = K_1 \hat{R} K_{1} \hat{R}\;,
\end{equation}
\noindent
since $det_qK=0$. This means that the
algebra generated by the  entries of $K$ in (\ref{dosap}) may be obtained by
restricting ${\cal M}_{q}^{(1)}$ (eq. (\ref{dosj})) by the condition
$det_{q}K=0$
so that  nothing is gained by considering (\ref{dosap}) as a separate case.
Notice that the same
reasonings applied to (\ref{dosj}) do not give any condition for $P_{\pm}
(\quad )P_{\pm}$, which is satisfied identically, and $P_{\pm} (\quad )P_{\mp}$
give, as for (\ref{dosk}), $P_{\pm} K_{1} \hat{R} K_{1} P_{\mp} =0$.

The factor
$q^{2}$ in (\ref{td}) is explained in a similar way; again, this algebra
corresponds to the too restrictive condition $det_{q} Y = 0.$

We now derive here the  expressions for
$K^{\epsilon}$ and $Y^{\epsilon}$ used in Sec. 5,
\begin{equation}\label{bh}
K_{1}^{\epsilon} = [2] tr_{q(2)} ( \hat{R}_{12} K_{1} P_{-\,12}) \quad,
\end{equation}
\begin{equation}\label{bi}
Y_{1}^{\epsilon} = [2] tr_{q(2)} (P_{-\,12} Y_{1} \hat{R}_{12}^{-1})\quad.
\end{equation}

\noindent
Obviously, the covariant vector $dK^{\epsilon}$ has an expression (\ref{ciu})
analogue (\ref{bh}). For instance, the r.h.s. of (\ref{bh}) reads
in  explicit component notation
$$
[2] (tr_{q(2)} (\hat{R}_{12} K_{1} P_{-\,12}))_{ik} = [2] D_{jb}
\hat{R}_{ib,am} K_{ac} P_{-cm, kj} \quad.
$$

\noindent
Using the expression (\ref{ah}) and
$D= \epsilon^{q} z \epsilon^{qt}$, the above expressions become equal to
\begin{equation}\label{bj}
 - \epsilon^{q}_{bk} \hat{R}_{ib,am} \epsilon^{q}_{cm} K_{ac} =
(\epsilon^{q})_{kb}^{t} \hat{R}_{ib,am} (\epsilon^{q})_{mc}^{t-1} K_{ac}
= \hat{R}^{\epsilon}  \, _{ik,ac} K_{ac} = K_{ik}^{\epsilon}
\end{equation}

\noindent
by (\ref{dosab})  and (\ref{dosaf}).
Eq. (\ref{bi}) is checked similarly by using
$\hat{R}_{12}^{\epsilon}$$=$$\hat{R}_{21}^{\epsilon}$.

\vspace{1\baselineskip}
\noindent
{\bf  B2 \hspace{0.5cm} Algebra isomorphisms for ${\cal M}_q^{(1)}$,
${\cal D}_q^{(1)}$}

Let us first  check that
\begin{equation}\label{ba}
K(q^{-1}) \equiv K' \approx K^{\epsilon} \;,\; \quad K'=
\left(
\begin{array}{ll}
\alpha' & \beta'\\
\gamma' & \delta'
\end{array}
\right) \quad,
\end{equation}

\noindent
i.e., that the commutation properties of the entries of $K'$ (which
generate the same algebra as those of $K$ but with
$q$  replaced by $q^{-1}$) are those of the elements of $K^{\epsilon}$.
The reflection equation for $K' \equiv K (q^{-1})$ is given by (cf. eq.
(\ref{dosj}))
\begin{equation}\label{bb}
\hat{R}_{12} (q^{-1}) K_{1}' \hat{R}_{12} (q^{-1}) K_{1}' = K_{1}' \hat{R}_{12}
(q^{-1}) K_{1}' \hat{R}_{12} (q^{-1})  \quad,
\end{equation}

\noindent
and  since $\hat{R}_{12} (q^{-1}) = \hat{R}_{21}^{-1} (q)$, this is equivalent
to
\begin{equation}\label{be}
\hat{R}_{21} K_{1}' \hat{R}_{21}^{-1} K_{1}' = K_{1}' \hat{R}_{21}^{-1} K_{1}'
\hat{R}_{21} \quad .
\end{equation}

\noindent
Now we notice that, due to the specific form of the $\hat{R}$ matrices,
a similarity transformation with
 $(\sigma^{1} \otimes \sigma^{1})$ is
equivalent to the action of the permutation operator ${\cal P},\; (\sigma^{1}
\otimes \sigma^{1}) \hat{R}_{12} (\sigma^{1} \otimes \sigma^{1}) =
\hat{R}_{21}$.
Also, $(\sigma^{1} \otimes \sigma^{1}) K_{i} (\sigma^{1} \otimes \sigma^{1})=
(\sigma^{1} K \sigma^{1})_{i}, i=1,2$. Thus, eq. (\ref{be}) gives
\begin{equation}\label{bd}
\hat{R}_{12} (\sigma^{1} K' \sigma^{1})_{1} \hat{R}_{12}^{-1} (\sigma^{1} K'
\sigma^{1})_{1} = (\sigma^{1} K' \sigma^{1})_{1} \hat{R}_{12}^{-1}
(\sigma^{1} K' \sigma^{1})_{1} \hat{R}_{12}\;.
\end{equation}

\noindent
Comparing with (\ref{tc}), which as we know is the same reflection equation
satisfied by $K^{\epsilon}$ because of its transformation properties
(\ref{dosaf2}), we find that there is an isomorphism among the algebras
generated
by $K$ and $K'$ given by
\begin{equation}\label{be2}
K \longmapsto K' = \sigma^{1} K^{\epsilon} \sigma^{1} \equiv \sigma^{1}
(\hat{R}^{\epsilon} K) \sigma^{1} \quad .
\end{equation}

\noindent
Since we also know that $K^{\epsilon} \approx Y$, we see that the linear
mappings which relate the entries of  $K',K$ and $Y$ define isomorphisms
between the algebras generated  by their entries. Specifically, the elements of
\begin{equation}\label{bf}
\begin{array}{ccc}
K'=
\left[
\begin{array}{cc}
\alpha' & \beta'\\
\gamma' & \delta'
\end{array}
\right]
\; , &\; \sigma^{1} K^{\epsilon} \sigma^{1} =
\left[
\begin{array}{cc}
-q^{-1} \alpha & q \gamma\\
q \beta & \lambda \alpha - q \delta
\end{array}
\right] \; , \; &
\sigma^{1} Y \sigma^{1}=
\left[
\begin{array}{cc}
z & w\\
v & u
\end{array}
\right] \;,  \\
  \, &   \,   &   \, \\
{\cal M}_{q^{-1}}^{(1)} & {\cal M}_{q}^{(1)}   &   {\cal D}_{q}^{(1)}
\end{array}
\end{equation}

\noindent
have the same commutation relations. An analogous argument shows the matrix
elements of
\begin{equation}\label{bg}
\begin{array}{ccc}
Y'=
\left[
\begin{array}{cc}
u' & v'\\
w' & z'
\end{array}
\right]
\; , \; &
\sigma^{1} Y^{\epsilon}\sigma^{1}=
\left[
\begin{array}{cc}
-q^{-1} u -\lambda z & q^{-1} w\\
q^{-1} v & -q z
\end{array}
\right] \; , \; &
\sigma^{1} K \sigma^{1}=
\left[
\begin{array}{cc}
\delta & \gamma\\
\beta & \alpha
\end{array}
\right] \,, \\
  \,   &   \,   &  \, \\
{\cal D}_{q^{-1}}^{(1)}    &   {\cal D}_{q}^{(1)}    &   {\cal M}_{q}^{(1)}
\end{array}
\end{equation}

\noindent
where $Y' \equiv Y (q^{-1}),$ also have the same commutation properties.
Identifying $K$ with the $q$-Minkowski coordinates and $Y$ with the
derivatives, eqs. (\ref{bf}) and (\ref{bg}) show that
${\cal M}_{q^{-1}}^{(1)} \approx {\cal M}_{q}^{(1)} \approx {\cal D}_{q}^{(1)}
 \approx {\cal D}_{q^{-1}}^{(1)}.$

\vspace{1\baselineskip}
\noindent
{\bf  B3 \hspace{0.5cm} $q$-trace and $q$-determinant for ${\cal M}_q^{(i)}$}

We give here a general expression for $tr_q$ and $det_q$ which apply to all
the cases considered. The $q$-trace of a matrix $B$ is defined by
(cf. \cite{FRT1,ZU-MPX})
\begin{equation}\label{gtr}
tr_q(B)=tr(DB) \quad , \quad
D=q^2tr_{(2)}({\cal P}(( (R^{(1)})^{t_1})^{-1})^{t_1}) \quad,
\end{equation}
\noindent
where $q$ is the deformation parameter in $R^{(1)}$ (see (\ref{dosal})).
This $q$-trace is invariant under the quantum group coaction
$B \mapsto MBM^{-1}$
(as well as under the coaction $C \mapsto \tilde{M}C\tilde{M}^{-1}$
since $R^{(1)}$=$R^{(4)\,t}$=${\cal P}R^{(4)}{\cal P}$, for $q$ real;
this condition is required for the consistency of (\ref{dosal}) with
$\tilde{M}^{-1}=M^{\dagger}$).

Supposing that $\hat{R}^{(1)}$=${\cal P}R^{(1)}$ has a spectral
decomposition like (\ref{af3})
with a rank three projector $P_+$ and a rank one projector $P_-$, and
that $det_qM$ (\ref{adet}) and $det_q \tilde{M}$ are central (which is true
in all our  cases), the $q$-determinant of the 2$\times$2 matrix $K$ is
given by the expression
\begin{equation}\label{gdet}
(det_qK)P_-= -q  P_- K_1 \hat{R}^{(3)}K_1 P_- \quad.
\end{equation}
\noindent
Substituting $\hat{R}_{12}$, ${\cal P}$ or $\hat{V}$ for $\hat{R}^{(3)}$
in (\ref{gdet}) we obtain the square of the $q$-Minkowski length
for ${\cal M}_q^{(1)}$ [(\ref{dosx})], ${\cal M}_q^{(2)}$ [(\ref{dosaq4})]
and ${\cal M}_q^{(3)}$ [below (\ref{dosau})] respectively
(recall that the $q$ in (\ref{gdet}) is the parameter in $R^{(1)}$ and
that $R^{(3)}$ may depend on other parameters).

As we saw, it is possible to write the invariant scalar product
as the $q$-trace of the matrix product of contravariant and covariant
matrices (vectors). The relation with the previous expressions in terms of
the $q$-determinant in the 2$\times$2  case is given by the normalized
`eigenvector' $\varepsilon_{ij}$ of $P_-$ with eigenvalue 1,
$(P_-)_{ij,kl}=\varepsilon_{ij}\varepsilon_{kl}$; then $D \propto
\varepsilon^{-1}\varepsilon^{t}$ where in this expression $\varepsilon$
is treated as a 2$\times $2 matrix (for the $SL_q(2)$ $R$-matrix,
$\varepsilon \propto \epsilon^q$, eq. (\ref{dosn}), and $P_-$ is given
by (\ref{ah})). The contravariant vector $K^{\epsilon}$ is written in
general as in eq. (\ref{dosaf}) with $\hat{R}^{\epsilon}=
(1 \otimes \varepsilon^t) \hat{R}^{(3)} (1 \otimes (\varepsilon^{-1})^t)$.


\begin{thebibliography}{99}

\bibitem{DRINF} V.G. Drinfel'd, in  Proc. of the 1986
{\it Int. Congr. of Mathematicians}, MSRI Berkeley, vol. {\bf I}, 798 (1987)
(A. Gleason, ed.)
\bibitem{JIM} M. Jimbo,  Lett. Math. Phys. {\bf 10}, 63 (1985); ibid {\bf 11},
247 (1986)

\bibitem{FRT1} L.D. Faddeev, N. Yu. Reshetikhin and L.A. Takhtajan,
Alg. i Anal. {\bf 1}, 178 (1989) (Leningrad Math. J. {\bf 1}, 193 (1990))


\bibitem{WOR1} S. L. Woronowicz,  Publ. RIMS Kyoto Univ. {\bf 23},
117 (1987)

\bibitem{POWO} P. Podle\'s  and S. Woronowicz, Commun. Math. Phys. {\bf 130},
381 (1990)

\bibitem{WA-ZPC48} U. Carow-Watamura, M. Schlieker, M. Scholl and
S Watamura, Z.Phys. {\bf C48}, 159 (1990)

\bibitem{WA-A6} U. Carow-Watamura, M. Schlieker, M. Scholl and
S. Watamura, Int. J. Mod. Phys. {\bf A6}, 3081 (1991)

\bibitem{SWZ-ZPC52} W. Schmidke, J. Wess and B. Zumino, Z. Phys.
{\bf C52}, 471 (1991)

\bibitem{OSWZ-CMP} O. Ogievetsky, W.B. Schmidke, J. Wess and B. Zumino, Commun.
Math. Phys. {\bf 150},  495 (1992)

\bibitem{SONG} X-C. Song,  Z. Phys.  {\bf C55}, 417 (1992)

\bibitem{VZW}S.P. Vokos, B. Zumino
and J. Wess, Z. Phys. {\bf C48}, 65 (1990)

\bibitem{DAVID} E. Corrigan, D.B. Fairlie, P. Fletcher and R. Sasaki,
J. Math. Phys. {\bf 31}, 776 (1990)

\bibitem{FADDEEV} L.D. Faddeev, Sov. Sci. Rev., Sec. C (Math. Phys.) {\bf 1},
107 (1980);
P.P. Kulish and E.K. Sklyanin, Lect. Notes in Phys. {\bf 151}, 61 (1981);
A.G. Izergin and V.E. Korepin, Fisika EChAYa {\bf 13}, 501 (1982);
F.A. Smirnov, {\it Form factors in completely integrable models},
World Sci. (1992);
E.K. Sklyanin, in   {\it Quantum groups and quantum integrable systems},
(M.-L. Ge ed.), World Sci. (1992), p. 63

\bibitem{TAK} L. A. Takhtajan, {\it Lectures in quantum groups}, in Nankai
Lectures
in Math. Phys. (M.-L. Ge and B.-H. Zhao eds.),
World Sci. (1990), p. 69

\bibitem{PC} V. Chari and A. Pressley, Nucl. Phys. (Proc. Suppl.) {\bf 18A},
207 (1990); {\it A guide to quantum groups}, CUP (1994)

\bibitem{MAJID}  S. Majid, Int. J. Mod. Phys. {\bf A5}, 1 (1990)

\bibitem{DOE} H. D. Doebner, J. D. Henning and W. L\"ucke, {\it
Mathematical guide to quantum groups},  in Proc. of the Clausthal
Int. Workshop on Math. Phys., H.-D.
Doebner and J.-D. Henning Eds., Springer-Verlag (1990), p.29


\bibitem{MANIN} Yu. I. Manin, Commun. Math. Phys. {\bf 123}, 163 (1989);
{\it Topics in non-commutative geometry},
Princeton Univ. Press (1991)

\bibitem{WZ} J. Wess and B. Zumino, Nucl. Phys. (Proc. Suppl.) {\bf 18B},
302-312 (1990);
\\ See also the contributions
of J. Wess and  B. Zumino  to
the Proc. of the {\it XIX Int. Coll. on Group Theor. Methods in Phys.},
Salamanca, 1992
(M.A. del Olmo , M. Santander  and J. Mateos Guilarte eds.), Anales
de F\'{\i}sica: Monograf\'{\i}as {\bf 1}, vol. I, pp. 33 and 41
respectively (1993)

\bibitem{K-SKL} P.P. Kulish  and  E.K. Sklyanin, J. Phys. {\bf A25},
5963 (1992)

\bibitem{KS} P.P. Kulish and R. Sasaki,
Progr. Theor. Phys. {\bf 89}, 741 (1993)

\bibitem{K2} P.P. Kulish, Theor. and Math. Phys. {\bf 94}, 137 (1993)
(Teor. i Mat. Fiz. {\bf 94}, 193 (1993))


\bibitem{CHERED} I. Cherednik, Theor. and Math. Phys. {\bf 61}, 55
(1984)

\bibitem{SKL} E. Sklyanin, J. Phys. {\bf A21}, 2375 (1988)

\bibitem{LN} L. Mezincescu and R. Nepomechie, J. Phys. {\bf A24}, L17 (1991)

\bibitem{MAJ-LNM} Majid S., in {\it Quantum Groups}, Lect. Notes Math.
{\bf 1510}, 79 (1992) (P.P. Kulish, ed.);  J. Math. Phys. {\bf 32}, 3246 (1991)

\bibitem{SMBR2} S. Majid, J. Math. Phys. {\bf 34}, 1176  (1993); J. Geom.
Phys. {\bf 13}, 169 (1994)

\bibitem{FM} L. Freidel and J. M. Maillet, Phys. Lett. {\bf B262}, 278 (1991)

\bibitem{MSC} G. Mack and V. Schomerus, Nucl. Phys. {\bf B370}, 185 (1992)

\bibitem{RSM} N. Yu. Reshetikhin and F. Smirnov, Commun. Math. Phys. {\bf 131},
157 (1990)

\bibitem{PSW} M. Pillin, W.B. Schmidke and J. Wess, Nucl. Phys.
{\bf B403}, 223 (1993); {\it Proceed. XXVI Inter. Symp. Theory of
Elem. Particles}, DESY 93-013,72, 99 (1993)


\bibitem{PILLIN} M. Pillin, J. Math. Phys. {\bf 35}, 2804 (1994)

\bibitem{SCHI-MPI93} A. Schirrmacher, Lett. Math. Phys. {\bf 29}, 329 (1993)

\bibitem{WEICH} W. Weich, {\it Quantum mechanics with
$SO_q(3)$-symmetry},  M\"unchen   LMU-TPW 93-27


\bibitem{WA-ZPC49} U. Watamura, M. Schlieker and S. Watamura,
Z. Phys. {\bf C49}, 439 (1991)

\bibitem{OZ-MPI25} O. Ogievetsky and B. Zumino,  Lett. Math. Phys. {\bf 25},
121 (1992)

\bibitem{AKRREL} J.A. de Azc\'{a}rraga, P.P. Kulish and F. Rodenas,
{\it On the physical contents of $q$-deformed Minkowski spaces},
FTUV-94-64, IFIC-94-61, hep-th 9411121, to appear in Phys. Lett. {\bf B} (1995)


\bibitem{CESAR} C. G\'omez and G. Sierra, Phys. Lett. {\bf B255}, 51 (1991)

\bibitem{DOB} V. K. Dobrev, {\it $q$-deformations of non-compact Lie
(Super-) Algebras: the examples of $q$-deformed Lorentz, Weyl,
Poincar\'{e} and (Super-) conformal algebra}, in
Proc. of the II Wigner Symposium, Goslar (1991),  Springer Verlag;
Anales de  F\'{\i}sica (Monograf\'{\i}as), {\bf 1} vol. 1, 91 (1993)

\bibitem{CAR} M. Chaichian, J. A. De Azc\'arraga
and F. Rodenas, {\it $q$-Fock space representations of the
$q$-Lorentz algebra and irreducible tensors},
in {\it Symmetries  in Science VI }, B.Gruber ed., Plenum Press (1994),
p. 157

\bibitem{PUSZ} W. Pusz, Commun. Math. Phys. {\bf 152}, 591 (1993)

\bibitem{CAST} L. Castellani, Phys. Lett. {\bf B298}, 335 (1993)

\bibitem{SMPoinc} S. Majid, J. Math. Phys. {\bf 34}, 2045 (1993)

\bibitem{BHOS} A. Ballesteros, F. J. Herranz, M. A. del Olmo and
M. Santander, J. Math. Phys. {\bf 35}, 4928 (1994); J. Phys. {\bf A27},
1283 (1994)

\bibitem{PWOR} P. Podle\'s and S. L. Woronowicz, {\it On the
classification of quantum Poincar\'e groups}, March 1995, hep-th 9412059

\bibitem{FIR} E. Celeghini, R. Giachetti, E. Sorace and M. Tarlini,
J. Math. Phys. {\bf 31}, 2548 (1990); ibid. {\bf 32}, 1155, 1159 (1991)

\bibitem{LNRT} J. Lukierski,  A. Nowicki, H. Ruegg  and V.N. Tolstoi,
Phys. Lett. {\bf B264}, 331 (1991);
J. Lukierski,  A. Nowicki and  H. Ruegg,
Phys. Lett. {\bf B293}, 344 (1992); J. Geom. Phys. {\bf 11}, 425 (1993)

\bibitem{RUEGG} H. Ruegg, {\it $q$-deformation of semisimple and simple
algebras}, in {\it Integrable systems, quantum groups and quantum field
theories} (Salamanca 1992), L. A. Ibort and M. A. Rodriguez eds., NATO-ASI
{\bf 409}, Kluwer (1993) p. 45

\bibitem{LRZ} J. Lukierski, H. Ruegg and W. J. Zakrzewski, {\it Classical
and quantum mechanics of free $\kappa$-relativistic systems},
hep-th 9312153, to appear in Ann. Phys.

\bibitem{MARU} S. Majid and H. Ruegg, Phys. Lett. {\bf B334}, 348 (1994)

\bibitem{SITARZ} A. Sitarz, Phys. Lett. {\bf B349}, 42 (1995)

\bibitem{AZPB} J. A. de Azc\'arraga and J. C. P\'erez Bueno, {\it
Relativistic and non-relativistic $\kappa$-spacetimes}, FTUV/95-12,
IFIC/95-12, April 1995, q-alg/9505004

\bibitem{WOZA} S.L. Woronowicz and S. Zakrzewski, Compositio Math.
{\bf 94}, 211 (1994)

\bibitem{MeyerM}  U. Meyer, {\it The $q$-Lorentz group and braided
coaddition on $q$-Minkowski space}, DAMTP 93-45 (revised, Jan. 1994),
to appear in Commun. Math. Phys.

\bibitem{AKR} J.A. de Azc\'{a}rraga, P.P. Kulish and F. Rodenas,
Lett. Math. Phys. {\bf 32}, 173 (1994)

\bibitem{SM2}  S. Majid and U. Meyer, Z. Phys. {\bf C63}, 357 (1994)

\bibitem{ZU-MPX} B. Zumino, {\it Introduction to the differential geometry
of  quantum groups},
in  {\it Mathematical Physics X},  K.  Schm\"udgen ed.,
Springer-Verlag (1992), p. 20


\bibitem{SONG-JPA} X-C. Song, J. Phys. {\bf A25}, 2929 (1992)

\bibitem{DRSWZ} B. Drabant, M. Schlieker, W. Weich and B. Zumino,
Commun. Math. Phys. {\bf 147}, 625 (1992)

\bibitem{PRAGA} J.A. de Azc\'{a}rraga, P.P. Kulish and F. Rodenas,
Czech. J. Phys. {\bf 44}, 981 (1994)

\bibitem{MajEucl} S. Majid, J. Math. Phys. {\bf 35}, 5025 (1994)

\bibitem{MAJCLA} S. Majid, {\it Some remarks of the $q$-Poincar\'e algebra in
$R$-matrix form}, DAMTP/95-08, $q$-alg/9502014

\bibitem{CHA-DEM} M. Chaichian and A. Demichev, Phys. Lett. {\bf B304},
220 (1993); J. Math. Phys. {\bf 36}, 398 (1995)
\bibitem{LRT} J. Lukierski, H. Ruegg, V.N. Tolstoi and A. Nowicki,
J. Phys. {\bf A27}, 2389 (1994)

\bibitem{DTW} V. Drinfel'd, Alg. i Anal. {\bf 1}, 30 (Leningrad Math. J.
{\bf 1}, 1419 (1990))

\bibitem{RTW} N. Reshetikhin, Lett. Math. Phys. {\bf 20}, 331 (1990)

\bibitem{SLWO} S. L. Woronowicz, Commun. Math. Phys. {\bf 122}, 125 (1989)

\bibitem{DIFF} P. Schupp, P. Watts and B. Zumino, Lett. Math. Phys.
{\bf 25}, 139 (1992)

\bibitem{MANIN2} Yu. I. Manin, Theor. and Math. Phys. {\bf 92}, 997 (1993)
(Theor. and Math Fiz. {\bf 92}, 425 (1992))

\bibitem{SUD} A. Sudbery, Phys. Lett. {\bf B284}, 61 (1992)

\bibitem{FP} L. D. Faddeev and P. N. Pyatov, {\it The differential calculus
on quantum linear groups},  preprint (1993), hep-th 9402070

\bibitem{K-POMI} P.P. Kulish, Zap. Nauch. Semin. POMI {\bf 205}, 66 (1993)

\bibitem{BITAR} L. C. Biedenharn and M. Tarlini, Lett. Math. Phys.
 {\bf 20}, 271 (1990);  L. C. Biedenharn, {\it A $q$-boson
realization of the quantum group
$su (2)_{q}$ and the theory of $q$-tensor
operators}, in Proc. of the Clausthal
Int. Workshop on Math. Physics, H.-D.
Doebner and J.-D. Henning eds., Springer-Verlag (1990), p.67


\bibitem{RITS} V. Rittenberg and  M. Scheunert,
J. Math. Phys. {\bf 33}, 436 (1992)


\bibitem{FENG} Feng Pan, J. Phys. {\bf A24}, L803 (1991);
L. K. Hadjiivanov, R. R.  Paunov and I. T. Todorov,  J. Math. Phys. {\bf 33},
1379 (1992)

\bibitem{SCHI-MPI92} A. Schirrmacher, {\it Quantum groups, quantum space-time
and Dirac equation}, MPI-PTh/92-92 (1992). Talk given at the workshop on
{\it Low dimensional topology and quantum field theory}, Newton Inst.
(Cambridge), Sep. 1992

\bibitem{MEQ} U. Meyer, {\it Wave equations on $q$-Minkowski spaces},
DAMTP 94-10 (1994), hep-th 9404054

\bibitem{ABR} U. H. Niederer and
L. O'Raifeartaigh, Forts. der Phys. {\bf 22},  111, 131  (1974);
J.A. de Azc\'{a}rraga and L. J. Boya, J. Math. Phys. {\bf 9}, 1689 (1968)

\bibitem{JW} J. Wess, {\it $q$-deformed quantization}, LMU
M\"unchen preprint (1994)

\bibitem{K-TMF} P.P. Kulish, Teor. Matem. Fiz. {\bf 86}, 157 (1991)
(Theor. Math. Phys. {\bf 86}, 108 (1991))

\bibitem{RID} G. Rideau, Lett. Math. Phys. {\bf 24}, 147 (1992)


\bibitem{OSWZ-MPI51} O. Ogievetskii, W.B. Schmidke, J. Wess and B. Zumino,
Lett. Math. Phys. {\bf 23}, 233 (1991)

\bibitem{PPK} P.P. Kulish,  Alg. i Anal. {\bf 6}, 195  (1994)

\bibitem{PODL} P. Podle\'s, Lett. Math. Phys. {\bf 14}, 193 (1987)

\bibitem{NOUMI} M. Noumi and K. Mimachi, Duke Math. Jour. {\bf 63}, 65 (1991)

\bibitem{ALEK} A. Alekseev, L.D. Faddeev and M.A. Semenov-Tian-Shansky, in
{\it Quantum Groups}, Lect. Notes Math. {\bf 1510}, 148 (1992)
(P.P. Kulish, ed.)
\bibitem{ISVL} A. P. Isaev and A. A. Vladimirov, {\it $GL_q(N)$-covariant
braided differential bialgebras}, JINR Dubna (1993), hep-th 9402024

\bibitem{GER} E. Cremmer and J. -L. Gervais, Commun. Math. Phys. {\bf 144},
279 (1991)

\bibitem{DRH} V.G. Drinfel'd, Sov. Math. Dokl. {\bf 27}, 68 (1983)

\bibitem{SEM} M.A. Semenov-Tian-Shansky, Publ. RIMS {\bf 21}, 1237 (1985)


\bibitem{FLATO} M. Flato and D. Sternheimer, Lett. Math. Phys. {\bf 22}, 155
(1991); Alg. i
Anal. {\bf 6}, (1994)

\bibitem{ZUMF} C. Chrissomalakos and B. Zumino, {\it Translations,
integrals and Fourier transform in the quantum plane}, UCB-PTH 93/30

\bibitem{KEMPF} A. Kempf and S. Majid, J. Math. Phys. {\bf 35}, 6802 (1994)

\end{thebibliography}
\end{document}